\documentclass[a4paper,10pt]{article}
\pdfoutput=1 

\usepackage{jheppub} 

\usepackage[T1]{fontenc} 
\usepackage{macros_JHEP}
\usepackage{ytableau}
\Yboxdim{6pt}

\graphicspath{{./PT4Figures/}}

\title{\boldmath The 4-fold Pandharipande--Thomas vertex and Jeffrey--Kirwan residue}

\author[a]{Taro Kimura,}
\author[b,c]{Go Noshita}

\affiliation[a]{Université Bourgogne Europe, CNRS, IMB UMR 5584, Dijon, France }
\affiliation[b]{Department of Physics, Institute of Science Tokyo, Tokyo, Japan }
\affiliation[c]{Center for High Energy Physics, Peking University, Beijing, China}

\emailAdd{taro.kimura@ube.fr }
\emailAdd{gnoshita969hep@gmail.com}

\abstract{ We present a contour integral formalism for computing the K-theoretic equivariant Pandharipande--Thomas (PT) 4-vertex. Within the Jeffrey--Kirwan (JK) residue framework, we show that the PT 4-vertex can be obtained from the same integrand as the Donaldson--Thomas (DT) 4-vertex by choosing a different reference vector. We illustrate the formalism through examples involving curves and surfaces on the 4-fold. Furthermore, we investigate the DT/PT correspondence for the 4-fold setting together with its higher rank and supergroup-like generalizations.}
\makeatletter
\gdef\@fpheader{\phantom{}}
\makeatother
\preprint{TIT/HEP-707}

\begin{document} 
\maketitle
\flushbottom

\section{Introduction}
In the study of supersymmetric gauge theories and string theory, enumerative invariants of Calabi–Yau (CY) varieties play an essential role in counting certain stable bound states of D-branes, i.e., the BPS bound states. In the case of Calabi–Yau 3-folds, Donaldson--Thomas (DT) theory \cite{Thomas:1998uj} and Pandharipande--Thomas (PT) theory \cite{Pandharipande:2007kc,Pandharipande:2007sq} provide a mathematical way to count D6-D2-D0 bound states. Recently, there has been growing interest in extending these ideas to Calabi–Yau 4-folds \cite{Cao:2014bca,Borisov:2017GT,Cao:2017swr,Cao:2019tnw,Cao:2019tvv,Oh:2020rnj,Monavari:2022rtf,Nekrasov:2017cih,Nekrasov:2018xsb,Nekrasov:2023nai,Bao:2024ygr,Szabo:2023ixw}, where new phenomena emerge due to the richer mathematical structure and physical perspective.

The \textit{magnificent four} \cite{Nekrasov:2017cih,Nekrasov:2018xsb,Billo:2021xzh} is a physical setup whose partition function is given by the generating function of the enumerative invariants of $\mathbb{C}^{4}$. Such a setup was introduced along a series of studies \cite{Nekrasov:2015wsu,Nekrasov:2016qym,Nekrasov:2016ydq,Nekrasov:2017rqy,Nekrasov:2017gzb}, where higher-dimensional generalizations and supersymmetric gauge theories arising from intersecting D-branes were introduced, i.e. the \textit{gauge origami} framework \cite{Nekrasov:2016gud,Nekrasov:2016qym,Rapcak:2018nsl,Rapcak:2020ueh,Pomoni:2021hkn,Pomoni:2023nlf,Fasola:2023ypx,Nekrasov:2017cih,Nekrasov:2018xsb}. In particular, in type IIA string theory, the magnificent four system arises from D8-branes wrapping the $\mathbb{C}^{4}$ with D0-branes probing it. The flavored Witten index of the supersymmetric quantum mechanics of the D0-brane worldvolume theory then gives the partition function of the theory. Remarkably, this partition function is the generating function of solid partitions, the four-dimensional analogue of plane partitions, weighted by equivariant parameters corresponding to the $\Omega$-background on $\mathbb{C}^{4}$.

Extending this correspondence to general toric Calabi–Yau 4-folds requires a local building block analogous to the topological vertex \cite{Maulik:2003rzb,Maulik:2004txy,Okounkov:2003sp,Iqbal:2007ii,Aganagic:2003db,Awata:2008ed,Nekrasov:2014nea} of the 3-fold case. In the DT4 theory, such a building block is called the equivariant DT4 vertex \cite{Monavari:2022rtf,Nekrasov:2023nai,Piazzalunga:2023qik,DelZotto:2021gzy}, which encodes the contributions of solid partitions with prescribed asymptotics along the four complex coordinate axes of a local patch. The global enumerative invariants are then obtained by gluing these vertices according to the combinatorics of the toric diagram. Compared to the toric CY3 case, we can associate two types of asymptotics to the four directions. We can consider plane partitions as asymptotics of the four edges of the vertex or Young diagrams as asymptotics of the six faces of the vertex. Physically, the former correspond to considering D8-D2-D0 bound states, and the latter correspond to D8-D4-D0 bound states. One can combine both of them to consider the D8-D4-D2-D0 bound states.


In parallel, the equivariant PT4 vertex, based on stable pair theory for 4-folds, provides an alternative but equivalent description of the same geometry, differing in the underlying moduli problem while sharing the gluing formalism. The PT theory was first introduced to study the curve counting for 3-folds, showing the analytically better behavior and the numerical efficiency compared with the DT theory~\cite{Pandharipande:2007kc,Pandharipande:2007sq}. The PT theory was then formulated for 4-folds to study the curve counting as in the case of the 3-fold setups~\cite{Cao:2019fqq,Cao:2019tnw,Cao:2019tvv,Cao:2023lon,liu20234foldpandharipandethomasvertex}. Moreover, in particular to 4-fold setups, one may also consider the surface counting~\cite{Bae:2022pif,Bae:2024bpx}, which takes into account additional D4 branes for the D8-D2-D0 bound states in the context of string theory. Understanding the relationship between the DT and PT vertices is essential for a complete picture of curve counting and surface counting in complex dimension four.

The goal of this paper is to provide a method to compute both the equivariant DT4 and PT4 vertices within a systematical way using the Jeffrey--Kirwan (JK) residue formalism \cite{Jeffrey1993LocalizationFN,Szenes2003ToricRA,Brion1999ArrangementOH} (see also \cite{Benini:2013xpa,Benini:2013nda,Hori:2014tda,Hwang:2014uwa,Cordova:2014oxa}). Basically, this work is a 4-fold generalization of our previous work \cite{Kimura-Noshita-PT3}. As discussed in \cite{Nekrasov:2018xsb}, by studying the supersymmetric quantum mechanics of the D0-branes probing the D8-branes, the Witten index reduces to a finite-dimensional contour integral over the Cartan part of the corresponding gauge algebra. The JK-residue formalism organizes the poles of the integrand, which are naturally parametrized by solid partitions in this case. Moreover, the so-called \textit{sign rules}, which is a characteristic naturally occurring in 4-folds, are automatically determined by construction. From this view point, the JK-residue formalism is a systematic way to perform explicit computations of the equivariant vertices, which automatically fixes all the normalizations in a natural way.

In this paper, we consider both the leg and surface boundary conditions, corresponding to the curve and the surface degrees of freedom in the 4-fold setups, where plane partitions and Young diagrams appear as asymptotics of the solid partition. From the viewpoint of supersymmetric quantum mechanics, the contribution coming from the framing node is modified depending on such an asymptotics. We will give a way to determine this framing node contribution and write down the contour integral. We then move on to how to evaluate the integral using the JK-residue formalism. The additional datum $\eta$, which we call the reference vector, determines whether we consider the DT vertex or the PT vertex. We will see that the standard reference vector $\eta=(1,\ldots, 1)$ gives the equivariant DT4 vertex, while the alternative reference vector $\eta=(-1,\ldots,-1)$ gives the equivariant PT4 vertex. We perform explicit computations for the setups with both the leg and surface boundary conditions. A complete box-counting rule is yet to be clarified but our method gives an explicit way to study it. We also discuss the so-called DT/PT correspondence and have confirmed it for low levels. The existence of the sign rules so that DT/PT correspondence holds is an automatic consequence of our construction. Further generalizations, including the higher rank setup and the supergroup-like setup, of the DT/PT correspondence are also proposed.



The organization of this paper is as follows. We first review the magnificent four partition function and the JK-residue formalism in section~\ref{sec:MF-and-JKresidue}. We then discuss the DT4 counting with leg and surface boundary conditions using the JK-residue formalism in section~\ref{sec:DT4-JK}. We propose how to compute the equivariant PT4 vertex with leg boundary conditions using the JK-residue formalism in section~\ref{sec:PT4-JK-leg}. We study several examples for the one-leg (section~\ref{sec:PT4-JK-oneleg}), two-legs (section~\ref{sec:PT4-JK-twoleg}), three-legs (section~\ref{sec:PT4-JK-threeleg}), and the four-legs (section~\ref{sec:PT4-JK-fourleg}). Similar computations for the equivariant PT4 vertex with surface boundary conditions are given in section~\ref{sec:PT4-JK-surface}. Examples for the cases with one surface (section~\ref{sec:PT4onesurface}), two surface (section~\ref{sec:PT4twosurface}), and three surfaces (section~\ref{sec:PT4threesurface}) are discussed. The DT/PT correspondence and the higher rank generalizations are discussed in section~\ref{sec:DTPTcorrespondence}. In Appendix~\ref{app:JK-residue-degenerate}, we briefly review the JK-residue formalism for degenerate poles.

\section{Magnificent four and JK-residue}\label{sec:MF-and-JKresidue}
Let us briefly review general formulas regarding the computation of the Witten index of the 1d $\mathcal{N}=2$ supersymmetric quantum mechanics (SQM). See \cite{Benini:2013xpa,Benini:2013nda,Hori:2014tda,Hwang:2014uwa} for details.

The 1d $\mathcal{N}=2$ SQM is obtained from a dimensional reduction of the 2d $\mathcal{N}=(0,2)$ gauge theory. We have three basic multiplets: vector multiplet $V$, chiral multiplet $\Phi$, and the Fermi multiplet $\Lambda$. In this paper, we always consider unitary gauge groups. We also have the so-called $J$-term and $E$-terms. The $J$-term is a holomorphic function of the chiral fields coupled to the Fermi superfields. The $E$-term is also a holomorphic function of the chiral fields obeying $\overline{\mathcal{D}}\Lambda=E(\Phi)$, where $\overline{\mathcal{D}}$ is the gauge covariant superderivative. We note that we also have the condition $\Tr(E(\Phi)\cdot J(\Phi))=0$ for two supersymmetries.

Given the $\mathcal{N}=2$ SQM, the Witten index is
\bea
\mathcal{Z}=\frac{1}{|W_{G}|}\oint_{\text{JK}}\mathcal{Z}_{V}\prod_{i}\mathcal{Z}_{\Phi_i}\prod_{\alpha}\mathcal{Z}_{\Lambda_{\alpha}}
\eea
with 
\bea
\mathcal{Z}_{V}=\prod_{I=1}^{\text{rk}G}\frac{d\phi_I}{2\pi i}\prod_{\alpha\in G}\sh(\alpha\cdot \phi),\quad \mathcal{Z}_{\Phi_i}=\prod_{\rho\in V^{(i)}_{\text{chiral}}}\frac{1}{\sh(\rho\cdot \mu)},\quad \mathcal{Z}_{\Lambda_\alpha}=\prod_{\rho\in V^{(\alpha)}_{\text{Fermi}}}\sh(\rho\cdot \mu)
\eea
where the variable $\phi=(\phi_I)_{I=1}^{\text{rk}G}$ parametrizes the Cartan subalgebra of the gauge group $G$, for which we abuse the notation of the corresponding root system, the $V^{(i)}_{\text{chiral}},V^{(\alpha)}_{\text{Fermi}}$ denote the representation of the chiral superfield $\Phi_{i}$ and Fermi superfield $\Lambda_{\alpha}$, respectively, by which we denote the corresponding sets of the weights as well, and $\mu$ contains other parameters denoting the flavor fugacities of the theory.
We write
\bea
\sh(x)=2\sinh(x/2)=e^{x/2}-e^{-x/2}=[e^{x}].
\eea
For later use, we also introduce
\bea
\ch(x)=2\cosh(x/2)=e^{x/2}+e^{-x/2}.
\eea

\paragraph{Magnificent four}We denote the ten-dimensional space-time as $\mathbb{R}\times\mathbb{S}^{1}\times\mathbb{C}^{4}$. There are four complex one-planes and three-planes which we write as $\mathbb{C}_{a},\mathbb{C}^{3}_{\bar{a}}$ for $a\in\four$ and $\bar{a}\in\four^{\vee}$, where the four sets are defined as
\bea
\four=\{1,2,3,4\},\quad \four^{\vee}=\{123,124,134,234\}
\eea
and $\bar{S}$ is the complement of $S$ in $\four$. We have six complex two-planes $\mathbb{C}^{2}_{A}$ for
\bea
A \in \six=\{12,13,14,23,34,34\}.
\eea

The magnificent four system is a low energy limit of $n$ pairs of D8 and D8$'$ branes wrapping the $\mathbb{C}^{4}$ part.\footnote{In the original paper \cite{Nekrasov:2017cih}, the D8$'$-brane is denoted as the $\overline{\D8}$-brane. A physical realization of the magnificent four was also discussed in \cite{Billo:2021xzh} by introducing D8-branes with some constant background field and D8$'$-branes without any background field. We rather follow the notation of \cite{Billo:2021xzh} and call it the D8$'$-brane.  } We can also introduce D0-branes to the system. The brane configuration is summarized as
\bea\label{eq:2Bmagnificentfour}
\renewcommand{\arraystretch}{1.05}
\begin{tabular}{|c|c|c|c|c|c|c|c|c|c|c|}
\hline
& \multicolumn{2}{c|}{$\mathbb{C}_{1}$} & \multicolumn{2}{c|}{$\mathbb{C}_{2}$} & \multicolumn{2}{c|}{$\mathbb{C}_{3}$} & \multicolumn{2}{c|}{$\mathbb{C}_{4}$} & \multicolumn{2}{c|}{$\mathbb{R}\times \mathbb{S}^{1}$} \\
\cline{2-11}  & 1 & 2 & 3 & 4& 5 & 6 & 7 & 8 & 9& 0\\
\hline D0& $\bullet$ & $\bullet$  & $\bullet$  & $\bullet$  & $\bullet$  & $\bullet$   & $\bullet$  & $\bullet$  & $\bullet$   & $-$\\
\hline
$\D8 $& $-$ & $-$ & $-$ & $-$ & $-$ & $-$ & $-$ & $-$ & $\bullet$ & $-$ \\\hline
$\D8' $& $-$ & $-$ & $-$ & $-$ & $-$ & $-$ & $-$ & $-$ & $\bullet$ & $-$ \\
\hline
\end{tabular}
\eea
Generally, we may also consider situations with different number of D8 and D8$'$ branes \cite{Billo:2021xzh}, but we always keep it to be the same in this paper.

Including suitable $B$-field \cite{Witten:2000mf}, the low energy field theory of the $k$ D0-branes is described by a 1d $\mathcal{N}=2$ SQM whose quiver diagram is described as
\bea\label{eq:2SUSYquiver-magnificent}
\adjustbox{valign=c}{\begin{tikzpicture}[decoration={markings,mark=at position 0.7 with {\arrow{latex}}}]
 \tikzset{
        box/.style={draw, minimum width=0.7cm, minimum height=0.7cm, text centered,thick},
        ->-/.style={decoration={
        markings,mark=at position #1 with {\arrow[scale=1.5]{>}}},postaction={decorate},line width=0.5mm},
        -<-/.style={decoration={
        markings,
        mark=at position #1 with {\arrow[scale=1.5]{<}}},postaction={decorate},line width=0.5mm}    
    }
\begin{scope}[xshift=4cm]
    \draw[fill=black!10!white,thick](0,0) circle(0.4cm);
    \node at (0,0) {$k$};
    \node[box,fill=black!10!white] at (-0.5,1.6) {$n$};
    \node[box,fill=black!10!white] at (0.5,1.6) {$n$};
    \draw[postaction={decorate},thick] (-0.5,1.25)--(-0.1,0.4);
    \draw[postaction={decorate},thick,red] (0.5,1.25)--(0.1,0.4);
    
    \foreach \ang in {90,145,215,270} {
    \begin{scope}[rotate=\ang]
        \chiralarc[postaction={decorate},thick](0,0.5)(-45:225:0.22:0.65)
    \end{scope}
    }
    \foreach \ang in {90,145,270} {
    \begin{scope}[rotate=\ang]
    \fermiarc[postaction={decorate},thick](0,0.5)(-45:225:0.1:0.5)
    \end{scope}
    \node[left] at (-0.4,0.8) {$\mathsf{I}$};
    
    \node[right] at (0.4,0.8) {\textcolor{red}{$\Lambda_{\mathsf{I}}$}};
    \node[left] at (-1.5,0) {$\mathsf{B}_{2},\textcolor{red}{\Lambda_{2}}$};
    \node[right] at (1.6,0) {$\mathsf{B}_{1},\textcolor{red}{\Lambda_{1}}$};
    \node[below left] at (-0.9,-1){$\mathsf{B}_{3},\textcolor{red}{\Lambda_{3}}$};
    \node[below right] at (0.9,-1){$\mathsf{B}_{4}$};
    \draw[fill=black!10!white,thick](0,0) circle(0.4cm);
    \node at (0,0) {$k$};
    
    }
\end{scope}
\end{tikzpicture}}\quad\adjustbox{valign=c}{\begin{tikzpicture}[decoration={markings,mark=at position 0.7 with {\arrow{latex}}}]
 \tikzset{
        box/.style={draw, minimum width=0.7cm, minimum height=0.7cm, text centered,thick},
        ->-/.style={decoration={
        markings,mark=at position #1 with {\arrow[scale=1.5]{>}}},postaction={decorate},line width=0.5mm},
        -<-/.style={decoration={
        markings,
        mark=at position #1 with {\arrow[scale=1.5]{<}}},postaction={decorate},line width=0.5mm}    
    }
\begin{scope}[xshift=4cm]
    \draw[fill=black!10!white,thick](0,0) circle(0.4cm);
    \node at (0,0) {$k$};
    \node[box,fill=black!10!white] at (0,1.6) {$n|n$};
    \draw[postaction={decorate},thick] (-0.1,1.25)--(-0.1,0.4);
    \draw[postaction={decorate},thick,red] (0.1,1.25)--(0.1,0.4);
    
    \foreach \ang in {90,145,215,270} {
    \begin{scope}[rotate=\ang]
        \chiralarc[postaction={decorate},thick](0,0.5)(-45:225:0.22:0.65)
    \end{scope}
    }
    \foreach \ang in {90,145,270} {
    \begin{scope}[rotate=\ang]
    \fermiarc[postaction={decorate},thick](0,0.5)(-45:225:0.1:0.5)
    \end{scope}
    \node[left] at (-0.1,0.8) {$\mathsf{I}$};
    
    \node[right] at (0.1,0.8) {\textcolor{red}{$\Lambda_{\mathsf{I}}$}};
    \node[left] at (-1.5,0) {$\mathsf{B}_{2},\textcolor{red}{\Lambda_{2}}$};
    \node[right] at (1.6,0) {$\mathsf{B}_{1},\textcolor{red}{\Lambda_{1}}$};
    \node[below left] at (-0.9,-1){$\mathsf{B}_{3},\textcolor{red}{\Lambda_{3}}$};
    \node[below right] at (0.9,-1){$\mathsf{B}_{4}$};
    \draw[fill=black!10!white,thick](0,0) circle(0.4cm);
    \node at (0,0) {$k$};
    
    }
\end{scope}
\end{tikzpicture}}
\eea
The black arrows are the chiral superfields $B_a\,(a\in\four)$ and the red arrows are the Fermi superfields $\Lambda_i$. The D8-D0 strings correspond to the chiral superfields $\mathsf{I}$, while the D8$'$-D0 strings correspond to the Fermi superfields $\Lambda_{\mathsf{I}}$. For the left quiver, we distinguished the flavor nodes for the D8 and D8'-branes. For later use, we shortly draw the quiver as the right figure, but note that the flavor fugacities of the D8 and D8'-branes are generally different.

The $E,J$-terms are 
\bea\label{eq:2SUSYJEterm-magnificent}
    E_{i}=[\mathsf{B}_{4},\mathsf{B}_{i}]\quad J_{i}=\frac{1}{2}\varepsilon_{ijk4}[\mathsf{B}_{j},\mathsf{B}_{k}]
\eea
for $i=1,2,3$, where $\varepsilon_{ijk4}$ is the totally antisymmetric tensor with $\varepsilon_{1234}=1$. The system has a $U(1)^{3}$ flavor symmetry coming from the rotation of the $\mathbb{C}_{a}$-plane:
\bea
\mathsf{B}_{a}&\longrightarrow q_{a}\mathsf{B}_{a},\\
\mathsf{I},\Lambda_{\mathsf{I}}&\longrightarrow \mathsf{I},\Lambda_{\mathsf{I}},\\
\Lambda_{1,2,3}&\longrightarrow q_{4}q_{1,2,3}\Lambda_{1,2,3}
\eea
where $q_{1}q_{2}q_{3}q_{4}=1$ and $q_{a}=e^{\eps_a}$. For later use, we shortly write
\bea
q_{\mathcal{S}}=\prod_{a\in\mathcal{S}}q_{a},\quad \eps_{\mathcal{S}}=\sum_{a\in\mathcal{S}}\eps_{a}
\eea
where $\mathcal{S}$ is some set of indices.

The magnificent four partition function with $\D8\tbar\D8'$-branes is then
\bea
\mathcal{Z}_{n}^{\D8}[\fq,\{\mu_{\alpha}\}_{\alpha=1}^{n};q_{1,2,3,4}]=\sum_{k=0}^{\infty}\fq^{k}\mathcal{Z}^{\D8}[k]
\eea
where
\bea\label{eq:M4contourJK}
\mathcal{Z}^{\D8}[k]&=\frac{1}{k!}\left(\frac{\sh(-\epsilon_{14,24,34})}{\sh(-\epsilon_{1,2,3,4})}\right)^{k}\oint\prod_{I=1}^{k}\frac{d\phi_{I}}{2\pi i }\prod_{I=1}^{k}\mathcal{Z}^{\D8\tbar\D0}(\{\fra_{\alpha}\},\{\frb_{\alpha}\},\phi_I)\prod_{I<J}\mathcal{Z}^{\D0\tbar\D0}(\phi_I,\phi_J),
\eea
and
\bea
\,&\mathcal{Z}^{\D8\tbar\D0}(\{\fra_{\alpha}\},\{\frb_{\alpha}\},\phi_I)=\prod_{\alpha=1}^{n}\mathcal{Z}^{\D8\tbar\D0}(\fra_{\alpha},\phi_I)\mathcal{Z}^{\D8'\tbar\D0}(\frb_{\alpha},\phi_I),\\
&\mathcal{Z}^{\D8\tbar\D0}(\fra_{\alpha},\phi_I)=\frac{1}{\sh(\phi_I-\fra_{\alpha})},\quad \mathcal{Z}^{{\D8'}\tbar\D0}(\frb_{\alpha},\phi_I)=\sh(\phi_I-\frb_{\alpha}),\\
&\mathcal{Z}^{\D0\tbar\D0}(\phi_I,\phi_J)=\frac{\sh(\phi_{I}-\phi_{J})\sh(\phi_{I}-\phi_{J}-\epsilon_{14,24,34})}{\sh(\phi_{I}-\phi_{J}-\epsilon_{1,2,3,4})}\frac{\sh(\phi_{J}-\phi_{I})\sh(\phi_{J}-\phi_{I}-\epsilon_{14,24,34})}{\sh(\phi_{J}-\phi_{I}-\epsilon_{1,2,3,4})}
\eea
The parameters $\fra_{\alpha},\frb_{\alpha}$ denote the flavor fugacities of the framing node and we wrote $\mu_{\alpha}=e^{\frb_{\alpha}-\fra_{\alpha}}$.

\paragraph{Jeffrey--Kirwan (JK) residue}
Given the contour integral formula of the Witten index, we need to evaluate the contour integral. The correct way to evaluate it is known to be the Jeffrey--Kirwan residue formalism \cite{Jeffrey1993LocalizationFN,Szenes2003ToricRA,Brion1999ArrangementOH}(see als \cite{Benini:2013xpa,Benini:2013nda,Hori:2014tda,Hwang:2014uwa}). We basically follow the Appendix of \cite{Nawata:2023aoq,Kim:2024vci}.

Let the contour integral be
\bea
\mathcal{Z}=\oint_{\JK}\prod_{i=1}^{k}\frac{d\phi_i}{2\pi i }\mathcal{Z}(\phi)
\eea
where $\phi_i$ takes values in the Cartan subalgebra $\mathfrak{h}$ of the gauge group. The poles come from the zeros of them in the denominator:
\bea
Q_{i}(\phi)+f_{i}(\fra,\eps_{a})=0,\quad i=1,\ldots n
\eea
where $f_{i}$ is some function depending on the parameters of the theory. Each pole corresponds to a hyperplane $H_i$ in $\mathfrak{h}$ and the charge vector $Q_{i}\in\mathfrak{h}^{\ast}$ is associated with it. The integrand is singular at $\mathcal{M}_{\text{sing}}=\cup_i H_{i}$ and we denote $\mathcal{M}^{\ast}_{\text{sing}}$ to be the set of isolated points where $n\geq k$ linearly independent singular hyperplanes meet. When exactly $k$ hyperplanes meet a point $\phi_{\ast}\in\mathcal{M}^{\ast}_{\text{sing}}$, the intersection is called \textit{non-degenerate}, while when $n\geq k$ hyperplanes meet, the intersection is called \textit{degenerate}.

To perform the JK-residue formalism, we need an additional ingredient $\eta\in\mathfrak{h}^{\ast}$, which is called the reference vector. To explicitly show the reference vector dependence, we denote the contour integral as
\bea
\oint_{\eta}\prod_{i=1}^{k}\frac{d\phi_i}{2\pi i }\mathcal{Z}(\phi).
\eea
For a generic reference vector, the contour integral is then evaluated as
\bea
\oint _{\eta}\prod_{i=1}^{k}\frac{d\phi_i}{2\pi i}\mathcal{Z}(\phi)=\sum_{\phi_{\ast}\in \mathcal{M}_{\text{sing}}^{\ast}}\underset{\phi=\phi_{\ast}}{\text{JK-Res}}(Q(\phi_{\ast}),\eta)\mathcal{Z}(\phi).
\eea
If $\phi_{\ast}$ is a non-degenerate pole associated with the charge vectors $Q_{1},\ldots,Q_{k}$, the JK-residue is
\bea
\underset{\phi=\phi_{\ast}}{\text{JK-}\text{Res}}(Q(\phi_{\ast}),\eta)\mathcal{Z}(\phi)=\delta(Q,\eta)\frac{1}{|\det Q|}\underset{\delta_{k}=0}{\Res}\cdots \underset{\delta_{1}=0}{\Res}\left.\mathcal{Z}(\phi)\right|_{Q_{i}(\phi)+f_{i}(\fra,\eps_{a})=\delta_{i}}
\eea
where
\bea
\delta(Q,\eta)=\begin{dcases}
    1,\,\,\eta\in \text{Cone}(Q_{1},\ldots,Q_{k})\\
    0,\,\, \text{otherwise}
\end{dcases},\quad 
\text{Cone}(Q_{1},\ldots,Q_{k})=\left\{\sum_{i=1}^{k}\lambda_{i}Q_{i}=\eta\mid \lambda_{i}>0\right\}.
\eea
When the pole is a degenerate pole, we need more detailed analysis (see Appendix~\ref{app:JK-residue-degenerate}).

In this paper, two choices of reference vectors will be used frequently:
\bea
\eta_{0}=(1,\ldots,1),\quad \tilde{\eta}_{0}=(-1,\ldots,-1).
\eea

\paragraph{Magnificent four and solid partitions}
Let us use the JK-residue method to evaluate the contour integral of the rank one magnificent four partition function. The integrand is
\bea
\prod_{I=1}^{k}\mathcal{Z}^{\D8\tbar\D0}(\fra,\phi_I)\mathcal{Z}^{\D8'\tbar\D0}(\frb,\phi_I)\prod_{I<J}\mathcal{Z}^{\D0\tbar\D0}(\phi_I,\phi_J)
\eea
The denominators are
\bea
\sh(\phi_I-\fra),\quad \sh(\phi_I-\phi_J-\eps_{1,2,3,4})
\eea
and thus the poles are
\bea
\phi_I-\fra=0,\quad \phi_I-\phi_J-\eps_{1,2,3,4}=0.
\eea
We denote the unit vector pointing the $i$-th direction in $\mathbb{R}^{k}$ as $e_{i}$. The charge vectors are then associated to the poles as
\bea
\phi_{I}-\fra=0\quad &\longleftrightarrow \quad  e_{I}\\
\phi_{I}-\phi_{J}-\eps_{1,2,3,4}=0\quad &\longleftrightarrow \quad e_{I}-e_{J}.
\eea

Let us choose the reference vector to be $\eta=\eta_0$ and explicitly perform the JK-residue formalism. An interesting property is that the charge vectors need to form an oriented tree structure determined as follows.
\begin{itemize}
    \item We have $k$ vertices labeled by $\{e_{I}\}_{I=1}^{k}$ and arrows connecting them. Each vertex appears only once.
    \item The root vertex correspond to the charge vectors $+e_I$.
    \item The tree grows by adding a vertex $e_{K}$ to a vertex $e_{J}$ with an arrow $J\rightarrow K$. For each such arrow, the charge vector $e_{K}-e_{J}$.

\end{itemize}
A set of charge vectors chosen in this way obeys the condition that the reference vector lives in the cone $\eta\in \text{Cone}(Q_{1},\ldots, Q_{k})$. See \cite{Hwang:2014uwa} for a proof why this is true. We also note that if instead we choose the reference vector $\eta=\tilde{\eta}_0$, the charge vectors associated to the vertex $I$ and the arrow $J\rightarrow K$ will be $-e_{I}$ and $e_{J}-e_{K}$. Namely, the signs will be flipped.

Although the above tree structure gives the poles to evaluate, not all of them have non-zero JK-residues. In particular, for the one instanton level, the possible choice of charge vector is $\{e_{1}\}$ only and thus the pole picked up is $\phi_1-\fra=0$, which gives a non-zero JK-residue. For the second level, the possible choices are
\bea
\{e_{1},e_{2}\},&\quad \phi_1-\fra=0,\,\,\phi_2-\fra=0,\\
\{e_{1},-e_{1}+e_{2}\},&\quad \phi_1-\fra=0,\,\,\phi_2-\phi_1=\eps_{1,2,3,4},\\
\{e_{2},e_{1}-e_{2}\},&\quad \phi_2-\fra=0,\,\phi_1-\phi_2=\eps_{1,2,3,4}.
\eea
The second and third comes from the Weyl invariance and eventually cancels with the Weyl group factor $k!$ in the denominator. The first one actually cancels with the numerator $\sh(\phi_I-\phi_J)^{2}$ in the D0-D0 part. Omitting the degrees of freedom coming from the Weyl invariance, the poles giving non-zero JK-residues are
\bea
(\phi_1,\phi_2)=(\fra,\fra+\eps_{a}),\quad a\in\four.
\eea

Generally, the poles giving non-zero JK-residues are classified by a solid partition. A solid partition $\rho$ is a sequence of non-integers $\{\rho_{a,b,c}\}$ obeying the condition
\bea
\rho_{a,b,c}\geq \rho_{a+1,b,c},\quad 
\rho_{a,b,c}\geq \rho_{a,b+1,c},\quad 
\rho_{a,b,c}\geq \rho_{a,b,c+1}.
\eea
The size is defined as $|\rho|=\sum_{a,b,c}\rho_{a,b,c}$. The 4-cubes $\hcube$ in the solid partition are assigned coordinates in a natural way $(i,j,k,l)$ and it obeys
\bea
(i,j,k,l)\in\rho \Leftrightarrow 1\leq l \leq \rho_{i,j,k}.
\eea
Boxes in the solid partition obeys the melting rule.
\begin{condition}\label{cond:DT4}
    If any of 
    \bea
    (i+1,j,k,l),\quad (i,j+1,k,l),\quad (i,j,k+1,l),\quad (i,j,k,l+1)
    \eea
    are contained in the box configurations, then $(i,j,k,l)$ is also included. In other words, the boxes are stacked in a way as if the gravity is pointing the $(-1,-1,-1,-1)$ direction.
\end{condition}
The poles are classified by the boxes of the solid partition.
\begin{theorem}[{\cite{Nekrasov:2017cih,Nekrasov:2018xsb}}]
The poles of the rank one magnificent four giving non-zero JK-residue are classified as
\bea
\{\phi_I\}_{I=1}^{k} \longrightarrow \{c_{\four,\fra}(\hcube)\mid \hcube=(i,j,k,l)\in\rho\},\quad |\rho|=k
\eea
where we defined
\bea
c_{\four,\fra}(\hcube)=\fra+(i-1)\eps_1+(j-1)\eps_2+(k-1)\eps_3+(l-1)\eps_4.
\eea
For later use, we also introduce
\bea
\chi_{\four,v}(\hcube)=vq_{1}^{i-1}q_{2}^{j-1}q_{3}^{k-1}q_{4}^{l-1},\quad v = e^{\fra}.
\eea
\end{theorem}

Tuning the parameter $\frb=\fra+\eps_a$, the framing node contribution reduces to the D6-D0 configuration:
\bea\label{eq:D6-frame-def}
\mathcal{Z}^{\D6_{\bar{a}}\tbar\D0}(\fra,\phi_I)&=\frac{\sh(\phi_I-\fra-\eps_a)}{\sh(\phi_I-\fra)}.
\eea
Note that we also have the relation
\bea\label{eq:D6D8-relation}
\mathcal{Z}^{\D8\tbar\D0}(\fra,\phi_I)=\prod_{l=1}^{\infty}\mathcal{Z}^{\D6_{\bar{a}}\tbar\D0}(\fra+(l-1)\eps_a,\phi_I).
\eea
Generally setting $\frb=\fra+\sum_{a\in\four}n_{\bar{a}}\eps_{a}$, we obtain the tetrahedron instanton partition function with $n_{\bar{a}}$ D6$_{\bar{a}}$-branes.

\paragraph{Plethystic exponential (PE) formula}
The magnificent four partition function actually has a plethysic exponential (PE) formula\footnote{We also note that one can also obtain a PE formula for the case when we do not have the D8'-branes (for example see \cite[(4.4.88)-(4.4.93)]{Noshita:2025bzg}). } \cite{Nekrasov:2017cih,Nekrasov:2018xsb,Kool:2025qou}:
\bea\label{eq:D8-PEformula}
\mathcal{Z}^{\D8}_{n}[\fq,\{\mu_{\alpha}\}_{\alpha=1}^{n};q_{1,2,3,4}]=\PE\left[\frac{[q_{14}][q_{24}][q_{34}]}{[q_{1}][q_{2}][q_{3}][q_{4}]}\frac{\left[\prod\limits_{\alpha}\mu_{\alpha}\right]}{\left[\fq \prod\limits_{\alpha}\mu_{\alpha}^{-1/2}\right]\left[\fq \prod\limits_{\alpha}\mu_{\alpha}^{1/2} \right]}\right]
\eea
where the PE formula is defined as
\bea
\PE[f(x_{1},\ldots x_{n})]=\exp\left(\sum_{\ell=1}^{\infty}\frac{1}{\ell}f(x_{1}^{\ell},\ldots, x_{n}^{\ell})\right).
\eea
For the rank one case, we simply have
\bea\label{eq:D8-PEformula-rank1}
\mathcal{Z}^{\D8}[\fq,\mu;q_{1,2,3,4}]=\PE\left[\frac{[q_{14}][q_{24}][q_{34}]}{[q_{1}][q_{2}][q_{3}][q_{4}]}\frac{\left[\mu\right]}{\left[\fq \mu^{-1/2}\right]\left[\fq \mu^{1/2} \right]}\right].
\eea
For later use, we introduce
\bea\label{eq:MFdef}
\text{MF}[\mu]=\PE\left[\frac{[q_{14}][q_{24}][q_{34}]}{[q_{1}][q_{2}][q_{3}][q_{4}]}\frac{\left[\mu\right]}{\left[\fq \mu^{-1/2}\right]\left[\fq \mu^{1/2} \right]}\right]
\eea
and the rank $n$ magnificent four partition function is obtained by $\mu=\prod_{\alpha=1}^{n}\mu_{\alpha}$. 

An interesting property is that we have the following reflection formula
\bea\label{eq:MF-reflection}
\text{MF}[\mu^{-1}]&=\PE\left[\frac{[q_{14}][q_{24}][q_{34}]}{[q_{1}][q_{2}][q_{3}][q_{4}]}\frac{\left[\mu^{-1}\right]}{\left[\fq \mu^{+1/2}\right]\left[\fq \mu^{-1/2} \right]}\right]\\
&=\PE\left[-\frac{[q_{14}][q_{24}][q_{34}]}{[q_{1}][q_{2}][q_{3}][q_{4}]}\frac{\left[\mu\right]}{\left[\fq \mu^{+1/2}\right]\left[\fq \mu^{-1/2} \right]}\right]\\
&=\text{MF}[\mu]^{-1}
\eea
where we used $[\mu^{-1}]=-[\mu]$.

\section{DT4 counting and JK-residue}\label{sec:DT4-JK}
In this section, we review the JK-residue formalism of the equivariant Donaldson--Thomas (DT) 4-vertex \cite{Nekrasov:2023nai}. While the magnificent partition function is understood as a character of the infinite dimensional module parametrized by the solid partitions with nontrivial weights depending on the equivariant parameters $q_{1,2,3,4}$, the DT4 vertex is a weighted character of solid partitions with nontrivial boundary conditions. In particular, we will discuss two-types of boundary conditions, the leg boundary conditions and the surface boundary conditions. We also discuss what will happen for the case with both types of the boundary conditions. Physically, we are considering the partition function of the D8-D4-D2-D0 bound states.

\subsection{DT4 counting with boundary conditions}\label{sec:DT4vertexnotation}

\paragraph{Leg boundary conditions}
Let us start from the leg boundary condition case. A solid partition with leg boundary conditions is a configuration with asymptotic \emph{finite} plane partitions in the four legs of the solid partition. We denote the four plane partitions as $\pi_{1,2,3,4}$. The plane partition is described as a two-dimensional array of non-negative integers such as $\pi_{a}=\{\pi_{a,ij}\}$ obeying the conditions
\bea
\pi_{a,ij}\geq \pi_{a,i+1,j},\quad 
\pi_{a,ij}\geq \pi_{a,i,j+1}.
\eea
The orientation of the partition is given in a symmetric way. Namely, for the plane partition $\pi_{a}$, $\pi_{a,ij}$ is the number of boxes stacked in the $a+3$-direction and the two-dimensional base is the $(a+1,a+2)$-plane, where $a$ is understood modulo four.

To illustrate it, we decompose the solid partition into sequences of plane partitions in the fourth-direction.\footnote{This is the $(1,3)$-decomposition in \cite{Nekrasov:2017cih}} Under this decomposition, we obtain a non-increasing sequence of plane partitions with boundary conditions. The asymptotic plane partitions $\pi_{1,2,3}$ will be described as non-increasing sequence of Young diagrams, while the plane partition $\pi_{4}$ is a constant sequence of plane partition. The minimal solid partition with no additional boxes except the asymptotic plane partitions looks like
\bea\label{eq:minimalsolid_leg}
\adjustbox{valign=c}{\includegraphics[width=13cm]{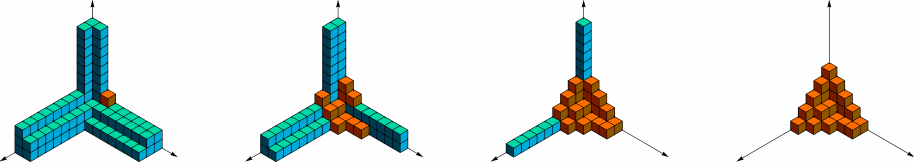}}\cdots
\eea
where the plane partitions $\pi_{1,2,3}$ are colored in light blue, and the plane partition $\pi_{4}$ is colored in orange. We first placed the plane partitions $\pi_{1,2,3}$ and the boxes of $\pi_{4}$ not included in $\pi_{1,2,3}$ are drawn in orange. Note that $\pi_{4}$ is extending semi-infinitely in the positive direction of the 4-axis.

For later use, we also identify the plane partitions $\pi_{1,2,3,4}$ with the corresponding character given by the box contents defined as
\bea
\pi_{a}=\sum_{i,j=1}^{\infty}\sum_{k=1}^{\pi_{a,ij}}q_{a+1}^{i-1}q_{a+2}^{j-1}q_{a+3}^{k-1}.
\eea
For example, the plane partition $\pi_{3}$ is 
\bea
\pi_{3}=(1+q_{1}+q_{2})+q_{4}(1+q_{1})+q_{4}^{2}.
\eea

\paragraph{Surface boundary conditions}
A solid partition with surface boundary conditions is a configuration with asymptotic \emph{finite} Young diagrams in the six surfaces of the solid partition. We denote the six Young diagrams as $\lambda_{A}\,(A\in\six)$ depending on which plane the Young diagram extends semi-infinitely. One may also impose an orientation on such Young diagrams but we omit the discussion.

Similar to the leg boundary conditions case, we illustrate such solid partitions by decomposing it in the fourth direction. This time, the boundary Young diagrams $\lambda_{14,24,34}$ are visualized as sequence of Young diagrams constant in the fourth-direction:
\bea\label{eq:minimalsolid_surface1}
\adjustbox{valign=c}{\includegraphics[width=12cm]{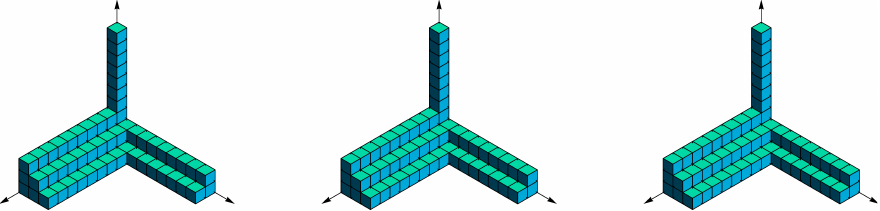}}
\eea
On the other hand, the Young diagrams $\lambda_{12,23,31}$ are visualized as non-increasing 1d partitions extending in the $3,1,2$ directions respectively:
\bea\label{eq:minimalsolid_surface2}
\adjustbox{valign=c}{\includegraphics[width=12cm]{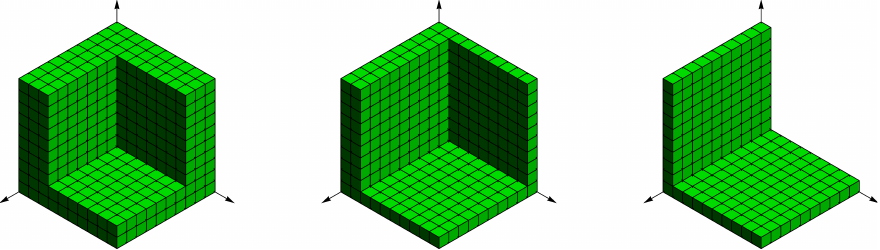}}
\eea
where we colored the surfaces in the 123-plane in light green.

For generic boundary conditions $\{\lambda_{A}\}_{A\in\six}$, the minimal solid partition looks like
\bea\label{eq:minimalsolid_surface3}
\adjustbox{valign=c}{\includegraphics[width=13cm]{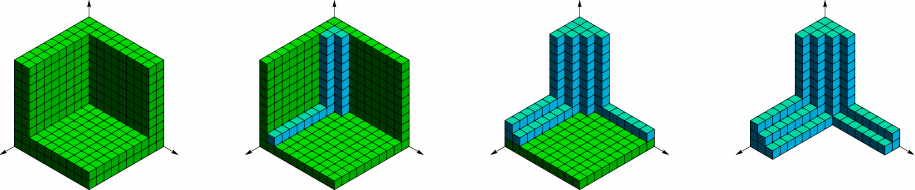}}
\eea
Note that the set of the light blue boxes are also extending semi-infinitely in the fourth-direction.

The Young diagrams $\lambda_{A}$ are identified with the character with the box contents
\bea
\lambda_{ab}=\sum_{(i,j)\in\lambda_{ab}}q_{a}^{i-1}q_{b}^{j-1}.
\eea
For example, for $\lambda_{24}$ in \eqref{eq:minimalsolid_surface3}, we have
\bea
\lambda_{24}=1+q_{2}+q_{3}.
\eea

We also note that the surface boundary conditions can be interpret as the leg boundary conditions with \textit{infinite size} of plane partitions at the four-legs. We denote such plane partitions as $\{\widetilde{\pi}_{a}\}_{a\in\four}$. For example, when we only have the surface $\lambda_{12}$, the plane partitions are
\bea
\widetilde{\pi}_{1}=\frac{\lambda_{12}}{1-q_{2}},\quad \widetilde{\pi}_{2}=\frac{\lambda_{12}}{1-q_{1}},\quad \widetilde{\pi}_{3}=\widetilde{\pi}_{4}=\emptyset.
\eea

\paragraph{Leg and surface boundary conditions}
Although we will not discuss this case in detail in this paper, we can also combine both the leg and surface boundary conditions\footnote{See also \cite{Nekrasov:2023nai} where a DT4 vertex was proposed for general boundary conditions. After normalizing the vertex term there so that it starts from $1$, it is the same with the vertex term defined here. } (see also a discussion in section~\ref{sec:conclusion}). Such conditions are also understood as solid partitions with infinite size of plane partitions at the four legs. For the example when we have one surface and one leg, the minimal solid partition will look like
\bea
\adjustbox{valign=c}{\includegraphics[width=13cm]{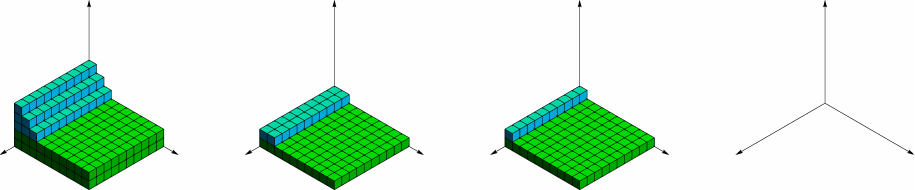}}
\eea
This time the corresponding character takes the form as
\bea
\widetilde{\pi}_{1}=\frac{\lambda_{12}}{1-q_{2}}+\pi_{1},\quad \widetilde{\pi}_{2}=\frac{\lambda_{12}}{1-q_{1}},\quad \widetilde{\pi}_{3}=\widetilde{\pi}_{4}=\emptyset
\eea
where $\pi_{1}$ correspond to a finite set of boxes. Note that $\pi_{1}$ itself is not a plane partition anymore. Generally, the character decomposes into a character corresponding to the surface leg boundary conditions and a finite part corresponding to the leg boundary conditions.

\subsection{Framing node and JK-residue}
In this section, we derive the contour integral formulas whose poles are classified by solid partitions with the leg and surface boundary conditions. As mentioned in section~\ref{sec:MF-and-JKresidue}, the information of the contour integrand and the reference vector controls how the integral is evaluated. In this section, we keep the reference vector $\eta=\eta_0$ and construct the framing node contribution producing solid partitions with boundary conditions. 

To construct the framing node contribution, one can start from analyzing the D8-D2-D0 or D8-D4-D0 bound states and construct a 1d $\mathcal{N}=2$ SQM. Additional superfields and $J,E$-terms will be added to the system and the Witten index of it gives the contour integral. See for example \cite{Galakhov:2021xum} and \cite[section~3]{Kimura-Noshita-PT3} for a discussion of this direction for the D6-D2-D0 case. We will not take this direction in this paper, because a detailed classification and derivation of the $J,E$-terms is essential. Instead, we will use the infinite product construction discussed in \cite[section~3.2]{Kimura-Noshita-PT3}.

\paragraph{Leg boundary conditions}
Let us start from the leg boundary conditions. We first introduce the following set of boxes
\bea
\mathcal{B}_{a,\pi_{a}}=\{(x_{1},x_{2},x_{3},x_{4})\mid x_{a}=1,\ldots,\infty, (x_{b},x_{c},x_{d})\in\pi_a,\,\,(b,c,d\neq a)\}
\eea
for $a\in\four$. They are boxes living at the four legs of the solid partition. We then define
\bea
\mathcal{B}_{\pi_{1}\pi_{2}\pi_{3}\pi_{4}}\coloneqq \sum_{a\in\four}\mathcal{B}_{a,\pi_{a}}-\sum_{ab\in\six}\mathcal{B}_{ab,\pi_{a}\cap\pi_{b}}+\sum_{\overline{(abc)}\in\four}\mathcal{B}_{abc,\pi_{a}\cap\pi_{b}\cap\pi_{c}}-\mathcal{B}_{\four,\pi_{1}\cap\pi_{2}\cap\pi_{3}\cap\pi_{4}}
\eea
where $\mathcal{B}_{ab,\pi_{a}\cap\pi_{b}}=\mathcal{B}_{a,\pi_{a}}\cap \mathcal{B}_{b,\pi_{b}},\,\,\mathcal{B}_{abc,\pi_{a}\cap \pi_{b}\cap\pi_{c}}=\bigcap\limits_{i=a,b,c}\mathcal{B}_{i,\pi_{a}} $ and $\mathcal{B}_{\four,\cap_{a\in\four}\pi_{a}}=\bigcap\limits_{a\in\four}\mathcal{B}_{a,\pi_{a}}$. The boxes in $\mathcal{B}_{\pi_1,\pi_2,\pi_3,\pi_4}$ are the boxes of the minimal solid partition with the leg boundary conditions.

Recalling the JK-residue prescription, whenever a box is added at $c_{\four,\fra}(\shcube)$ and the pole is evaluated, the contribution $\mathcal{Z}^{\D0\tbar\D0}(c_{\four,\fra}(\hcube))$ appears and modifies the contour integrand. To include boundary conditions, we need to include the boundary boxes $\mathcal{B}_{\pi_{1}\pi_{2}\pi_{3}\pi_{4}}$ and thus a natural candidate is
\bea
\mathcal{Z}^{\D8\tbar\D2\tbar\D0}_{\pi_{1}\pi_2\pi_3\pi_4}(\fra,\frb,\phi_I)=\mathcal{Z}^{\D8\tbar\D0}(\fra,\phi_I)\mathcal{Z}^{\D8'\tbar\D0}(\frb,\phi_I)\prod_{\shcube\in\mathcal{B}_{\pi_1\pi_2\pi_3\pi_4}}\mathcal{Z}^{\D0\tbar\D0}(c_{\four,\fra}(\hcube),\phi_I).
\eea
However, since the number of boxes in $\mathcal{B}_{\pi_1\pi_2\pi_3\pi_4}$ is infinite, we need to regularize the infinite product appearing here. 

To make the discussion concrete, let us consider the case $(\pi_1,\pi_2,\pi_3,\pi_4)=(\varnothing,\varnothing,\varnothing,\cube)$. The boundary boxes at $\mathcal{B}_{4,\pi_4}$ give
\bea
\prod_{l=1}^{\infty}\mathcal{Z}^{\D0\tbar\D0}(\fra+\eps_4(l-1),\phi_I)&=\frac{\sh(\phi_I-\fra)\sh(\phi_I-\fra+\eps_{41,42,43})}{\sh(\phi_I-\fra+\eps_4)\sh(\phi_I-\fra-\eps_{1,2,3})}\\
&=\mathcal{Z}^{\overline{\D2}_4\tbar\D0}(\fra,\phi_I).
\eea
We also have the relation
\bea\label{eq:D0D2-relation}
\frac{\mathcal{Z}^{\overline{\D2}_4\tbar\D0}(\fra,\phi_I)}{\mathcal{Z}^{\overline{\D2}_4\tbar\D0}(\fra+\eps_4,\phi_I)}=\mathcal{Z}^{\D0\tbar\D0}(\fra,\phi_I).
\eea
Using the quadrality, we also define
\bea
\mathcal{Z}^{\overline{\D2}_{a}\tbar\D0}(\fra,\phi_I)=\frac{\sh(\phi_I-\fra)}{\sh(\phi_I-\fra+\eps_a)}\prod_{i\in\bar{a}}\frac{\sh(\phi_I-\fra+\eps_a+\eps_i)}{\sh(\phi_I-\fra-\eps_i)}.
\eea
Combinatorially, these contributions correspond to a sequence of boxes extending semi-infinitely in the positive direction of the $a$-axis starting from $\fra$. We call such one-dimensional objects as \textit{rods} and specify the direction it extends as $a$-rod $(a\in\four)$. To obtain the regularized formula of general boundary conditions, one first decompose $\mathcal{B}_{\pi_1\pi_2\pi_3\pi_4}$ into non-intersecting one-dimensional rods and then associated to each of them the D2-D0 factors. For example, when we have at most one box at each leg, the framing node contributions are given as follows:
\begin{itemize}
    \item One-leg 
    \bea
    \mathcal{Z}_{\varnothing\varnothing\varnothing\,\scube}^{\D8\tbar\D2\tbar\D0}(\fra,\mathfrak{b},\phi_{I})&=\mathcal{Z}^{\D8'\tbar\D0}(\mathfrak{b},\phi_{I})\mathcal{Z}^{\D8\tbar\D0}(\fra,\phi_{I})\mathcal{Z}^{\overline{\D2}_{4}\tbar\D0}(\fra,\phi_{I})
    \eea
    \item Two-legs
    \bea
    \mathcal{Z}_{\scube\,\scube\varnothing\varnothing}^{\D8\tbar\D2\tbar\D0}(\fra,\mathfrak{b},\phi_{I})&=\mathcal{Z}^{\D8'\tbar\D0}(\mathfrak{b},\phi_{I})\mathcal{Z}^{\D8\tbar\D0}(\fra,\phi_{I})\mathcal{Z}^{\overline{\D2}_{1}\tbar\D0}(\fra,\phi_{I})\mathcal{Z}^{\overline{\D2}_{2}\tbar\D0}(\fra+\epsilon_{2},\phi_{I})
    \eea
    \item Three-legs
     \bea
    \mathcal{Z}_{\scube\,\scube\,\scube\varnothing}^{\D8\tbar\D2\tbar\D0}(\fra,\mathfrak{b},\phi_{I})&=\mathcal{Z}^{\D8'\tbar\D0}(\mathfrak{b},\phi_{I})\mathcal{Z}^{\D8\tbar\D0}(\fra,\phi_{I})\mathcal{Z}^{\overline{\D2}_{1}\tbar\D0}(\fra,\phi_{I})\\
    &\times \mathcal{Z}^{\overline{\D2}_{2}\tbar\D0}(\fra+\epsilon_{2},\phi_{I})\mathcal{Z}^{\overline{\D2}_{3}\tbar\D0}(\fra+\epsilon_{3},\phi_{I})
    \eea
    \item Four-legs
     \bea
    \mathcal{Z}_{\scube\,\scube\,\scube\,\scube}^{\D8\tbar\D2\tbar\D0}(\fra,\mathfrak{b},\phi_{I})&=\mathcal{Z}^{\D8'\tbar\D0}(\mathfrak{b},\phi_{I})\mathcal{Z}^{\D8\tbar\D0}(\fra,\phi_{I})\mathcal{Z}^{\overline{\D2}_{1}\tbar\D0}(\fra,\phi_{I})\mathcal{Z}^{\overline{\D2}_{2}\tbar\D0}(\fra+\epsilon_{2},\phi_{I})\\
    &\times \mathcal{Z}^{\overline{\D2}_{3}\tbar\D0}(\fra+\epsilon_{3},\phi_{I})\mathcal{Z}^{\overline{\D2}_{4}\tbar\D0}(\fra+\epsilon_{4},\phi_{I})
    \eea
\end{itemize}

\paragraph{Surface boundary conditions}
For the surface boundary conditions, we denote the set of boxes living at the six surfaces as
\bea
\mathcal{B}_{A,\lambda_A}=\{(x_{1},x_{2},x_{3},x_{4})\mid x_{a,b}=1,\ldots,\infty\,\,(a,b)\in A,\quad (x_{c},x_{d})\in\lambda_A\,(c,d\in\bar{A})\}
\eea
and the union of it as $\mathcal{B}_{\{\lambda_A\}_{A\in\six}}$. The framing node contribution is 
\bea
\mathcal{Z}^{\D8\tbar\D4\tbar\D0}_{\{\lambda_{A}\}}(\fra,\mathfrak{b},\phi_{I})&=\mathcal{Z}^{\D8\tbar\D0}(\fra,\phi_I)\mathcal{Z}^{\D8'\tbar\D0}(\frb,\phi_I)\prod_{\shcube\in\mathcal{B}_{\pi_1\pi_2\pi_3\pi_4}}\mathcal{Z}^{\D0\tbar\D0}(c_{\four,\fra}(\hcube),\phi_I).
\eea
Again, since the product is an infinite product, we need to regularize it properly. 

Let us consider the case when we have one surface $\lambda_{12}=1$. The boundary contributions are
\bea
\prod_{i=1}^{\infty}\prod_{j=1}^{\infty}\mathcal{Z}^{\D0\tbar\D0}(\fra+(i-1)\eps_1+(j-1)\eps_2,\phi_I)&=\frac{\sh(\phi_I-\fra)\sh(\phi_I-\fra-\eps_{34})}{\sh(\phi_I-\fra-\eps_3)\sh(\phi_I-\fra-\eps_4)}\\
&=\mathcal{Z}^{\overline{\D4}_{12}\tbar\D0}(\fra,\phi_I).
\eea
Note that we also have the relations
\bea\label{eq:D2D4-relation}
\mathcal{Z}^{\overline{\D4}_{12}\tbar\D0}(\fra,\phi_I)&=\prod_{i=1}^{\infty}\mathcal{Z}^{\overline{\D2}_{2}\tbar\D0}(\fra+(i-1)\eps_1,\phi_I),\\
\frac{\mathcal{Z}^{\overline{\D4}_{12}\tbar\D0}(\fra,\phi_I)}{\mathcal{Z}^{\overline{\D4}_{12}\tbar\D0}(\fra+\eps_2,\phi_I)}&=\mathcal{Z}^{\overline{\D2}_{2}\tbar\D0}(\fra,\phi_I),
\eea
and 
\bea\label{eq:D4D6-relation}
\mathcal{Z}^{\D6_{\bar{4}}\tbar\D0}(\fra+\eps_3,\phi_I)=\mathcal{Z}^{\overline{\D4}_{12}\tbar\D0}(\fra,\phi_I)\mathcal{Z}^{\D6_{\bar{4}}\tbar\D0}(\fra,\phi_I).
\eea
Using the quadrality symmetry, we define
\bea
\mathcal{Z}^{\overline{\D4}_{A}\tbar\D0}(\fra,\phi_I)=\frac{\sh(\phi_I-\fra)\sh(\phi_I-\fra-\eps_{\bar{A}})}{\sh(\phi_I-\fra-\eps_{c})\sh(\phi_I-\fra-\eps_{d})},\quad c,d\in\overline{A}.
\eea
Combinatorially, such contribution corresponds to adding a two-dimensional surface in the $A$-plane starting from the coordinate $\fra$.

For generic boundary conditions, we need to decompose the minimal solid partition into non-intersecting surfaces and use this formula. However, generally, the minimal solid partition is not decomposed only in the surfaces but one-dimensional rods are also necessary. For example, for the case with $\lambda_{12}=\lambda_{13}=1$, the minimal solid partition is
\bea
\adjustbox{valign=c}{\includegraphics[width=10cm]{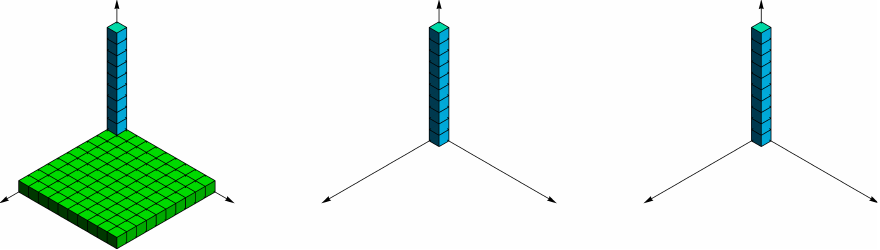}}\cdots
\eea
We can decompose it into a 12-surface with the coordinate $\fra$, a 34-surface with the coordinate $\fra+\eps_3$, and a one-dimensional 4-rod with the coordinate $\fra+\eps_4$. Thus, the framing node contribution is
\bea
\mathcal{Z}^{\D8\tbar\D4\tbar\D0}_{\{\lambda_{A}\}}(\fra,\frb,\phi_{I})&=\mathcal{Z}^{\D8\tbar\D0}(\fra,\phi_I)\mathcal{Z}^{\D8'\tbar\D0}(\frb,\phi_I)\mathcal{Z}^{\overline{\D4}_{12}\tbar\D0}(\fra,\phi_I)\\
&\times \mathcal{Z}^{\overline{\D4}_{34}\tbar\D0}(\fra+\eps_3,\phi_I)\mathcal{Z}^{\overline{\D2}_{4}\tbar\D0}(\fra+\eps_4,\phi_I).
\eea

\paragraph{Choice of contour}Although the above derivations determine the contour integrand, to evaluate the JK-residue, we need to specify which poles are picked up by the reference vector $\eta=\eta_0$. Such choice is related with whether the corresponding chiral multiplets are fundamental or anti-fundamental with respect to the D8-branes. To understand this property, one needs to study the 1d $\mathcal{N}=2$ SQM in detail. In this paper, instead we practically determine them in the following way. An observation is that the poles of the contour integrand can be classified into poles corresponding to the coordinates of the addable boxes of the minimal solid partitions and poles that do not. For poles that do not correspond to the addable boxes, we flip them using
\bea
\frac{1}{\sh(\phi_I-c_{\four,\fra}(\shcube))}\rightarrow -\frac{1}{\sh(-\phi_I+c_{\four,\fra}(\shcube))}.
\eea
Note that the numerator part does not affect how one performs the contour integral.

For example, for the one leg one box case, we have
\bea
\mathcal{Z}_{\varnothing\varnothing\varnothing\,\scube}^{\D8\tbar\D2\tbar\D0}(\fra,\mathfrak{b},\phi_{I})&=\frac{\sh(\phi_{I}-\mathfrak{b})}{\sh(\phi_{I}-\fra+\epsilon_{4})}\frac{\sh(\phi_{I}-\fra+\epsilon_{41,42,43})}{\sh(\phi_{I}-\fra-\epsilon_{1,2,3})}\\
&=\frac{\sh(-\phi_{I}+\mathfrak{b})}{\sh(-\phi_{I}+\fra-\epsilon_{4})}\frac{\sh(\phi_{I}-\fra+\epsilon_{41,42,43})}{\sh(\phi_{I}-\fra-\epsilon_{1,2,3})}
\eea
where we flipped the sign in the second line. In the first line the poles $\fra+\eps_{1,2,3}$ correspond to the addable boxes while the others do not and so we flip $\sh(\phi_I-\fra+\eps_4)$. 

For the one surface case, we have
\bea
\mathcal{Z}^{\D8\tbar\D4\tbar\D0}_{\{\lambda_{A}\}}(\fra,\mathfrak{b},\phi_{I})&=\frac{\sh(\phi_{I}-\mathfrak{b})\sh(\phi_{I}-\fra-\epsilon_{34})}{\sh(\phi_{I}-\fra-\epsilon_{3})\sh(\phi_{I}-\fra-\epsilon_{4})}
\eea
and all of the poles correspond to the addable boxes and thus we do not need to flip any term of the denominator.

In this paper, the framing node contribution is always implicitly written in this way. We then define the DT4 partition functions using this framing node contribution.
\begin{definition}
    The DT4 partition function or the equivariant DT4 vertex with leg boundary conditions is defined as
    \bea
\mathcal{Z}^{\DT\tbar\JK}_{\pi_1,\pi_2,\pi_3,\pi_4}[\fq,\mu;q_{1,2,3,4}]=\sum_{k=0}^{\infty}\fq^{k}\mathcal{Z}^{\DT\tbar\JK}_{\pi_1,\pi_2,\pi_3,\pi_4}[k],
    \eea
    where
    \bea
\mathcal{Z}^{\DT\tbar\JK}_{\pi_1,\pi_2,\pi_3,\pi_4}[k]=\frac{1}{k!}\left(\frac{\sh(-\epsilon_{14,24,34})}{\sh(-\epsilon_{1,2,3,4})}\right)^{k}\oint_{\eta_0}\prod_{I=1}^{k}\frac{d\phi_{I}}{2\pi i }\mathcal{Z}_{\pi_1\pi_2\pi_3\pi_4}^{\D8\tbar\D2\tbar\D0}(\fra,\mathfrak{b},\phi_{I})\prod_{I<J}\mathcal{Z}^{\D0\tbar\D0}(\phi_I,\phi_J).
    \eea
    The pole structure of the framing node contribution is obtained as discussed above.
\end{definition}

\begin{definition}
    The DT4 partition function with surface boundary conditions is defined as
    \bea
\mathcal{Z}^{\DT\tbar\JK}_{\{\lambda_A\}}[\fq,\mu;q_{1,2,3,4}]=\sum_{k=0}^{\infty}\fq^{k}\mathcal{Z}^{\DT\tbar\JK}_{\{\lambda_A\}}[k],
    \eea
    where
    \bea
\mathcal{Z}^{\DT\tbar\JK}_{\{\lambda_{A}\}}[k]=\frac{1}{k!}\left(\frac{\sh(-\epsilon_{14,24,34})}{\sh(-\epsilon_{1,2,3,4})}\right)^{k}\oint_{\eta_0}\prod_{I=1}^{k}\frac{d\phi_{I}}{2\pi i }\mathcal{Z}_{\{\lambda_A\}}^{\D8\tbar\D4\tbar\D0}(\fra,\mathfrak{b},\phi_{I})\prod_{I<J}\mathcal{Z}^{\D0\tbar\D0}(\phi_I,\phi_J).
    \eea
    Again, the pole structure of the framing node contribution is obtained as discussed above.
\end{definition}

Taking the reference vector $\eta=\eta_0$, one can show that the poles are indeed classified by solid partitions with the boundary conditions.

\section{PT4 counting with leg boundary conditions}\label{sec:PT4-JK-leg}
In this section, we explicitly perform the Pandharipande-Thomas (PT) counting for four-folds with leg boundary conditions for various situations. For the moment, a complete combinatorial rule for the PT4 counting seems to be not available yet especially for the four-legs case. See \cite{Cao:2019tnw,Cao:2019tvv,liu20234foldpandharipandethomasvertex} for reference.

In this paper, we follow the strategy given in \cite{Kimura-Noshita-PT3} and propose that PT4 counting is obtained by changing the reference vector from $\eta_0$ to $\tilde{\eta}_0$. The main claim of this section is the following proposal.
\begin{claim}
    The PT4 partition function or the equivariant PT4 vertex with leg boundary conditions is computed by
    \bea
\mathcal{Z}^{\PT\tbar\JK}_{\pi_1,\pi_2,\pi_3,\pi_4}[\fq,\mu;q_{1,2,3,4}]=\sum_{k=0}^{\infty}\fq^{k}\mathcal{Z}^{\PT\tbar\JK}_{\pi_1,\pi_2,\pi_3,\pi_4}[k],
    \eea
    where
    \bea
\mathcal{Z}^{\PT\tbar\JK}_{\pi_1,\pi_2,\pi_3,\pi_4}[k]=\frac{1}{k!}\left(\frac{\sh(-\epsilon_{14,24,34})}{\sh(-\epsilon_{1,2,3,4})}\right)^{k}\oint_{\tilde{\eta}_0}\prod_{I=1}^{k}\frac{d\phi_{I}}{2\pi i }\mathcal{Z}_{\pi_1\pi_2\pi_3\pi_4}^{\D8\tbar\D2\tbar\D0}(\fra,\mathfrak{b},\phi_{I})\prod_{I<J}\mathcal{Z}^{\D0\tbar\D0}(\phi_I,\phi_J).
    \eea
\end{claim}
The JK-residue formalism provides which poles to evaluate the residues and the non-zero JK-residue poles have a one-to-one correspondence with the PT4 box configurations (see for a similar analysis for the PT3 case in \cite{Kimura-Noshita-PT3}). For the moment, we do not have a complete classification of the poles and box-counting rules and we leave a complete description of them for future work.

We also note that when considering partition functions regarding D8-branes, one needs to deal with the so-called \textit{sign rules}. Such sign rule is a consequence of how one takes the JK-residue and compare it with index computations \cite{Nekrasov:2018xsb}. In this sense, the JK-residue formalism automatically fixes them in a natural way.



Before moving on to explicit computations, let us briefly discuss how we illustrate PT4 configurations. Following the description to derive PT3 configurations, we first start from a minimal solid partition
\bea
\adjustbox{valign=c}{\includegraphics[width=13cm]{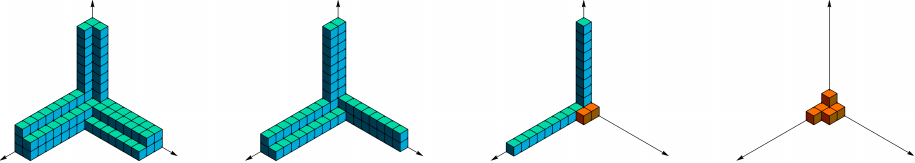}}\cdots
\eea
Note again that we are decomposing the solid partitions into slices of plane partitions. 

We then extend the four legs in the negative directions of each axis and then remove the boxes living in the positive quadrant, which forms a hollow structure:
\bea
\adjustbox{valign=c}{\includegraphics[width=9cm]{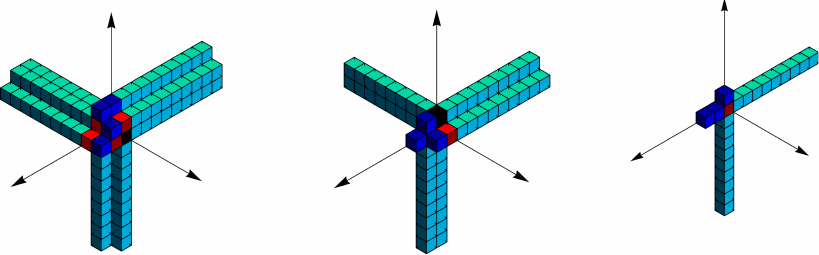}}\\
\cdots\adjustbox{valign=c}{\includegraphics[width=9cm]{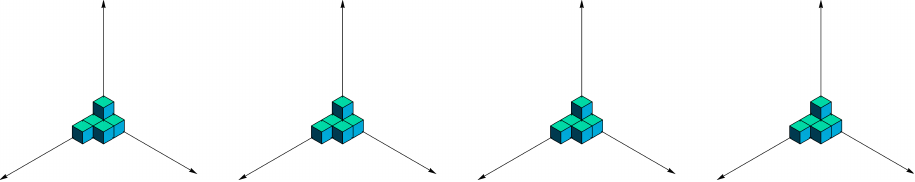}}
\eea
The first line has positive coordinates $x_{4}>0$ of the 4-axis and the second line has non-positive coordinates $x_{4}\leq 0$ of the 4-axis. Note that the second line extends semi-infinitely to the left. The \textbf{blue} boxes live at the intersection of exactly two legs, the \textbf{red} boxes live at the intersection of exactly three-legs, and the \textbf{black} boxes live at the intersection of the four-legs. Obviously, due to the existence of the four-legs, each three-dimensional slice do not look like a PT3 configuration anymore.

The boxes of the PT4 configurations will live inside the hollow structure. For the one-leg and two-legs cases, the boxes obey the following melting rule.
\begin{condition}\label{cond:PT4}
    If any of 
    \bea
    (i-1,j,k,l),\quad (i,j-1,k,l),\quad (i,j,k-1,l),\quad (i,j,k,l-1)
    \eea
    are contained in the box configurations, then $(i,j,k,l)$ is also included. In other words, the boxes are stacked in a way as if the gravity is pointing the $(1,1,1,1)$ direction.
\end{condition}
For the three-legs and four-legs cases, additional rules for the red boxes and black boxes will be necessary \cite{Pandharipande:2007sq,Pandharipande:2007kc,liu20234foldpandharipandethomasvertex}, but we will not discuss them in this paper.


In the JK-residue computations, we will see that for the one-leg and two-legs cases, the contour integrands have only first order poles and the poles correspond to the boxes obeying Cond.~\ref{cond:PT4} above. Namely, the PT4 counting rule is similar to the PT3 case. For the three-legs case, during the JK-residue procedure, second-order poles appear similar to the PT3 counting. On the other hand, for the four-legs case, the situation is much more complicated and during the JK-residue procedure, we will encounter third-order poles. This is the reason we need additional rules other than the previous condition. Such subtleties will be discussed elsewhere.


\subsection{One-leg}\label{sec:PT4-JK-oneleg}
\paragraph{One D2$_{3}$-brane}
Let us first study a well-known example, where the leg boundary conditions are
\bea\label{eq:PToneleg_ex1}
\pi_{1}=\pi_{2}=\pi_{4}=\varnothing,\quad \pi_{3}=1.
\eea
The minimal solid partition is illustrated as
\bea
\adjustbox{valign=c}{\includegraphics[width=8cm]{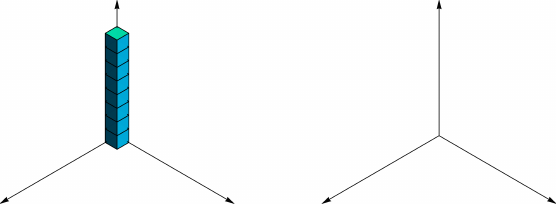}}
\eea
The framing node contribution is
\bea
\mathcal{Z}_{\varnothing\varnothing\,\scube\,\varnothing}^{\D8\tbar\D2\tbar\D0}(\fra,\mathfrak{b},\phi_{I})&=\mathcal{Z}^{\D8'\tbar\D0}(\mathfrak{b},\phi_{I})\mathcal{Z}^{\D8\tbar\D0}(\fra,\phi_{I})\mathcal{Z}^{\overline{\D2}_{3}\tbar\D0}(\fra,\phi_{I})\\
&=\frac{\sh(-\phi_{I}+\mathfrak{b})}{\sh(-\phi_{I}+\fra-\epsilon_{3})}\frac{\sh(\phi_{I}-\fra+\epsilon_{31,32,34})}{\sh(\phi_{I}-\fra-\epsilon_{1,2,4})}
\eea
and the contour integrand is
\bea
\prod_{I=1}^{k}\mathcal{Z}_{\varnothing\varnothing\,\scube\,\varnothing}^{\D8\tbar\D2\tbar\D0}(\fra,\mathfrak{b},\phi_{I})\prod_{I<J}\mathcal{Z}^{\D0\tbar\D0}(\phi_{I},\phi_{J}).
\eea

If we choose the reference vector to be $\eta=\eta_{0}=(1,\ldots,1)$, the poles picked up at the level one is
\bea
\phi_{1}=\fra+\eps_{1,2,4}
\eea
which correspond to the boxes that we can add to the minimal solid partition. Such choice of reference vector gives the DT4 counting.

Instead, let us choose the reference vector $\eta=\tilde{\eta}_{0}=(-1,\ldots,-1)$, the poles come from
\bea
-\phi_{I}+\fra-\eps_{3}=0,\quad \phi_{I}-\phi_{J}+\eps_{1,2,3,4}=0.
\eea
Using the Weyl-invariance, we can assume that the poles are picked up in the order $\phi_1,\ldots,\phi_k$ and
\bea
\phi_{1}=\fra-\eps_3,\quad \phi_{I}=\phi_J-\eps_{1,2,3,4}\quad (I>J).
\eea
For level one, we have
\bea
\phi_1=\fra-\eps_3.
\eea
For level two, the candidate of the poles are $\phi_{2}=\fra-\eps_3,\,\phi_{1}-\eps_{1,2,3,4}$. The pole $\phi_{2}=\fra-\eps_3=\phi_{1}$ is canceled by the numerator $\sh(\phi_I-\phi_J)$ of $\mathcal{Z}^{\D0\tbar\D0}(\phi_{I},\phi_{J})$. The poles   
\bea
\phi_2=\phi_1-\eps_{1,2,4}=\fra-\eps_{31,32,34}
\eea
are canceled by the numerator of $\mathcal{Z}_{\varnothing\varnothing\,\scube\,\varnothing}^{\D8\tbar\D2\tbar\D0}(\fra,\mathfrak{b},\phi_{I})$. Thus, the pole coming from $\phi_{2}=\fra-2\eps_{3}$ only remains.

Recursively, we can see that the poles 
\bea
\phi_{I}=\fra-I\eps_{3},\quad I=1,\ldots,k
\eea
only remain. The PT4 configuration is understood as stacking boxes in the hollow structure extending in the negative direction of the 3-direction:
\bea
\adjustbox{valign=c}{\includegraphics[width=3cm]{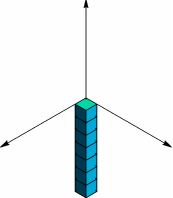}}
\eea

The JK-residue is then computed as
\bea
\mathcal{Z}^{\PT\tbar\JK}_{\varnothing\varnothing\,\scube\,\varnothing}[k]&=(-1)^{k}\left(\frac{\sh(-\eps_{14,24,34})}{\sh(-\eps_{1,2,3,4})}\right)^{k}\underset{\phi_{k}=\fra-k\eps_{3}}{\Res}\cdots \underset{\phi_{1}=\fra-\eps_{3}}{\Res}\prod_{I=1}^{k}\mathcal{Z}_{\varnothing\varnothing\,\scube\,\varnothing}^{\D8\tbar\D2\tbar\D0}(\fra,\mathfrak{b},\phi_{I})\prod_{I<J}\mathcal{Z}^{\D0\tbar\D0}(\phi_{I},\phi_{J})\\
&=\prod_{i=1}^{k}\frac{\sh(\mathfrak{b}-\fra+i\eps_{3})}{\sh(i\eps_3)}=\prod_{i=1}^{k}\frac{[\mu q_{3}^{i}]}{[q_{3}^{i}]}
\eea
where $\mu=e^{b-\fra}$. The PT4 partition function is then \cite[Lemma 6.5.1]{Monavari:2022umi}
\bea
\mathcal{Z}^{\PT\tbar\JK}_{\varnothing\varnothing\,\scube\,\varnothing}[\fq,\mu;q_{1,2,3,4}]=\sum_{k=0}^{\infty}\fq^{k}\mathcal{Z}^{\PT\tbar\JK}_{\varnothing\varnothing\,\scube\,\varnothing}[k]=\sum_{k=0}^{\infty}\fq^{k}\prod_{i=1}^{k}\frac{[\mu q_{3}^{i}]}{[q_{3}^{i}]}=\PE\left[\fq\frac{[\mu q_{3}]}{[q_{3}]}\right].
\eea
Note that setting $\mu=q_{4}$, we simply obtain the PT3 vertex with one-leg and one-box.

\paragraph{Multiple D2$_{3}$-branes}
As a nontrivial example, let us consider the case with 
\bea\label{eq:PT4oneleg_ex2}
\pi_{1}=\pi_{2}=\pi_{4}=\varnothing,\quad \pi_{3}=1+q_{1}+q_{2}+q_{4}.
\eea
Starting from the minimal solid partition, extending it in the negative direction and removing the boxes in the positive quadrant gives
\bea
\adjustbox{valign=c}{\includegraphics[width=6cm]{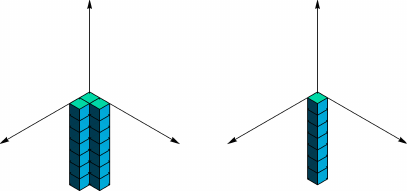}}
\eea
The boxes are then stacked inside this hollow structure in a way such that the gravity is pointing the $(1,1,1,1)$ direction. Let us list down the configurations following this melting rule up to the second level. For level one, we can place boxes at $-\eps_{3}+\eps_{1,2,4}$, where we identified the coordinates with the $\eps$-coordinates and omit the origin $\fra$ for simplicity. For level two, we have the following configurations
\bea\label{eq:PT4oneleg_JK_ex2}
(-\eps_{3}+\eps_{4},\eps_{4}-2\eps_{3}),&\quad (-\eps_{3}+\eps_{1},\eps_{1}-2\eps_{3}),\quad (\eps_{2}-\eps_{3},\eps_{2}-2\eps_{3}) \\
(-\eps_{3}+\eps_{1},-\eps_{3}+\eps_{4}),&\quad  (-\eps_{3}+\eps_{2},-\eps_{3}+\eps_{4}),\quad (-\eps_{3}+\eps_{1},-\eps_{3}+\eps_{2}).
\eea
Each slice of the PT4 configuration is a PT3 configuration and due to the gravity also pointing the $(0,0,0,1)$ direction, boxes need to be supported from the positive $4$-direction.

Let us check this using the JK-residue formalism. First, the framing node contribution is
\bea
\mathcal{Z}_{\pi_{1},\pi_{2},\pi_{3},\pi_{4}}^{\D8\tbar\D2\tbar\D0}(\fra,\mathfrak{b},\phi_{I})&=\frac{\sh(b-\phi_{I})\sh(\phi_{I}-\fra-\eps_1-2\eps_2)\sh(\phi_I-\fra-2\eps_{1}-\eps_2)\sh(\phi_{I}-\fra-\eps_1-2\eps_4)}{\sh(\fra+\eps_1-\eps_3-\phi_{I})\sh(\fra+\eps_2-\eps_3-\phi_{I})\sh(\fra+\eps_4-\eps_3-\phi_{I})}\\
&\times \frac{\sh(\phi_{I}-\fra-\eps_2-2\eps_4)\sh(\phi_I-\fra-2\eps_1-\eps_4)\sh(\phi_I-\fra-2\eps_2-\eps_4)\sh(\phi_I-\fra-\eps_{124})^{2}}{\sh(\phi_{I}-\fra-2\eps_{1,2})\sh(\phi_I-\fra-\eps_{1,2}-\eps_{4})\sh(\phi_{I}-\fra-2\eps_{4})\sh(\phi_{I}-\fra-2\eps_4)}.
\eea
Choosing the reference vector to be $\eta=\tilde{\eta}_{0}$, at the first level, the poles picked up are
\bea
\phi_{1}=\fra-\eps_{3}+\eps_{1,2,4}
\eea
which indeed matches with the configuration discussed above. For the next level, the poles picked up are
\bea
\phi_{I}=\fra-\eps_{3}+\eps_{1,2,4},\quad \phi_{2}=\phi_{1}-\eps_{1,2,3,4}.
\eea
We have two classes of poles. The first class of poles comes from picking two out of the three $\fra-\eps_{3}+\eps_{1,2,4}$. All of them do not vanish and we have
\bea
(\phi_{1},\phi_2)=(\fra-\eps_{3}+\eps_{1},\fra-\eps_{3}+\eps_{2}),\quad (\fra-\eps_{3}+\eps_{2},\fra-\eps_{3}+\eps_{4}),\quad (\fra-\eps_{3}+\eps_{1},\fra-\eps_{3}+\eps_{4})
\eea
up to Weyl invariance. 

The other class of poles comes from $\phi_{2}=\phi_1-\eps_{1,2,3,4}$ which extends the growth of the crystal. Due to the triality between $1,2,4$, let us focus on the case with $\phi_{1}=\fra-\eps_3+\eps_1$. The next pole comes from $\phi_2=\phi_1-\eps_{1,2,3,4}=\fra-\eps_3,\fra+\eps_1-\eps_{23,34},\fra+\eps_{1}-2\eps_{3}$. The pole $\phi_2=\fra-\eps_{3}$ vanishes from the $\sh(\phi_I-\fra-\eps_{124})^{2}$ coming from the numerator of the framing node contribution. The poles $\phi_{2}=\fra+\eps_{1}-\eps_{23,34}=\fra+2\eps_{1}+\eps_{4,2}$ are canceled by the numerators
\bea
\sh(\phi_{I}-\fra-2\eps_{1}-\eps_{2}),\quad 
\sh(\phi_{I}-\fra-2\eps_{1}-\eps_{4}).
\eea
Thus, only $\phi_{2}=\fra+\eps_{1}-2\eps_{3}$ gives non-zero JK-residues. Using the triality symmetry the following contributions remain
\bea
(\phi_1,\phi_2)=(\fra+\eps_{1,2,4}-\eps_{3},\fra+\eps_{1,2,4}-2\eps_{3}).
\eea
Therefore, we indeed obtain the configurations in \eqref{eq:PT4oneleg_JK_ex2}.

\paragraph{Multiple D2$_{4}$-branes} Since we are using the $(1,3)$-decomposition, the PT4 configuration looks different when the leg is extending in the 4-direction. For the example \eqref{eq:PT4oneleg_ex2}, considering the minimal solid partition, extending it in the negative direction and removing the boxes in the positive quadrant, we obtain
\bea
\cdots \cdots\adjustbox{valign=c}{\includegraphics[width=14cm]{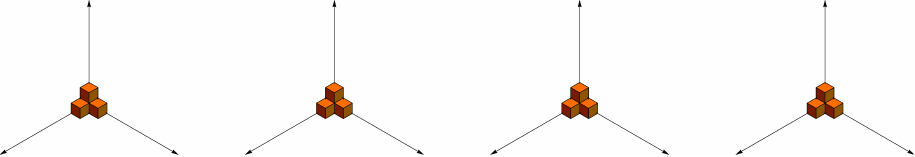}}
\eea
Note that the rightmost slice has the coordinate $-\eps_{4}$ and the slices extends semi-infinitely to the left. The PT4 configurations are stack of boxes inside this hollow structure. In this description, each slice of the PT4 configuration looks like a reversed plane partition, and boxes are stacked from the right to the left in the 4-direction.

\subsection{Two-legs}\label{sec:PT4-JK-twoleg}
Let us then move on to the two-legs case. For this case, the situation is similar to the one-leg case and the PT4 box-counting rule is simple. The resulting PT4 configurations obey only the melting rule with the gravity pointing the $(1,1,1,1)$ direction.

Let us first study the case with the boundary conditions
\bea
\pi_{1}=\pi_{2}=1,\quad \pi_{3}=\pi_{4}=\varnothing.
\eea
See the computation for the three-fold version in \cite[secton~3.4.2]{Kimura-Noshita-PT3}. 

The hollow structure is given as
\bea
\adjustbox{valign=c}{\includegraphics[width=3cm]{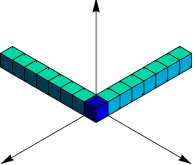}}
\eea
The PT configurations are similar to the three-fold case. For level one, we have a box at the origin. For generic levels, we have multiple boxes stacked in the negative directions of the 1,2-axes.

Let us derive these PT configurations using the JK-residue formalism. The framing node contribution is
\bea
\mathcal{Z}_{\scube\,\scube\,\varnothing\varnothing}^{\D8\tbar\D2\tbar\D0}(\fra,\mathfrak{b},\phi_{I})&=\mathcal{Z}^{\D8'\tbar\D0}(\mathfrak{b},\phi_{I})\mathcal{Z}^{\D8\tbar\D0}(\fra,\phi_{I})\mathcal{Z}^{\overline{\D2}_{1}\tbar\D0}(\fra,\phi_{I})\mathcal{Z}^{\overline{\D2}_{1}\tbar\D0}(\fra+\epsilon_{2},\phi_{I})\\
&=\frac{\sh(-\phi_{I}+\mathfrak{b})\sh(\phi_{I}-\fra+\epsilon_{12})\sh(\phi_{I}-\fra+\epsilon_{3,4})}{\sh(-\phi_{I}+\fra)\sh(\phi_{I}-\fra-\epsilon_{3,4})\sh(\phi_{I}-\fra-\epsilon_{12})}.
\eea
Choosing the reference vector $\eta=\tilde{\eta}_{0}$, for level one, the pole picked up is
\bea
\phi_{1}=\fra.
\eea
The JK-residue is evaluated as
\bea
\mathcal{Z}^{\PT\tbar\JK}_{\scube\,\scube\,\varnothing\varnothing}[1]&=-\left(\frac{\sh(-\eps_{14,24,34})}{\sh(-\eps_{1,2,3,4})}\right)\underset{\phi_1=\fra}{\Res}\mathcal{Z}_{\scube\,\scube\varnothing\varnothing}^{\D8\tbar\D2\tbar\D0}(\fra,\mathfrak{b},\phi_{1})=-\frac{[\mu][q_{12}][q_{13}][q_{23}]}{[q_{1}][q_{2}][q_{3}][q_{4}]}.
\eea

For the second level, the poles come from $\phi_{2}=\phi_{1}-\eps_{1,2,3,4}$. The poles $\phi_{2}=\phi_{1}-\eps_{3,4}=\fra-\eps_{3,4}$ are canceled by the numerators $\sh(\phi_{I}-\fra+\eps_{3,4})$ from the numerators of the framing node contribution. The non-zero contributions come from
\bea
\phi_{2}=\fra-\eps_{1,2}.
\eea

Recursively, at the generic level, one can show that the configurations giving non-zero JK-residues are classified by two non-negative integers $m,n\geq 0$:
\bea
\{\phi_{I}\}_{I=1}^{k}=\{\fra,\,\,\fra-i\eps_{1},\,\,\fra-j\eps_{2}\mid i=1,\ldots, m,\,j=1,\ldots, n\},
\eea
where $m,n$ are the number of boxes extending in the 1,2-directions respectively. Denoting such configuration as $[m,n]$, the JK-residue is evaluated as
\bea
\mathcal{Z}^{\PT\tbar\JK}_{\scube\,\scube\,\varnothing\varnothing}[[m,n]]&=(-1)^{m+n+1}\left(\frac{\sh(-\eps_{14,24,34})}{\sh(-\eps_{1,2,3,4})}\right)^{m+n+1}\underset{\phi=\phi_{[m,n]}}{\Res}\prod_{I}
\mathcal{Z}_{\scube\,\scube\varnothing\varnothing}^{\D8\tbar\D2\tbar\D0}(\fra,\mathfrak{b},\phi_{I})\prod_{I<J}\mathcal{Z}^{\D0\tbar\D0}(\phi_{I},\phi_{J})\\
&=\frac{[\mu][q_{34}]}{[q_{4}][q_{3}]}\times \frac{[q_{3}q_{1}^{m+1}q_{2}^{-n}][q_{4}q_{1}^{m+1}q_{2}^{-n}]}{[q_{1}^{m+1}q_{2}^{-n}][q_{1}^{m}q_{2}^{-n-1}]}\times \prod_{i=1}^{m}\frac{[\mu q_{1}^{i}]}{[q_{1}^{i}]}\prod_{j=1}^{n}\frac{[\mu q_{2}^{j}]}{[q_{2}^{j}]},
\eea
where we simply denoted the iterative residue as
\bea
\underset{\phi=\phi_{[m,n]}}{\Res}=\underset{\phi_{m+n+1}=\fra-n\epsilon_{2}}{\Res}\cdots\underset{\phi_{m+1}=\fra-\epsilon_{2}}{\Res}\underset{\phi_{m}=\fra-m\epsilon_{1}}{\Res}\cdots\underset{\phi_{1}=\fra}{\Res}.
\eea
Note that this formula is also applicable to $m=n=0$. The PT4 vertex is then
\bea
\mathcal{Z}^{\PT\tbar\JK}_{\scube\,\scube\,\varnothing\varnothing}[\fq,\mu;q_{1,2,3,4}]=1+\sum_{m,n=0}^{\infty}\mathfrak{q}^{m+n+1}\mathcal{Z}^{\PT\tbar\JK}_{\scube\,\scube\varnothing\varnothing}[[m,n]].
\eea

Tuning $\mu=q_{4}$, this reduces to the PT3 vertex in \cite[Section~3.4.2]{Kimura-Noshita-PT3}. Note that if we take $\mu=q_{3}$, we instead obtain the PT3 vertex using the D6$_{\bar{3}}$-branes.

\paragraph{Generic case}
The above example is an example where the minimal solid partition is restricted to a three-dimensional subspace and the boundary plane partitions are two-dimensional Young diagrams. For such cases, the PT4 vertex will be a one parameter ($\mu$) deformation of the PT3 vertex and tuning the parameter $\mu$ reproduces it. 

Let us consider an example where the boundary plane partitions are purely three-dimensional:
\bea
\pi_{1}=1+q_{2}+q_{4},\quad \pi_{2}=1+q_{1}+q_{3}+q_{4},\quad \pi_{3}=\pi_{4}=\varnothing
\eea
We then obtain the hollow structure
\bea
\adjustbox{valign=c}{\includegraphics[width=7cm]{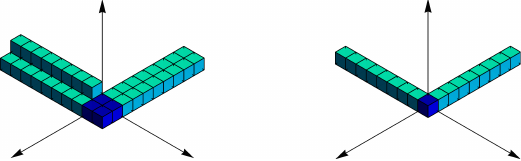}}
\eea
Each slice looks like the hollow structure of the PT3 vertex.

The framing node contribution is given as
\bea
&\mathcal{Z}_{\pi_1,\pi_2,\pi_3,\pi_4}^{\D8\tbar\D2\tbar\D0}(\fra,\mathfrak{b},\phi_{I})=\frac{\sh(\phi_I-\fra)\sh(b-\phi_I)\sh(\phi_I-\fra-\eps_1-2\eps_3)\sh(\phi_I-\fra+2\eps_4+\eps_3)\sh(\phi_I-\fra+\eps_3-\eps_4)}{\sh(\fra+\eps_{12}-\phi_I)\sh(\fra+\eps_4-\phi_I)\sh(\fra+\eps_3-\eps_2-\phi_I)}\\
&\times \frac{\sh(\phi_I-\fra-\eps_3-2\eps_4)\sh(\phi_I-\fra+2\eps_3+\eps_4)\sh(\phi_I-\fra-2\eps_3-\eps_4)\sh(\phi_I-\fra+\eps_2)}{\sh(\phi_I-\fra-2\eps_{12})\sh(\phi_I-\fra-2\eps_3)\sh(\phi_I-\fra-\eps_{13})\sh(\phi_I-\fra-2\eps_4)\sh(\phi_I-\fra-\eps_{124})\sh(\phi_I-\fra-\eps_{34})}.
\eea
Choosing the reference vector $\phi=\tilde{\eta}_{0}$, at level one, the poles picked up are
\bea
\phi_1=\fra+\eps_{1}+\eps_{2},\quad \fra+\eps_3-\eps_2,\quad \fra+\eps_4.
\eea
The first two-poles are actually the boxes at level one for the PT3 vertex $\mathcal{Z}^{\PT\tbar\JK}_{\bar{4};\lambda\mu\nu}[\fq,q_{1,2,3,4}]$ with boundary conditions $\lambda=\yng(1,1),\,\mu=\yng(2,1),\,\nu=\varnothing$, and the last pole is the box at level one for the case $\lambda=\yng(1),\,\mu=\yng(1),\,\nu=\varnothing$. 

Studying higher levels, the PT4 configurations can be understood as slices of PT3 configurations. Additionally, each PT3 slices need to obeying the melting rule Cond.~\ref{cond:PT4} in the fourth direction. In other words, the boxes need to be also supported from the positive direction of the 4-axis.

If instead one of the two-legs point the fourth direction, the illustration will look like
\bea
\cdots \adjustbox{valign=c}{\includegraphics[width=15cm]{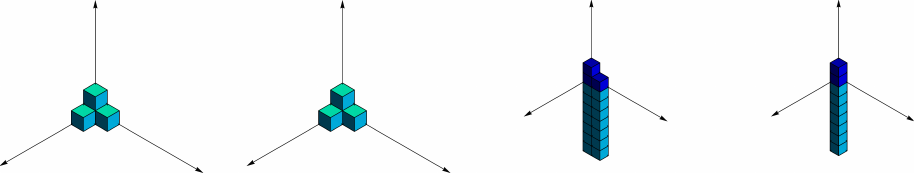}}
\eea
where $\pi_{1}=\pi_{2}=\varnothing$ and $\pi_{3}=1+q_{2}+q_{4}, \pi_{4}=1+q_{1}+q_{2}+q_{3}$. Instead of coloring the boxes of $\pi_{4}$ in orange, we colored it with light-blue as other boxes in the boundary conditions. The second slice from the right has the coordinate $0$ and the right-most one has $\eps_{4}$. Note again that the plane partition $\pi_{4}$ extends semi-infinitely to the left. 

In this description, each slice no more looks like a PT3 configuration anymore. Even so, the rules to stack boxes are the same as Cond.~\ref{cond:PT4} and one can list down possible PT4 configurations easily. For example, one can stack boxes at $\eps_{4}+\eps_{3}$, $\eps_{2}$, $-\eps_4+\eps_1$ at the first level, because all these boxes do not need to be supported from the positive direction. 

\subsection{Three-legs}\label{sec:PT4-JK-threeleg}
When there are three-legs, similar to the PT3 vertex, the contour integrand contains maximally second-order poles. For an example, we only consider the case
\bea
\pi_{1}=\pi_{2}=\pi_{3}=1,\quad \pi_4=\varnothing.
\eea
Other examples when the legs are purely plane partitions, but not Young diagrams, can be discussed in a similar way but we will focus only on this example.

The hollow structure is
\bea
\adjustbox{valign=c}{\includegraphics[width=4cm]{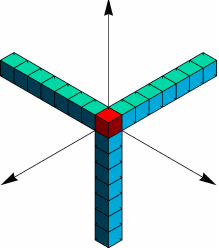}}
\eea
For this example, the PT4 configurations are the same as the PT3 configurations and the partition function is a one-parameter deformation of the PT3 computation. 

Let us derive the PT4 vertex for low levels using the JK-residue formalism. The discussion here is a 4-fold version of the one given in \cite[section~3.4.3]{Kimura-Noshita-PT3}. We reproduce the discussion again here.

The flavor node contribution is
\bea
\mathcal{Z}_{\scube\,\scube\,\scube\,\varnothing}^{\D8\tbar\D2\tbar\D0}(\fra,\mathfrak{b},\phi_{I})&=\frac{\sh(\phi_I-\frb)\sh(-\phi_I+\fra-\eps_4)^2\sh(\phi_I-\fra+\eps_{1,2,3})}{\sh(\fra-\phi_I)^{2}\sh(\phi_I-\fra-\eps_{12,13,23})\sh(\phi_{I}-\fra-\eps_4)}.
\eea
Note that tuning $\frb=\fra+\eps_4$, it reduces to the framing node contribution coming from the D6-brane (see \eqref{eq:D6-frame-def}). 
\paragraph{Level one}
Choosing the reference vector $\eta=\tilde{\eta}_0$, the poles picked up at level one is
\bea
\phi_1=\fra
\eea
and it is a second order pole. The JK-residue is
\bea
\mathcal{Z}^{\PT\tbar\JK}_{\scube\,\scube\,\scube\,\varnothing}[1]&=-\left(\frac{\sh(-\eps_{14,24,34})}{\sh(-\eps_{1,2,3,4})}\right)\underset{\phi_1=\fra}{\Res}\mathcal{Z}_{\scube\,\scube\,\scube\,\varnothing}^{\D8\tbar\D2\tbar\D0}(\fra,\mathfrak{b},\phi_{1})\\
&=-\frac{1}{2}\ch(\frb-\fra)-\frac{3}{2}\sh(\frb-\fra)\frac{\ch(\eps_{1}+\eps_{2}+\eps_3)}{\sh(\eps_1+\eps_2+\eps_3)}\\
&+\frac{1}{2}\sh(\frb-\fra)\left(\frac{\ch(\eps_1)}{\sh(\eps_1)}+\frac{\ch(\eps_2)}{\sh(\eps_2)}+\frac{\ch(\eps_3)}{\sh(\eps_3)}+\frac{\ch(\eps_1+\eps_2)}{\sh(\eps_1+\eps_2)}+\frac{\ch(\eps_1+\eps_3)}{\sh(\eps_1+\eps_3)}+\frac{\ch(\eps_3+\eps_2)}{\sh(\eps_3+\eps_2)}\right)
\eea
where the residue is evaluated as
\bea
\underset{\phi_1=\fra}{\Res}\mathcal{Z}_{\scube\,\scube\,\scube\,\varnothing}^{\D8\tbar\D2\tbar\D0}(\fra,\mathfrak{b},\phi_{I})=\lim_{\phi_1\rightarrow \fra}\frac{\partial}{\partial \phi_1}\left(\frac{\sh(\phi_I-\frb)\sh(\phi_I-\fra+\eps_4)^2\sh(\phi_I-\fra+\eps_{1,2,3})}{\sh(\phi_I-\fra-\eps_{12,13,23})\sh(\phi_{I}-\fra-\eps_4)}\right).
\eea

If we tune the parameter $\fra=\fra+\eps_4$, we obtain
\bea
\mathcal{Z}^{\PT\tbar\JK}_{\scube\,\scube\,\scube\,\varnothing}[1]\xlongrightarrow{b\rightarrow \fra+\eps_4}\mathcal{Z}_{\bar{4};\,\Bbox\Bbox\Bbox}^{\PT\tbar\JK}[1],
\eea
which reduces to the PT3 counting result in \cite[Section~3.4.3]{Kimura-Noshita-PT3}

\paragraph{Level two}
Let us next consider the second level. Again, choosing $\eta=(-1,\ldots,-1)$, the poles picked up come from 
\bea
-\phi_{I}+\fra=0,\quad \phi_{I}-\phi_{J}=-\epsilon_{1,2,3,4}.
\eea
Using the Weyl invariance, let us focus on the following cases 
\bea
(\phi_{1},\phi_{2})&=(\fra,\fra),\quad (\phi_1,\phi_2)=(\fra,\fra-\epsilon_{1,2,3,4}).
\eea

Let us check which poles remain non-zero. We start first from the pole $(\phi_{1},\phi_{2})=(\fra,\fra)$. In usual instanton computations, contributions such as this disappear because of the numerator $\prod_{I\neq J}\sh(\phi_{I}-\phi_{J})$. One would expect that a similar cancellation happens here but this is not the case.\footnote{This phenomenon occurs also in the PT3 computation (see \cite[section~3.4.3]{Kimura-Noshita-PT3}). } The contour integrand is schematically written as
\bea\label{eq:4leg-level2-1-def}
\prod_{I=1}^{2}\mathcal{Z}_{\scube\,\scube\,\scube\,\varnothing}^{\D8\tbar\D2\tbar\D0}(\fra,\mathfrak{b},\phi_{I})\mathcal{Z}^{\D0\tbar\D0}(\phi_{1},\phi_{2})=\frac{\sh(\phi_{1}-\phi_{2})^{2}f_{0}(\phi_{1},\phi_{2})}{\sh(\phi_{1}-\fra)^{2}\sh(\phi_{2}-\fra)^{2}}
\eea
where $f_{0}(\phi_{1},\phi_{2})$ is a function with no zeros nor poles at $\phi_{1}=\phi_{2}=\fra$. The residue is then computed as
\bea\label{eq:4leg-level2-1}
\underset{\phi_{1}=\fra}{\Res}\prod_{I=1}^{2}\mathcal{Z}_{\scube\,\scube\,\scube\,\varnothing}^{\D8\tbar\D2\tbar\D0}(\fra,\mathfrak{b},\phi_{I})\mathcal{Z}^{\D0\tbar\D0}(\phi_{1},\phi_{2})&=\left.\partial_{\phi_{1}}\left(\frac{\sh(\phi_{1}-\phi_{2})^{2}f_{0}(\phi_{1},\phi_{2})}{\sh(\phi_{2}-\fra)^{2}}\right)\right|_{\phi_{1}=\fra}\\
&=\partial_{\phi_{1}}f_{0}(\phi_{1},\phi_{2})|_{\phi_{1}=\fra}-\frac{f_{0}(\fra,\phi_{2})\ch(\fra-\phi_{2})}{\sh(\phi_{2}-\fra)}\\
&\xrightarrow{\phi_{2}\simeq \fra} -\frac{2f_{0}(\fra,\fra)}{\sh(\phi_{2}-\fra)}
\eea
where we used $\sh(x)'=\ch(x)/2$ and $\ch(0)=2$. Obviously, we still have a pole at $\phi_{2}=\fra$ and thus the residue is non-zero.

Let us next consider the poles at $(\phi_{1},\phi_{2})=(\fra,\fra-\epsilon_{1,2,3})$:
\bea\label{eq:4leg-level2-2-def}
&\prod_{I=1}^{2}\mathcal{Z}_{\scube\,\scube\,\scube\,\varnothing}^{\D8\tbar\D2\tbar\D0}(\fra,\mathfrak{b},\phi_{I})\mathcal{Z}^{\D0\tbar\D0}(\phi_{1},\phi_{2})\\
=&\frac{f_{1}(\phi_{1},\phi_{2})}{\sh(\phi_{1}-\fra)^{2}}\frac{\sh(\phi_{2}-\fra+\epsilon_{1})}{\sh(\phi_{2}-\phi_{1}+\epsilon_{1})}=\frac{f_{2}(\phi_{1},\phi_{2})}{\sh(\phi_{1}-\fra)^{2}}\frac{\sh(\phi_{2}-\fra+\epsilon_{2})}{\sh(\phi_{2}-\phi_{1}+\epsilon_{2})}=\frac{f_{3}(\phi_{1},\phi_{2})}{\sh(\phi_{1}-\fra)^{2}}\frac{\sh(\phi_{2}-\fra+\epsilon_{3})}{\sh(\phi_{2}-\phi_{1}+\epsilon_{3})}
\eea
where $f_{1,2,3}(\phi_{1},\phi_{2})$ are functions with no zeros and poles at $\phi_{2}=\fra-\epsilon_{1,2,3}$, respectively. Using the triality, let us focus on the $\epsilon_{3}$ case. Looking at the integrand, since there is $\sh(\phi_{2}-\fra+\epsilon_{3})$, one might expect that the residue at $\phi_{2}=\fra-\epsilon_{3}$ vanishes. Again, this is not the case because when evaluating the residue at $\phi_{1}=\fra$, derivatives of the integrand appears:
\bea\label{eq:4leg-level2-2}
\underset{\phi_{1}=\fra}{\Res}\prod_{I=1}^{2}\mathcal{Z}_{\scube\,\scube\,\scube\,\varnothing}^{\D8\tbar\D2\tbar\D0}(\fra,\mathfrak{b},\phi_{I})\mathcal{Z}^{\D0\tbar\D0}(\phi_{1},\phi_{2})&=\left.\partial_{\phi_{1}}\left(f_{3}(\phi_{1},\phi_{2})\frac{\sh(\phi_{2}-\fra+\epsilon_{3})}{\sh(\phi_{2}-\phi_{1}+\epsilon_{3})}\right)\right|_{\phi_{1}=\fra}\\
&=\frac{f_{3}(\fra,\phi_{2})}{\sh(\phi_{2}-\fra+\epsilon_{3})}+\partial_{\phi_{1}}f_{3}(\fra,\phi_{2}).
\eea
We still have a single pole appearing in the denominator and thus this pole gives a non-zero residue.

Finally, let us consider the poles at $(\phi_{1},\phi_{2})=(\fra,\fra-\epsilon_{4})$:
\bea
&\prod_{I=1}^{2}\mathcal{Z}_{\scube\,\scube\,\scube\,\varnothing}^{\D8\tbar\D2\tbar\D0}(\fra,\mathfrak{b},\phi_{I})\mathcal{Z}^{\D0\tbar\D0}(\phi_{1},\phi_{2})=\frac{f_{4}(\phi_{1},\phi_{2})\sh(\phi_{2}-\fra+\epsilon_{4})^{2}}{\sh(\phi_{1}-\fra)^{2}\sh(-\phi_{1}+\phi_{2}+\epsilon_{4})}
\eea
where $f_{4}(\phi_{1},\phi_{2})$ is a function with no zeros and poles. The residue at $\phi_{1}=\fra$ is 
\bea
\underset{\phi_{1}=\fra}{\Res}\prod_{I=1}^{2}\mathcal{Z}_{\scube\,\scube\,\scube\,\varnothing}^{\D8\tbar\D2\tbar\D0}(\fra,\mathfrak{b},\phi_{I})\mathcal{Z}^{\D0\tbar\D0}(\phi_{1},\phi_{2})=\sh(\phi_{2}-\fra+\epsilon_{4})\partial_{\phi_{1}}f_{4}(\fra,\phi_2)+f_{4}(\fra,\phi_{2}).
\eea
Thus, this gives no nontrivial residue.

Denoting the residue of the contour integrand without the Weyl group factor as
\bea
\mathcal{Z}_{(x,y)}\coloneqq\underset{\phi_{2}=\fra+y}{\Res}\,\,\underset{\phi_{1}=\fra+x}{\Res}\left(\frac{\sh(-\epsilon_{14,24,34})}{\sh(-\epsilon_{1,2,3,4})}\right)^{2}\prod_{I=1}^{2}\mathcal{Z}_{\scube\,\scube\,\scube\,\varnothing}^{\D8\tbar\D2\tbar\D0}(\fra,\mathfrak{b},\phi_{I})\mathcal{Z}^{\D0\tbar\D0}(\phi_{1},\phi_{2})
\eea
the PT4 partition function at two-instanton level is
\bea\label{eq:4leglevel2}
\mathcal{Z}^{\PT\tbar\JK}_{\scube\,\scube\,\scube\,\varnothing}[2]&=\frac{1}{2}\left(\mathcal{Z}_{(0,0)}+2\sum_{i=1}^{3}\mathcal{Z}_{(0,-\epsilon_{i})}\right),\\
\frac{1}{2}\mathcal{Z}_{(0,0)}&=\frac{\sh\left(\epsilon _1+\epsilon _2\right)^2 \sh\left(\epsilon _1+\epsilon
   _3\right)^2 \sh\left(\epsilon _2+\epsilon _3\right)^2
   \sh(\frb-\fra)^2}{\sh\left(\epsilon _1\right)^2 \sh\left(\epsilon _2\right)^2
   \sh\left(\epsilon _3\right)^2 \sh\left(\epsilon _1+\epsilon _2+\epsilon
   _3\right)^2},\\
\mathcal{Z}_{(0,-\epsilon_{1})}&=-\frac{\sh\left(-\epsilon _1+\epsilon _2+\epsilon _3\right) \sh\left(2 \epsilon
   _1+\epsilon _2+\epsilon _3\right) \sh(\frb-\fra) \sh\left(-\fra+\frb+\epsilon _1\right)}{2
   \sh\left(\epsilon _1\right) \sh\left(2 \epsilon _1\right) \sh\left(\epsilon
   _2+\epsilon _3\right) \sh\left(\epsilon _1+\epsilon _2+\epsilon _3\right)},\\
\mathcal{Z}_{(0,-\epsilon_{2})}&=-\frac{\sh\left(\epsilon _1-\epsilon _2+\epsilon _3\right) \sh\left(\epsilon _1+2 \epsilon
   _2+\epsilon _3\right) \sh(\frb-\fra) \sh\left(-\fra+\frb+\epsilon _2\right)}{2
   \sh\left(\epsilon _2\right) \sh\left(2 \epsilon _2\right) \sh\left(\epsilon
   _1+\epsilon _3\right) \sh\left(\epsilon _1+\epsilon _2+\epsilon _3\right)},\\
\mathcal{Z}_{(0,-\epsilon_{3})}&=\frac{\sh\left(-\epsilon _1-\epsilon _2+\epsilon _3\right) \sh\left(\epsilon _1+\epsilon
   _2+2 \epsilon _3\right) \sh(\frb-\fra) \sh\left(-\fra+\frb+\epsilon _3\right)}{2
   \sh\left(\epsilon _1+\epsilon _2\right) \sh\left(\epsilon _3\right) \sh\left(2
   \epsilon _3\right) \sh\left(\epsilon _1+\epsilon _2+\epsilon _3\right)}.
\eea
The overall factor is the Weyl group factor $1/2$. For the configurations coming from the poles $(\fra,\fra-\epsilon_{1,2,3})$, we have two situations in taking the residue $(\phi_{1},\phi_{2})=(\fra,\fra-\epsilon_{i}),\,(\fra-\epsilon_{i},\fra)$ and the Weyl group factor is canceled out by such multiplicity.

\paragraph{Level three}
For the next level, let us assume $(\phi_{1},\phi_{2})=(\fra,\fra),(\fra,\fra-\epsilon_{1,2,3})$ and evaluate the pole at $\phi_{3}$. For the case $(\phi_{1},\phi_{2})=(\fra,\fra)$, using \eqref{eq:4leg-level2-1-def} and \eqref{eq:4leg-level2-1}, we have
\bea
&\underset{\phi_{2}=\fra}{\Res}\,\underset{\phi_{1}=\fra}{\Res}\prod_{I=1}^{3}\mathcal{Z}_{\scube\,\scube\,\scube\,\varnothing}^{\D8\tbar\D2\tbar\D0}(\fra,\mathfrak{b},\phi_{I})\prod_{I<J}\mathcal{Z}^{\D0\tbar\D0}(\phi_{I},\phi_{J})\\
=&\underset{\phi_{2}=\fra}{\Res}\left(\mathcal{Z}_{\scube\,\scube\,\scube\,\varnothing}^{\D8\tbar\D2\tbar\D0}(\fra,\mathfrak{b},\phi_{3})\mathcal{Z}^{\D0\tbar\D0}(\phi_{2},\phi_{3})\partial_{\phi_{1}}\left(f_{0}(\phi_{1},\phi_{2})\mathcal{Z}^{\D0\tbar\D0}(\phi_{1},\phi_{3})\right)\right.\\
-&\left.\left.\frac{f_{0}(\fra,\phi_{2})\ch(\fra-\phi_{2})}{\sh(\phi_{2}-\fra)}\mathcal{Z}_{\scube\,\scube\,\scube\,\varnothing}^{\D8\tbar\D2\tbar\D0}(\fra,\mathfrak{b},\phi_{3})\mathcal{Z}^{\D0\tbar\D0}(\phi_{1},\phi_{3})\mathcal{Z}^{\D0\tbar\D0}(\phi_{2},\phi_{3})
  \right)\right|_{\phi_{1}=\fra}\\
  =&-2f_{0}(\fra,\fra)\mathcal{Z}_{\scube\,\scube\,\scube\,\varnothing}^{\D8\tbar\D2\tbar\D0}(\fra,\mathfrak{b},\phi_{3})\mathcal{Z}^{\D0\tbar\D0}(\fra,\phi_{3})^{2}.
\eea

Using the triality, we can focus on $(\phi_{1},\phi_{2})=(\fra,\fra-\epsilon_{3})$. From JK-formalism, the candidates are
\bea
\phi_{3}=\phi_{2}-\epsilon_{1,2,3,4}=\fra-2\epsilon_{3},\fra-\epsilon_{31,32,34}\quad \text{or} \quad \phi_3=\fra.
\eea
From \eqref{eq:4leg-level2-2-def} and \eqref{eq:4leg-level2-2} and attributing
\bea
f_{3}(\phi_{1},\phi_{2})\longrightarrow f_{3}(\phi_{1},\phi_{2})\mathcal{Z}_{\scube\,\scube\,\scube\,\varnothing}^{\D8\tbar\D2\tbar\D0}(\fra,\mathfrak{b},\phi_{3})\mathcal{Z}^{\D0\tbar\D0}(\phi_{1},\phi_{3})\mathcal{Z}^{\D0\tbar\D0}(\phi_{2},\phi_{3})
\eea
we have
\bea
&\underset{\phi_{2}=\fra-\epsilon_{3}}{\Res}\,\underset{\phi_{1}=\fra}{\Res}\prod_{I=1}^{3}\mathcal{Z}_{\scube\,\scube\,\scube\,\varnothing}^{\D8\tbar\D2\tbar\D0}(\fra,\mathfrak{b},\phi_{I})\prod_{I<J}\mathcal{Z}^{\D0\tbar\D0}(\phi_{I},\phi_{J})\\
=&f_{3}(\fra,\fra-\epsilon_{3})\mathcal{Z}_{\scube\,\scube\,\scube\,\varnothing}^{\D8\tbar\D2\tbar\D0}(\fra,\mathfrak{b},\phi_{3})\mathcal{Z}^{\D0\tbar\D0}(\fra,\phi_{3})\mathcal{Z}^{\D0\tbar\D0}(\fra-\epsilon_{3},\phi_{3}).
\eea
The poles $\phi_{3}=\fra-\epsilon_{31,32,34}$ all cancel with the numerator of $\mathcal{Z}^{\D0\tbar\D0}(\phi_{1},\phi_{3})|_{\phi_{1}=\fra}$, while the pole $\phi_{3}=\fra-2\epsilon_{3}$ is non-zero. Moreover, the pole at $\phi_3=\fra$ also gives non-zero contribution.

Summarizing, for the level three, the non-zero contributions come from the poles
\bea
(\fra,\fra,\fra-\epsilon_{i}),\quad (\fra,\fra-\epsilon_{i},\fra-2\epsilon_{i}),\quad i=1,2,3.
\eea

Denoting the residue as
\bea
\mathcal{Z}_{(x,y,z)}\coloneqq (-1)^{3}\underset{\phi_{3}=\fra+z}{\Res}\,\underset{\phi_{2}=\fra+y}{\Res}\,\,\underset{\phi_{1}=\fra+x}{\Res}\left(\frac{\sh(-\epsilon_{14,24,34})}{\sh(-\epsilon_{1,2,3,4})}\right)^{3}\prod_{I=1}^{3}\mathcal{Z}_{\scube\,\scube\,\scube\,\varnothing}^{\D8\tbar\D2\tbar\D0}(\fra,\mathfrak{b},\phi_{I})\prod_{I<J}\mathcal{Z}^{\D0\tbar\D0}(\phi_{I},\phi_{J})
\eea
we have
\bea\label{eq:4leglevel3}
\mathcal{Z}_{\bar{4};\,\Cbox\,\Cbox\,\Cbox}^{\PT\tbar\JK}[3]&=\frac{1}{6}\left(3\sum_{i=1}^{3}\mathcal{Z}_{(0,0,-\epsilon_{i})}+6\sum_{i=1}^{3}\mathcal{Z}_{(0,-\epsilon_{i},-2\epsilon_{i})}   \right),\\
\frac{1}{2}\mathcal{Z}_{(0,0,-\epsilon_{1})}&=\frac{\sh\left(\epsilon _1+\epsilon _2\right) \sh\left(2 \epsilon _1+\epsilon _2\right)
   \sh\left(\epsilon _1+\epsilon _3\right) \sh\left(2 \epsilon _1+\epsilon _3\right)
   \sh\left(-\epsilon _1+\epsilon _2+\epsilon _3\right)^2 \sh(\frb-\fra)^2
   \sh\left(-\fra+\frb+\epsilon _1\right)}{\sh\left(\epsilon _1\right) \sh\left(2 \epsilon
   _1\right)^2 \sh\left(\epsilon _2\right) \sh\left(\epsilon _2-\epsilon _1\right)
   \sh\left(\epsilon _3\right) \sh\left(\epsilon _3-\epsilon _1\right)
   \sh\left(\epsilon _1+\epsilon _2+\epsilon _3\right)^2},\\
   \frac{1}{2}\mathcal{Z}_{(0,0,-\epsilon_{2})}&= -\frac{\sh\left(\epsilon _1+\epsilon _2\right) \sh\left(\epsilon _1+2 \epsilon _2\right)
   \sh\left(\epsilon _1-\epsilon _2+\epsilon _3\right)^2 \sh\left(\epsilon _2+\epsilon
   _3\right) \sh\left(2 \epsilon _2+\epsilon _3\right) \sh(\frb-\fra)^2
   \sh\left(-\fra+\frb+\epsilon _2\right)}{\sh\left(\epsilon _1\right) \sh\left(\epsilon
   _2\right) \sh\left(2 \epsilon _2\right)^2 \sh\left(\epsilon _2-\epsilon _1\right)
   \sh\left(\epsilon _3\right) \sh\left(\epsilon _3-\epsilon _2\right)
   \sh\left(\epsilon _1+\epsilon _2+\epsilon _3\right)^2},\\
   \frac{1}{2}\mathcal{Z}_{(0,0,-\epsilon_{3})}&=\frac{\sh\left(\epsilon _1+\epsilon _3\right) \sh\left(-\epsilon _1-\epsilon _2+\epsilon
   _3\right)^2 \sh\left(\epsilon _2+\epsilon _3\right) \sh\left(\epsilon _1+2 \epsilon
   _3\right) \sh\left(\epsilon _2+2 \epsilon _3\right) \sh(\frb-\fra)^2
   \sh\left(-\fra+\frb+\epsilon _3\right)}{\sh\left(\epsilon _1\right) \sh\left(\epsilon
   _2\right) \sh\left(\epsilon _3\right) \sh\left(2 \epsilon _3\right)^2
   \sh\left(\epsilon _3-\epsilon _1\right) \sh\left(\epsilon _3-\epsilon _2\right)
   \sh\left(\epsilon _1+\epsilon _2+\epsilon _3\right)^2},\\
   \mathcal{Z}_{(0,-\epsilon_{1},-2\epsilon_{1})}&=-\frac{\sh\left(-2 \epsilon _1+\epsilon _2+\epsilon _3\right) \sh\left(3 \epsilon
   _1+\epsilon _2+\epsilon _3\right) \sh(\frb-\fra) \sh\left(-\fra+\frb+\epsilon _1\right)
   \sh\left(-\fra+\frb+2 \epsilon _1\right)}{\sh\left(2 \epsilon _1\right)^2 \sh\left(3
   \epsilon _1\right) \sh\left(\epsilon _2+\epsilon _3\right) \sh\left(\epsilon _1+\epsilon
   _2+\epsilon _3\right)},\\
   \mathcal{Z}_{(0,-\epsilon_{2},-2\epsilon_{2})}&=-\frac{\sh\left(\epsilon _1-2 \epsilon _2+\epsilon _3\right) \sh\left(\epsilon _1+3
   \epsilon _2+\epsilon _3\right) \sh(\frb-\fra) \sh\left(-\fra+\frb+\epsilon _2\right)
   \sh\left(-\fra+\frb+2 \epsilon _2\right)}{\sh\left(2 \epsilon _2\right)^2 \sh\left(3
   \epsilon _2\right) \sh\left(\epsilon _1+\epsilon _3\right) \sh\left(\epsilon _1+\epsilon
   _2+\epsilon _3\right)},\\
   \mathcal{Z}_{(0,-\epsilon_{3},-2\epsilon_{3})}&=\frac{\sh\left(-\epsilon _1-\epsilon _2+2 \epsilon _3\right) \sh\left(\epsilon _1+\epsilon
   _2+3 \epsilon _3\right) \sh(\frb-\fra) \sh\left(-\fra+\frb+\epsilon _3\right) \sh\left(-\fra+\frb+2
   \epsilon _3\right)}{\sh\left(\epsilon _1+\epsilon _2\right) \sh\left(2 \epsilon
   _3\right)^2 \sh\left(3 \epsilon _3\right) \sh\left(\epsilon _1+\epsilon _2+\epsilon
   _3\right)}.
\eea

\paragraph{Generic level $k\geq 2$}Similar to the PT3 case, two classes of poles only remain:
\bea
(0,-\epsilon_{i},\ldots, -(k-1)\epsilon_{i})
\eea
and
\bea
(0,0,-\epsilon_{1},\ldots,-n_{1}\epsilon_{1},-\epsilon_{2},\ldots,-n_{2}\epsilon_{2},\epsilon_{3},\ldots,-n_{3}\epsilon_{3}),\quad n_{1}+n_{2}+n_{3}=k-2.
\eea
We thus have $\binom{k}{2}+3$ possible configurations. The JK-residue schematically takes the form as
\bea\label{eq:4leglevelgeneric}
\mathcal{Z}^{\PT\tbar\JK}_{\bar{4};\,\Cbox\,\Cbox\,\Cbox}[k]&=\frac{1}{k!}\left(k!\sum_{i=1}^{3}\mathcal{Z}_{(0,-\epsilon_{i},\ldots,-(k-1)\epsilon_{i})}\right.\\
&\left.+\frac{k!}{2!}\sum_{n_{1}+n_{2}+n_{3}=k-2}\mathcal{Z}_{(0,0,-\epsilon_{1},\ldots,-n_{1}\epsilon_{1},-\epsilon_{2},\ldots,-n_{2}\epsilon_{2},\epsilon_{3},\ldots,-n_{3}\epsilon_{3})}\right),
\eea
where each term is understood as the JK-residue of the contour integrand without the Weyl group factor for each configuration. 

\subsection{Four-legs}\label{sec:PT4-JK-fourleg}
Let us move on to the nontrivial example where there are all four legs. As an example, we only discuss the example:
\bea
\pi_{1}=\pi_{2}=\pi_{3}=\pi_{4}=1.
\eea
After extending the minimal solid partition and removing the boxes in the positive quadrant, we obtain
\bea
\cdots\adjustbox{valign=c}{\includegraphics[width=12cm]{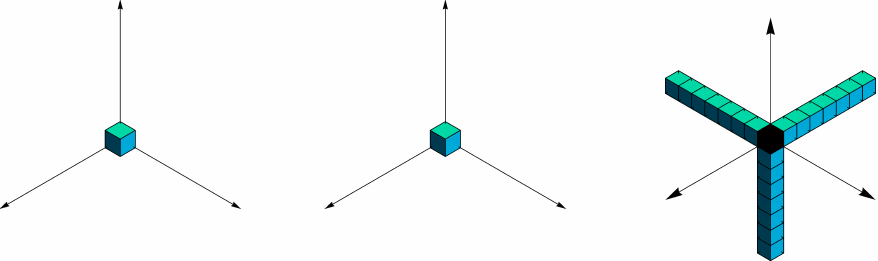}}
\eea
Note that there are boxes at the origin extending semi-infinitely from the right to the left.

The framing node contribution is
\bea
\mathcal{Z}_{\scube\,\scube\,\scube\,\scube}^{\D8\tbar\D2\tbar\D0}(\fra,\mathfrak{b},\phi_{I})&=\frac{\sh(-\phi_I+\frb)\prod_{a\in\four}\sh(\phi_I-\fra+\eps_a)^{2}}{\sh(\fra-\phi_I)^{3}\prod_{A\in\six}\sh(\phi_I-\fra-\eps_{A})}
\eea
Choosing the reference vector $\eta=\tilde{\eta}_{0}$ picks the pole
\bea
-\phi_{1}+\fra=0
\eea
and it is a third-order pole. The existence of this third-order pole is a new perspective of the PT4 counting. The JK-residue is evaluated as
\bea
\mathcal{Z}^{\PT\tbar\JK}_{\scube\,\scube\,\scube\,\scube}[1]&=-\left(\frac{\sh(-\eps_{14,24,34})}{\sh(-\eps_{1,2,3,4})}\right)\underset{\phi_1=\fra}{\Res}\mathcal{Z}_{\scube\,\scube\,\scube\,\scube}^{\D8\tbar\D2\tbar\D0}(\fra,\mathfrak{b},\phi_{1})
\eea
and the residue is performed as
\bea
\underset{\phi_1=\fra}{\Res}\mathcal{Z}_{\scube\,\scube\,\scube\,\scube}^{\D8\tbar\D2\tbar\D0}(\fra,\mathfrak{b},\phi_{1})=\frac{1}{2!}\lim_{\phi_1\rightarrow \fra}\frac{\partial^{2}}{\partial^{2}\phi_1}\left(\frac{\sh(\phi_1-\frb)\prod_{a\in\four}\sh(\phi_1-\fra+\eps_a)^{2}}{\prod_{A\in\six}\sh(\phi_1-\fra-\eps_{A})}\right).
\eea
We then obtain
\bea
\mathcal{Z}^{\PT\tbar\JK}_{\scube\,\scube\,\scube\,\scube}[1]&=\frac{1}{2}\frac{\ch(\frb-\fra)}{\sh(\eps_{12})\sh(\eps_{13})\sh(\eps_{23})}\sum_{a\in\four}\ch(\eps_{a})\prod_{i\in\bar{a}}\sh(\eps_i)\\
&-\frac{1}{4}\frac{\sh(\frb-\fra)}{\sh(\eps_{12})\sh(\eps_{23})\sh(\eps_{13})}\sum_{a\in\four}\frac{\ch(\eps_a)^{2}}{\sh(\eps_{a})}\prod_{i\in\bar{a}}\sh(\eps_i)\\
&-\frac{1}{4}\frac{\sh(\frb-\fra)\sh(\eps_1)\sh(\eps_2)\sh(\eps_3)\sh(\eps_4)}{\sh(\eps_{12})\sh(\eps_{23})\sh(\eps_{31})}\left(1+\frac{\ch(\eps_{23})^{2}}{\sh(\eps_{23})^{2}}+\frac{\ch(\eps_{12})^{2}}{\sh(\eps_{12})^{2}}+\frac{\ch(\eps_{13})^{2}}{\sh(\eps_{13})^{2}}\right)\\
&-\frac{\sh(\frb-\fra)}{\sh(\eps_{12})\sh(\eps_{13})\sh(\eps_{23})}\sum_{A\in\six}\prod_{a\in A}\ch(\eps_{a})\prod_{a\in\bar{A}}\sh(\eps_{a}).
\eea
Note that the partition function is quadrality invariant.

\paragraph{Level two}Let us consider the next level. The integrand is
\bea
\prod_{I=1}^{2}\mathcal{Z}_{\scube\,\scube\,\scube\,\scube}^{\D8\tbar\D2\tbar\D0}(\fra,\mathfrak{b},\phi_{I})\mathcal{Z}^{\D0\tbar\D0}(\phi_1,\phi_2).
\eea
The JK-prescription picks up the poles
\bea
(\phi_1,\phi_2)=(\fra,\fra),\quad (\fra,\fra-\eps_{1,2,3,4}).
\eea
Let us consider which poles give non-zero contributions.

For $(\phi_1,\phi_2)=(\fra,\fra)$, the integrand schematically takes the form as
\bea
\prod_{I=1}^{2}\mathcal{Z}_{\scube\,\scube\,\scube\,\scube}^{\D8\tbar\D2\tbar\D0}(\fra,\mathfrak{b},\phi_{I})\mathcal{Z}^{\D0\tbar\D0}(\phi_1,\phi_2)=\frac{\sh(\phi_1-\phi_2)^{2}g_{0}(\phi_1,\phi_2)}{\sh(\phi_1-\fra)^{3}\sh(\phi_2-\fra)^{3}}
\eea
where again we only extracted the zeros and poles associated with $\phi_1=\phi_2=\fra$ and $g_{0}(\phi_1,\phi_2)$ has no zeros there. The residue at $\phi_1=\fra$ is
\bea
&\underset{\phi_1=\fra}{\Res}\prod_{I=1}^{2}\mathcal{Z}_{\scube\,\scube\,\scube\,\scube}^{\D8\tbar\D2\tbar\D0}(\fra,\mathfrak{b},\phi_{I})\mathcal{Z}^{\D0\tbar\D0}(\phi_1,\phi_2)\\
=&\frac{\ch(\phi_2-\fra)^{2}g_{0}(\fra,\phi_2)}{4\sh(\phi_2-\fra)^{3}}+\frac{g_{0}(\fra,\phi_2)}{4\sh(\phi_2-\fra)}-\frac{\ch(\phi_2-\fra)g^{(1,0)}_{0}(\fra,\phi_2)}{\sh(\phi_2-\fra)^{2}}+\frac{g_{0}^{(2,0)}(\fra,\phi_2)}{2\sh(\phi_2-\fra)}
\eea
where $g_{0}^{(m,n)}(\phi_1,\phi_2)$ denotes the $m$ ($n$) derivative of $\phi_1$ $(\phi_2)$, respectively. Obviously, there is a pole at $\phi_2=\fra$ and thus, we still can take the residue at $\phi_2=\fra$. Similar to the three-legs case, we have the PT4 configuration coming from $\phi_1=\phi_2=\fra$.

Let us then move on to $(\phi_1,\phi_2)= (\fra,\fra-\eps_{1,2,3,4})$. Using the quadrality, we can focus on $\phi_2=\fra-\eps_1$. The contour integrand schematically takes the form
\bea
\prod_{I=1}^{2}\mathcal{Z}_{\scube\,\scube\,\scube\,\scube}^{\D8\tbar\D2\tbar\D0}(\fra,\mathfrak{b},\phi_{I})\mathcal{Z}^{\D0\tbar\D0}(\phi_1,\phi_2)=\frac{\sh(\phi_2-\fra+\eps_1)^{2}g_{1}(\phi_1,\phi_2)}{\sh(\phi_1-\fra)^{3}\sh(\phi_2-\phi_1+\eps_1)}.
\eea
After evaluating the residue at $\phi_1=\fra$, we have
\bea
\underset{\phi_1=\fra}{\Res}\prod_{I=1}^{2}\mathcal{Z}_{\scube\,\scube\,\scube\,\scube}^{\D8\tbar\D2\tbar\D0}(\fra,\mathfrak{b},\phi_{I})\mathcal{Z}^{\D0\tbar\D0}(\phi_1,\phi_2)&=\frac{\ch(\phi_2-\fra+\eps_1)^{2}g_{1}(\fra,\phi_2)}{4\sh(\phi_2-\fra+\eps_1)}-\frac{1}{8}g_{1}(\fra,\phi_{2})\sh(\phi_{2}-\fra+\eps_{1})\\
+\frac{1}{2}\text{ch}(\phi_{2}-\fra+\eps_1)g_{1}^{(1,0)}(\fra,\phi_2)&+\frac{1}{2}\sh(\phi_2-\fra+\eps_1)g^{(2,0)}_{1}(\fra,\phi_2)
\eea
which gives a single pole. Thus, the pole $\phi_1=\fra,\phi_2=\fra-\eps_1$ gives non-zero JK-residue contribution. Summarizing, the poles giving the non-zero JK-residues are
\bea
(\fra,\fra),\,\,(\fra,\fra-\eps_{1,2,3,4}).
\eea
Explicitly, the JK-residues are given as
\bea
\frac{1}{2}\mathcal{Z}_{(0,0)}&=\frac{1}{4}\ch(\frb-\fra)^{2}+\frac{1}{4}\sh(\frb-\fra)^{2}-\frac{1}{2}\sh(\frb-\fra)\ch(\frb-\fra)\sum_{a\in\four}\frac{\ch(\eps_{a})}{\sh(\eps_a)}\\
&+\frac{1}{2}\sh(\frb-\fra)^{2}\left(\frac{\ch(\eps_{12})^{2}}{\sh(\eps_{12})^{2}}+\frac{\ch(\eps_{13})^{2}}{\sh(\eps_{13})^{2}}+\frac{\ch(\eps_{23})^{2}}{\sh(\eps_{23})^{2}}\right)+\sh(\frb-\fra)^{2}\sum_{A\in\six}\prod_{a\in A}\frac{\ch(\eps_a)}{\sh(\eps_a)},\\
\mathcal{Z}_{(0,-\eps_1)}&=\frac{\sh(\frb-\fra)\sh(\frb-\fra+\eps_1)\sh(\eps_2-\eps_1)\sh(\eps_3-\eps_1)\sh(\eps_1-\eps_4)}{\sh(\eps_1)\sh(2\eps_1)\sh(\eps_{12})\sh(\eps_{32})\sh(\eps_{13})},\\
\mathcal{Z}_{(0,-\eps_2)}&=-\frac{\sh(\frb-\fra)\sh(\frb-\fra+\eps_2)\sh(\eps_2-\eps_1)\sh(\eps_3-\eps_2)\sh(\eps_2-\eps_4)}{\sh(\eps_2)\sh(2\eps_2)\sh(\eps_{12})\sh(\eps_{32})\sh(\eps_{13})},\\
\mathcal{Z}_{(0,-\eps_3)}&=\frac{\sh(\frb-\fra)\sh(\frb-\fra+\eps_3)\sh(\eps_3-\eps_1)\sh(\eps_3-\eps_2)\sh(\eps_3-\eps_4)}{\sh(\eps_3)\sh(2\eps_3)\sh(\eps_{12})\sh(\eps_{32})\sh(\eps_{13})},\\
\mathcal{Z}_{(0,-\eps_4)}&=\frac{\sh(\frb-\fra)\sh(\frb-\fra-\eps_4)\sh(-\eps_4+\eps_1)\sh(\eps_2-\eps_4)\sh(\eps_3-\eps_4)}{\sh(\eps_4)\sh(2\eps_4)\sh(\eps_{12})\sh(\eps_{32})\sh(\eps_{13})}
\eea
and the PT4 partition function at second level is
\bea
\mathcal{Z}^{\PT\tbar\JK}_{\scube\,\scube\,\scube\,\scube}[2]&=\frac{1}{2}\left(\mathcal{Z}_{(0,0)}+2\sum_{a\in\four}\mathcal{Z}_{(0,-\epsilon_{a})}\right).
\eea

\paragraph{Level three}
Let us assume $(\phi_1,\phi_2)=(\fra,\fra)$ first and consider the pole at $\phi_3$. JK-residue pickes the poles at
\bea
-\fra+\phi_3=0,\quad \phi_I-\phi_3=\eps_{1,2,3,4}.
\eea
Let us first study the pole coming from $\phi_3=\fra$. The integrand schematically takes the form as
\bea
\prod_{I=1}^{3}\mathcal{Z}_{\scube\,\scube\,\scube\,\scube}^{\D8\tbar\D2\tbar\D0}(\fra,\mathfrak{b},\phi_{I})\prod_{I<J}\mathcal{Z}^{\D0\tbar\D0}(\phi_I,\phi_J)=\frac{\sh(\phi_1-\phi_2)^{2}\sh(\phi_1-\phi_3)^{2}\sh(\phi_2-\phi_3)^{2}h_{0}(\phi_1,\phi_2,\phi_3)}{\sh(\phi_1-\fra)^{3}\sh(\phi_2-\fra)^3\sh(\phi_3-\fra)^3}.
\eea
Performing the residue at $(\phi_1,\phi_2)=(\fra,\fra)$ gives
\bea
&\underset{\phi_2=\fra}{\Res}\underset{\phi_1=\fra}{\Res}\prod_{I=1}^{3}\mathcal{Z}_{\scube\,\scube\,\scube\,\scube}^{\D8\tbar\D2\tbar\D0}(\fra,\mathfrak{b},\phi_{I})\prod_{I<J}\mathcal{Z}^{\D0\tbar\D0}(\phi_I,\phi_J)\\
=&-\frac{3 \text{ch}\left(\phi _3-\fra\right){}^2 h_0\left(\fra,\fra,\phi _3\right)}{2 \text{sh}\left(\phi _3-\fra\right)}+\text{ch}\left(\phi
   _3-\fra\right) h_0{}^{(0,1,0)}\left(\fra,\fra,\phi _3\right)+\text{ch}\left(\phi _3-\fra\right) h_0{}^{(1,0,0)}\left(\fra,\fra,\phi
   _3\right)\\
&+\frac{3}{4} \text{sh}\left(\phi _3-\fra\right) h_0\left(\fra,\fra,\phi _3\right)+\frac{1}{2} \text{sh}\left(\phi _3-\fra\right)
   h_0{}^{(0,2,0)}\left(\fra,\fra,\phi _3\right)-2 \text{sh}\left(\phi _3-\fra\right) h_0{}^{(1,1,0)}\left(\fra,\fra,\phi _3\right)\\
   &+\frac{1}{2}
   \text{sh}\left(\phi _3-\fra\right) h_0{}^{(2,0,0)}\left(\fra,\fra,\phi _3\right)
\eea
where $h_{0}^{(a,b,c)}(\phi_1,\phi_2,\phi_3)$ denotes the $a,b,c$ derivatives of the first, second, third variables, respectively. We still have a single pole here and the residue is still non-zero at $\phi_3=\fra$:
\bea
(\phi_1,\phi_2,\phi_3)=(\fra,\fra,\fra).
\eea
This is a new phenomenon not appearing in the PT3 counting or the three-legs examples discussed before and we can place three boxes at the same position. Actually this phenomenon is what one would have expected because we had a third order pole at the origin. For this case, the non-zero JK-residue is computed as
\bea
\frac{1}{6}\mathcal{Z}_{(0,0,0)}=\left(\frac{\sh(\frb-\fra)\sh(\eps_{12})\sh(\eps_{13})\sh(\eps_{23})}{\sh(\eps_1)\sh(\eps_2)\sh(\eps_3)\sh(\eps_{123})}\right)^{3}.
\eea

Let us consider the other poles with $\phi_3=\fra-\eps_{1,2,3,4}$. Using the quadrality symmetry, it is enough to consider the case when $\phi_3=\fra-\phi_1$:
\bea
\prod_{I=1}^{3}\mathcal{Z}_{\scube\,\scube\,\scube\,\scube}^{\D8\tbar\D2\tbar\D0}(\fra,\mathfrak{b},\phi_{I})\prod_{I<J}\mathcal{Z}^{\D0\tbar\D0}(\phi_I,\phi_J)&=\frac{\sh(\phi_1-\phi_2)^{2}\sh(\phi_3-\fra+\eps_1)^{2}h_{1}(\phi_1,\phi_2,\phi_3)}{\sh(\phi_1-\fra)^{3}\sh(\phi_2-\fra)^3\sh(\phi_3-\phi_1+\eps_1)\sh(\phi_3-\phi_2+\eps_1)}.
\eea
Taking the residue gives
\bea
&\underset{\phi_2=\fra}{\Res}\underset{\phi_1=\fra}{\Res}\prod_{I=1}^{3}\mathcal{Z}_{\scube\,\scube\,\scube\,\scube}^{\D8\tbar\D2\tbar\D0}(\fra,\mathfrak{b},\phi_{I})\prod_{I<J}\mathcal{Z}^{\D0\tbar\D0}(\phi_I,\phi_J)\\
=&-\frac{h_1{}^{(0,1,0)}\left(\fra,\fra,\phi _3\right) \text{ch}\left(-\fra+\epsilon _1+\phi _3\right)}{2 \text{sh}\left(-\fra+\epsilon _1+\phi
   _3\right)}-\frac{h_1{}^{(1,0,0)}\left(\fra,\fra,\phi _3\right) \text{ch}\left(-\fra+\epsilon _1+\phi _3\right)}{2
   \text{sh}\left(-\fra+\epsilon _1+\phi _3\right)}\\
   &+\frac{1}{2} h_1{}^{(0,2,0)}\left(\fra,\fra,\phi _3\right)-2 h_1{}^{(1,1,0)}\left(\fra,\fra,\phi
   _3\right)+\frac{1}{2} h_1{}^{(2,0,0)}\left(\fra,\fra,\phi _3\right)
\eea
and we have a single pole at $\phi_3=\fra-\eps_{1}$. Thus, the following poles give non-zero JK-residues
\bea
(\phi_1,\phi_2,\phi_3)=(\fra,\fra,\fra-\eps_{1,2,3,4}).
\eea
The residues are complicated so we omit the explicit expression.

Let us next assume $(\phi_1,\phi_2)=(\fra,\fra-\eps_{1,2,3,4})$. Using the quadrality symmetry, it is enough to consider $(\phi_1,\phi_2)=(\fra,\fra-\eps_1)$. JK-residue picks the poles at
\bea
\phi_3=\fra,\quad \phi_3=\phi_2-\eps_{1,2,3,4}.
\eea
The pole coming from $\phi_3=\fra$ actually reduces to the previous computation, so let us focus on $\phi_3=\phi_2-\eps_{1,2,3,4}=\fra-2\eps_{1},\fra-\eps_{12,13,14}$. Focusing at the $\phi_3=\fra-2\eps_1$, the integrand schematically takes the form as
\bea
\prod_{I=1}^{3}\mathcal{Z}_{\scube\,\scube\,\scube\,\scube}^{\D8\tbar\D2\tbar\D0}(\fra,\mathfrak{b},\phi_{I})\prod_{I<J}\mathcal{Z}^{\D0\tbar\D0}(\phi_I,\phi_J)&=\frac{\sh(\phi_2-\fra+\eps_1)^{2}h_{2}(\phi_1,\phi_2,\phi_3)}{\sh(\phi_1-\fra)^{3}\sh(\phi_2-\phi_1+\eps_1)\sh(\phi_3-\phi_2+\eps_1)}
\eea
and after taking the residue at $\phi_1=\fra,\phi_2=\fra-\eps_1$, we have
\bea
\underset{\phi_2=\fra-\eps_1}{\Res}\underset{\phi_1=\fra}{\Res}\prod_{I=1}^{3}\mathcal{Z}_{\scube\,\scube\,\scube\,\scube}^{\D8\tbar\D2\tbar\D0}(\fra,\mathfrak{b},\phi_{I})\prod_{I<J}\mathcal{Z}^{\D0\tbar\D0}(\phi_I,\phi_J)=\frac{h_{2}(\fra,\fra,\phi_3)}{\sh(\phi_3-\fra+2\eps_1)}
\eea
and thus the pole is non-zero. After a similar analysis, one will see that poles such as $\phi_3=\fra-\eps_{12,13,14}$ disappear.

Summarizing, the poles giving non-zero JK-residues at level three are
\bea
(\fra,\fra,\fra),\quad (\fra,\fra,\fra-\eps_{1,2,3,4}),\quad (\fra,\fra-\eps_{a},\fra-2\eps_{a})\,\,(a\in\four).
\eea

\section{PT4 counting with surface boundary conditions}\label{sec:PT4-JK-surface}

Using the contour integrand in the DT surface counting, we can set the $\eta=(-1,-1,\ldots,-1)$ and evaluate it. This gives the PT surface counting. We explicitly perform computations for some examples up to three surfaces. Generalizations to situations when we have four, five, and six surfaces is straight forward, but we omit the explicit analysis.

\subsection{One surface}\label{sec:PT4onesurface}
Using the symmetry, we can focus on the case when the surfaces span the 12-plane.
Let us first consider the case when the surface boundary conditions are
\bea
\lambda_{12}=1,\quad \lambda_{A}=\varnothing \quad (A\neq 12)
\eea
and the minimal solid partition is
\bea
\adjustbox{valign=c}{\includegraphics[width=8cm]{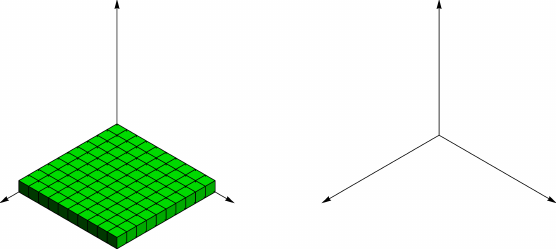}}
\eea
The framing node contribution is 
\bea
\mathcal{Z}^{\D8\tbar\D4\tbar\D0}_{\{\lambda_{A}\}}(\fra,\mathfrak{b},\phi_{I})&=\frac{\sh(\phi_{I}-\mathfrak{b})\sh(\phi_{I}-\fra-\epsilon_{34})}{\sh(\phi_{I}-\fra-\epsilon_{3})\sh(\phi_{I}-\fra-\epsilon_{4})}.
\eea
Obviously for this case, choosing the reference vector $\eta=\tilde{\eta}_{0}$ picks no pole and thus the PT4 partition function is trivial:
\bea
\mathcal{Z}^{\PT\tbar\JK}_{\{\lambda_{A}\}}[\fq,\mu;q_{1,2,3,4}]=1.
\eea

Actually, for the general case when we have multiple D4$_{12}$-branes, the PT4 partition function is always trivial. For the case
\bea
\lambda_{12}=\sum_{i=1}^{\ell(\lambda_{12})}\sum_{j=1}^{\lambda_{12,i}}q_{4}^{i-1}q_{3}^{j-1}
\eea
where we have $\lambda_{12,i}$ surfaces for each layer in the 4-direction. The framing node is computed as
\bea
\mathcal{Z}^{\D8\tbar\D4\tbar\D0}_{\{\lambda_{A}\}}(\fra,\frb,\phi_{I})&=\frac{\sh(\phi_I-\frb)}{\sh(\phi_I-\fra)}\prod_{i=1}^{\ell(\lambda_{12})}\prod_{j=1}^{\lambda_{12,i}}\mathcal{Z}^{\overline{\D4}_{12}\tbar\D0}(\fra+(i-1)\eps_4+(j-1)\eps_{3},\phi_I)\\
&=\sh(\phi_I-\frb)\frac{\prod\limits_{(i,j)\in R(\lambda_{12})}\sh(\phi_I-\fra-i\eps_4-j\eps_3)}{\prod\limits_{(i,j)\in A(\lambda_{12})}\sh(\phi_I-\fra-(i-1)\eps_4-(j-1)\eps_3)},
\eea
where $A(\lambda_{12}),R(\lambda_{12})$ are the addable and removable boxes of the Young diagram, respectively. The denominator is identified with the addable boxes of the minimal solid partition. Thus, all of the poles are picked up when the reference vector is $\eta=\eta_{0}$ and no poles are picked up when $\eta=\tilde{\eta}_0$. 

\begin{proposition}
    When only one of the surface boundary conditions is non-trivial, the PT4 partition function is trivial:
    \bea
\mathcal{Z}^{\PT\tbar\JK}_{\{\lambda_{A}\}}[\fq,\mu;q_{1,2,3,4}]=1.
    \eea
\end{proposition}

\subsection{Two surfaces}\label{sec:PT4twosurface}

Let us move on to the situation when we have two surfaces. Using the symmetry, we can choose one of the surfaces to be the $12$-surface. We have two situations depending on the type of the other surface.
\begin{itemize}
    \item \textbf{Case 1:} The intersection of the two surfaces is a stack of one-dimensional objects:
    \bea
        \lambda_{23},\,\,\lambda_{31},\,\,\lambda_{14},\,\,\lambda_{24}.
    \eea
   For this situation, we say that the intersection is one-dimensional. We call this \textbf{D2-brane like}, shortly D2-like.
    \item \textbf{Case 2:} The intersection of the two surfaces is a stack of zero-dimensional objects:
    \bea
    \lambda_{34}.
    \eea
    For this situation, we say that the intersection is zero-dimensional. We call this \textbf{D0-brane like}, shortly D0-like.
\end{itemize}

For case 1, using the quadrality symmetry again, we can focus on the pairs of surfaces $(\lambda_{12},\lambda_{23})$. Let us denote the Young diagrams as
\bea
\lambda_{12}=(\lambda_{12,1},\lambda_{12,2},\ldots,),\quad \lambda_{23}=(\lambda_{23,1},\lambda_{23,2},\ldots)
\eea
where $\lambda_{12,l},\lambda_{23,l}$ are the numbers of the 12, 34-surfaces at the $l$-th layer in the fourth direction. The minimal solid partition of such situation is illustrated as
\bea\label{eq:PT4_twosurface_1}
\adjustbox{valign=c}{\includegraphics[width=13cm]{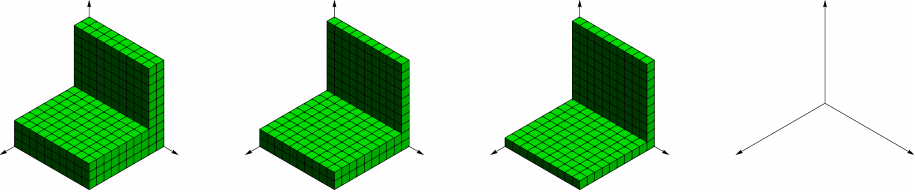}}
\eea
For each layer, the the intersection of the 12, 23 surfaces is a one-dimensional object extending in the 2-direction. More specifically, they are one-dimensional rows of boxes with coordinates
\bea
(i-1)\eps_1+(k-1)\eps_3,\quad 1\leq i\leq \lambda_{23,l},\quad 1\leq k\leq \lambda_{12,l}.
\eea
For other choices of surfaces, the illustration might change due to the $(1,3)$ decomposition but one can pick a nice frame where the figures look like the one above.

For case 2, the minimal solid partition is illustrated as
\bea\label{eq:PT4_twosurface_2}
\adjustbox{valign=c}{\includegraphics[width=13cm]{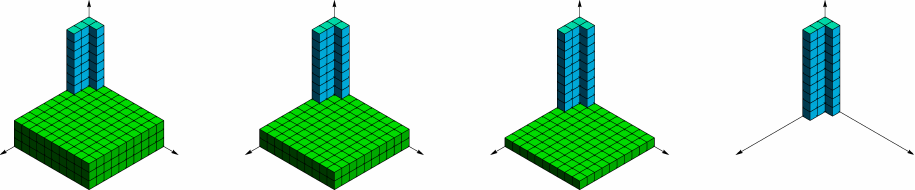}}\cdots
\eea
For each layer, we have $\lambda_{12,l}\,\,(l=1,2,\ldots)$ surfaces spanning the 12-surface and a leg boundary condition $\lambda_{34}$ extending in the 3-direction. The intersection are then stack of boxes whose coordinates are
\bea\label{eq:PT4-twosurface-zerodim-intersection}
(i-1)\eps_{1}+(j-1)\eps_2+(k-1)\eps_3+(l-1)\eps_4,\quad (i,j)\in\lambda_{34},\quad (k,l)\in\lambda_{12}.
\eea
This is the reason why we are calling it a zero-dimensional intersection. Note that we only have \textit{finite} number of such boxes. In particular, we have $|\lambda_{12}|\times |\lambda_{34}|$ boxes.

\subsubsection{Case 1} \label{sec:twosurface-case1}
Let us study the PT4 partition function of the case 1. The framing node is determined by
\bea
&\mathcal{Z}^{\D8\tbar\D4\tbar\D0}_{\{\lambda_{A}\}}(\fra,\frb,\phi_{I})=\mathcal{Z}^{\D8\tbar\D0}(\fra,\phi_I)\mathcal{Z}^{\D8'\tbar\D0}(\frb,\phi_I)\\
&\times\prod_{l=1}^{\infty}\left(\prod_{k=1}^{\lambda_{12,l}}\mathcal{Z}^{\overline{\D4}_{12}\tbar\D0}(\fra+(l-1)\eps_4+(k-1)\eps_3,\phi_I)\prod_{i=1}^{\lambda_{23,l}}\mathcal{Z}^{\overline{\D4}_{23}\tbar\D0}(\fra+(l-1)\eps_4+(i-1)\eps_1+\lambda_{12,l}\eps_3,\phi_I)\right),
\eea
where we shifted the coordinates of the $23$-surfaces to avoid the one-dimensional intersection. Using the infinite product expression of the D8-brane \eqref{eq:D6D8-relation}, we can rewrite the framing node contribution as
\bea\label{eq:PT4-twosurface-framing-1}
\mathcal{Z}^{\D8\tbar\D4\tbar\D0}_{\{\lambda_{A}\}}(\fra,\frb,\phi_{I})&=\mathcal{Z}^{\D8'\tbar\D0}(\frb,\phi_I)\mathcal{Z}^{\D8\tbar\D0}(\fra+N\eps_4,\phi_I)\\
&\times \prod_{l=1}^{N}\mathcal{Z}^{\D6_{\bar{4}}\tbar\D0}(\fra+(l-1)\eps_4+\lambda_{23,l}\eps_1+\lambda_{12,l}\eps_3,\phi_I)\\
&=\frac{\sh(\phi_I-\frb)}{\sh(\phi_I-\fra-N\eps_4)}\prod_{l=1}^{N}\frac{\sh(\phi_I-\fra-l\eps_4-\lambda_{23,l}\,\eps_1-\lambda_{12,l}\,\eps_3)}{\sh(\phi_I-\fra-(l-1)\eps_4-\lambda_{23,l}\,\eps_1-\lambda_{12,l}\,\eps_3)}
\eea
where $N=\max \left(\ell(\lambda_{12}),\ell(\lambda_{23})\right)$ and we used the \eqref{eq:D4D6-relation} to simplify the expression. 

Looking at the poles, one can see that all of the poles are positions of the addable boxes of the minimal solid partition. Thus, choosing the reference vector $\eta=\eta_{0}$ picks up all the poles in the denominator, while choosing the reference vector $\eta=\tilde{\eta}_{0}$ picks no poles. Therefore, similar to the one-surface case in section~\ref{sec:PT4onesurface} the PT4 partition function is trivial:
\bea
\mathcal{Z}^{\PT\tbar\JK}_{\{\lambda_{A}\}}[\fq,\mu;q_{1,2,3,4}]=1.
\eea

Actually, one can interpret the expression \eqref{eq:PT4-twosurface-framing-1} as a stack of $N$ D6$_{\bar{4}}$-branes with the Coulomb branch parameters $(\fra+(l-1)\eps_{4}+\lambda_{23,l}\eps_{1}+\lambda_{12,l}\eps_{3})_{l=1}^{N}$ and a pair of D8-D8$'$ brane with $\fra+N\eps_{4}$ and $\frb$, with no nontrivial boundary conditions. For a general stack of $\D6_{\bar{4}}$-branes with $\fra_1,...\fra_N$, the partition function does not depend on them. Thus, tuning the parameters to be $\fra_l=\fra+(l-1)\eps_4+\lambda_{23,l}\eps_1+\lambda_{12,l}\eps_3$ does not change the partition function and we still have the rank $N$ D6 partition function.\footnote{In general, the result obtained by first evaluating the JK residue and then tuning the parameters does not coincide with the result obtained by first tuning the parameters and then evaluating the JK residue. However, for the examples discussed in this paper, the two procedures give the same result.} Since the $N$ D6-branes can be understood as a result of tachyon condensation of $N$ pairs of D8-D8$'$ with the distance $q_{4}$, the combined system with the pair of D8-D8$'$ is obtained by setting $\mu_{1}=\mu_{2}=\cdots =\mu_{N}=q_{4}$ and $\mu_{N+1}=\mu q_{4}^{-N}$ in \eqref{eq:D8-PEformula}, where $\mu=e^{\frb-\fra}$. Therefore, the DT partition function after choosing the reference vector $\eta_{0}$ is 
\bea
\mathcal{Z}^{\DT\tbar\JK}_{\{\lambda_{A}\}}[\fq,\mu;q_{1,2,3,4}]&=\PE\left[\frac{[q_{14}][q_{24}][q_{34}]}{[q_{1}][q_{2}][q_{3}][q_{4}]}\frac{[(\mu q_{4}^{-N}) q_{4}^{N}]}{[\fq (\mu^{1/2} q_{4}^{-N/2}) q_{4}^{N/2}][\fq (\mu^{-1/2} q_{4}^{N/2}) q_{4}^{-N/2}]}\right]\\
&=\PE\left[\frac{[q_{14}][q_{24}][q_{34}]}{[q_{1}][q_{2}][q_{3}][q_{4}]}\frac{[\mu ]}{[\fq \mu^{1/2}][\fq \mu^{-1/2}]}\right]=\mathcal{Z}^{\D8}[\fq,\mu;q_{1,2,3,4}],
\eea
which is just the rank one magnificent four partition function.



Due to the triviality of the PT4 partition function, one can observe the following relation
\bea\label{eq:twosurface-DTPT}
\mathcal{Z}^{\DT\tbar\JK}_{\{\lambda_{A}\}}[\fq,\mu;q_{1,2,3,4}]=\mathcal{Z}^{\PT\tbar\JK}_{\{\lambda_{A}\}}[\fq,\mu;q_{1,2,3,4}]\mathcal{Z}^{\D8}[\fq,\mu;q_{1,2,3,4}].
\eea
Actually, this is the consequence of the DT/PT correspondence of surface counting.

Note that although we did not mention in the previous section, the one-surface case is also included in this situation and the DT/PT correspondence is trivially satisfied.

\begin{proposition}
    When only two of the surface boundary conditions are nontrivial and they share a one-dimensional intersection, the PT4 partition function is trivial:
    \bea
\mathcal{Z}^{\PT\tbar\JK}_{\{\lambda_{A}\}}[\fq,\mu;q_{1,2,3,4}]=1.
\eea

\end{proposition}

\subsubsection{Case 2}
For the case 2, the framing node is determined as
\bea
&\mathcal{Z}^{\D8\tbar\D4\tbar\D0}_{\{\lambda_{A}\}}(\fra,\frb,\phi_{I})=\mathcal{Z}^{\D8\tbar\D0}(\fra,\phi_I)\mathcal{Z}^{\D8'\tbar\D0}(\frb,\phi_I)\\
&\times\prod_{l=1}^{\infty}\left(\prod_{k=1}^{\lambda_{12,l}}\mathcal{Z}^{\overline{\D4}_{12}\tbar\D0}(\fra+(l-1)\eps_4+(k-1)\eps_3,\phi_I)\prod_{(i,j)\in\lambda_{34}}\mathcal{Z}^{\overline{\D2}_{3}\tbar\D0}(\fra+(i-1)\eps_1+(j-1)\eps_2+\lambda_{12,l}\eps_3)\right),
\eea
where we decomposed the 34-surface into one-dimensional objects extending in the 3-direction. Again using the D8-D6 relation and the D6-D4 relation, we can rewrite this as
\bea
\mathcal{Z}^{\D8\tbar\D4\tbar\D0}_{\{\lambda_{A}\}}(\fra,\frb,\phi_{I})=\mathcal{Z}^{\D8'\tbar\D0}(\frb,\phi_I)\prod_{l=1}^{\infty}\mathcal{Z}^{\D6_{\bar{4}}\tbar\D2\tbar\D0}_{\varnothing\,\varnothing\,\lambda_{34}}(\fra+(l-1)\eps_4+\lambda_{12,l}\eps_3,\phi_{I}).
\eea
Namely, each layer is understood as a minimal plane partition with a one-leg boundary condition but the origin is shifted by $\lambda_{12,l}\eps_3$. Thus, the PT4 counting of this situation should be understood as multiple layers of one-leg PT3 counting whose origin is shifted in a specific way. Moreover, similar to the leg boundary condition cases, the boxes are stacked in way such that the gravity points the $(1,1,1,1)$-direction.

To make the discussion more concrete, let us first study the following examples. Such examples were originally studied in \cite{Bae:2022pif,Bae:2024bpx}.

\paragraph{Example 1: $\lambda_{12}=\lambda_{34}=1$}The minimal solid partition is
\bea
\adjustbox{valign=c}{\includegraphics[width=10cm]{PT4Figures/surface_ex1.pdf}}\cdots
\eea
The framing node contribution is 
\bea
\mathcal{Z}^{\D8\tbar\D4\tbar\D0}_{\{\lambda_{A}\}}(\fra,\mathfrak{b},\phi_{I})=\frac{\sh(-\phi_{I}+\mathfrak{b})}{\sh(-\phi_{I}+\fra)}\frac{\sh(\phi_{I}-\fra+\epsilon_{1,2,3,4})}{\sh(\phi_{I}-\fra-\epsilon_{13,23,14,24})}
\eea
Choosing the eta vector as $\eta=\tilde{\eta}_{0}$, the poles picked are
\bea
-\phi_{I}+\fra=0,\quad \phi_{I}-\phi_{J}=-\epsilon_{1,2,3,4}.
\eea
For the one instanton case, we have
\bea
\phi_{1}=\fra.
\eea
For the next level, the poles are chosen as
\bea
\phi_{2}=\phi_{1}-\epsilon_{1,2,3,4}=\fra-\epsilon_{1,2,3,4}.
\eea
All of the poles are canceled by the numerator of the framing node contribution
\bea
\prod_{I}\sh(\phi_{I}-\fra+\epsilon_{1,2,3,4})
\eea
and thus the only nontrivial residue is
\bea
\mathcal{Z}^{\PT\tbar\JK}_{\{\lambda_{A}\}}[1]&=-\frac{\sh(-\epsilon_{14,24,34})}{\sh(-\epsilon_{1,2,3,4})}\underset{\phi_{1}=\fra}{\Res}\frac{\sh(-\phi_{I}+\mathfrak{b})}{\sh(-\phi_{I}+\fra)}\frac{\sh(\phi_{I}-\fra+\epsilon_{1,2,3,4})}{\sh(\phi_{I}-\fra-\epsilon_{13,23,14,24})}\\
&=\frac{\sh(-\epsilon_{14,24,34})}{\sh(-\epsilon_{1,2,3,4})}\frac{\sh(\mathfrak{b}-\fra)\sh(\epsilon_{1,2,3,4})}{\sh(-\epsilon_{13,23,14,24})}\\
&=\frac{\sh(\mathfrak{b}-\fra)\sh(\epsilon_{12})}{\sh(\epsilon_{13})\sh(\epsilon_{23})}=\frac{[\mu][q_{12}]}{[q_{13}][q_{23}]}.
\eea
We finally have
\bea
\mathcal{Z}^{\PT\tbar \JK}_{\{\lambda_{A}\}}[\mathfrak{q},\mu;q_{1,2,3,4}]=1+\mathfrak{q}\frac{[\mu][q_{12}]}{[q_{13}][q_{23}]},
\eea
which indeed matches with \cite[Example~1.9, 6.18.(1)]{Bae:2024bpx}. This PT vertex was called the PT$_{0}$ vertex there.

An observation here is that the poles picked up from the JK-residue formalism is the poles of the boxes living the intersection of the two surfaces. For this case, we only have one box at $\fra$. Moreover, the PT4 partition function terminates and we only have finite number of terms.

\paragraph{Example 2: $\lambda_{12}=1+q_{3},\,\lambda_{34}=1$}The minimal solid partition is
\bea
\adjustbox{valign=c}{\includegraphics[width=10cm]{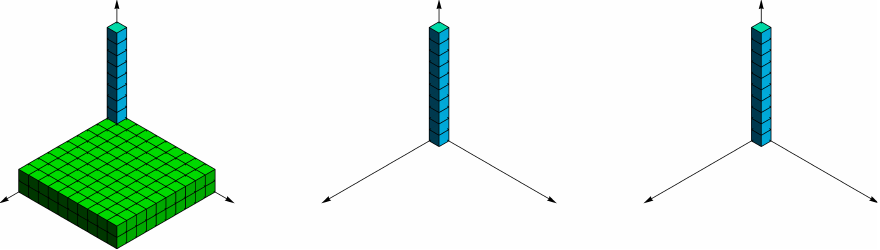}}\cdots
\eea
The framing node contribution is
\bea
\mathcal{Z}^{\D8\tbar\D4\tbar\D0}_{\{\lambda_{A}\}}(\fra,\mathfrak{b},\phi_{I})=\frac{\sh(\mathfrak{b}-\phi_{I})}{\sh(\fra+\epsilon_{3}-\phi_{I})}\frac{\sh(\phi_{I}-\fra+\epsilon_{3})\sh(\phi_{I}-\fra-\epsilon_{3}+\epsilon_{1,2})\sh(\phi_{I}-\fra+\epsilon_{4}-\epsilon_{3})}{\sh(\phi_{I}-\fra-\epsilon_{14})\sh(\phi_{I}-\fra-\epsilon_{24})\sh(\phi_{I}-\fra-\epsilon_{13^{2}})\sh(\phi_{I}-\fra-\epsilon_{23^{2}})}
\eea
With $\eta=\tilde{\eta}_{0}$, the poles are
\bea
\phi_{I}=\fra+\epsilon_{3},\quad \phi_{I}-\phi_{J}=-\epsilon_{1,2,3,4}.
\eea
For the first level, we have
\bea
\phi_{1}=\fra+\epsilon_{3}
\eea
giving the residue
\bea
\mathcal{Z}^{\PT\tbar\JK}_{\{\lambda_{A}\}}[1]&=-\frac{\sh(-\epsilon_{14,24,34})}{\sh(-\epsilon_{1,2,3,4})}\underset{\phi_{1}=\fra+\epsilon_{3}}{\Res}\mathcal{Z}^{\D8\tbar\D4\tbar\D0}_{\{\lambda_{A}\}}(\fra,\mathfrak{b},\phi_{1})=\frac{[q_{3}^{-1}\mu][q_{12}][q_{3}^{2}]}{[q_{3}][q_{1}q_{3}^{2}][q_{2}q_{3}^{2}]}.
\eea

For the second level, we have the poles
\bea
\phi_{2}=\phi_{1}-\epsilon_{1,2,3,4}=\fra,\fra+\epsilon_{3}-\epsilon_{1,2,4}.
\eea
The poles $\phi_{2}=\fra+\epsilon_{3}-\epsilon_{1,2,4}$ are canceled by the numerator
\bea
\prod_{I=1}^{k}\sh(\phi_{I}-\fra-\epsilon_{3}+\epsilon_{1,2,4})
\eea
and the non-zero JK-residue is evaluated as
\bea
\mathcal{Z}^{\PT\tbar\JK}_{\{\lambda_{A}\}}[2]&=\left(\frac{\sh(-\epsilon_{14,24,34})}{\sh(-\epsilon_{1,2,3,4})}\right)^{2} \underset{\phi_{2}=\fra}{\Res}\,\underset{\phi_{1}=\fra+\epsilon_{3}}{\Res}\prod_{I=1}^{2}\mathcal{Z}^{\D8\tbar\D4\tbar\D0}_{\{\lambda_{A}\}}(\fra,\mathfrak{b},\phi_{I})\mathcal{Z}^{\D0\tbar\D0}(\phi_{1},\phi_{2}) \\
&=\frac{[\mu][q_{3}^{-1}\mu][q_{12}][q_{12}q_{3}^{-1}]}{[q_{13}][q_{23}][q_{1}q_{3}^{2}][q_{2}q_{3}^{2}]}.
\eea 
Therefore, the PT4 partition function is given as
\bea
\mathcal{Z}^{\PT\tbar \JK}_{\{\lambda_{A}\}}[\mathfrak{q},\mu;q_{1,2,3,4}]=1+\mathfrak{q}\frac{[q_{3}^{-1}\mu][q_{12}][q_{3}^{2}]}{[q_{3}][q_{1}q_{3}^{2}][q_{2}q_{3}^{2}]}+\mathfrak{q}^{2}\frac{[\mu][q_{3}^{-1}\mu][q_{12}][q_{12}q_{3}^{-1}]}{[q_{13}][q_{23}][q_{1}q_{3}^{2}][q_{2}q_{3}^{2}]},
\eea
which indeed matches with \cite[Example 6.18.(3)]{Bae:2024bpx}.

Similar to the previous case, the poles at $\fra+\eps_3,\fra$ belong to the intersection of the two surfaces and the PT4 partition function terminates at level two, eventually giving only finite number of terms. Moreover, for this case, we have a \textit{melting rule} such that we can place a box at $\fra+\eps_3$ and then place a box at $\fra$. This is interpreted as a part of the PT3 counting rule with one leg extending at the three direction whose origin is $\fra+2\eps_3$.

\paragraph{Example 3: $\lambda_{12}=1+q_{3}+q_{4},\lambda_{34}=1$}The minimal solid partition is
\bea
\adjustbox{valign=c}{\includegraphics[width=10cm]{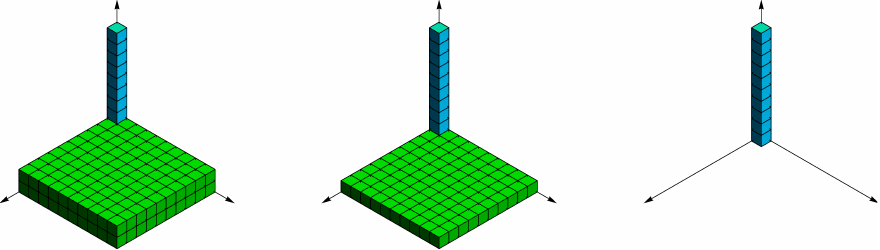}}\cdots
\eea
The framing node contribution is
\bea
&\mathcal{Z}^{\D8\tbar\D4\tbar\D0}_{\{\lambda_{A}\}}(\fra,\mathfrak{b},\phi_{I})=\frac{\sh\left(\phi_I-\fra\right) \sh\left(\phi_I-\frb\right) \sh\left(-\fra+\eps_4- \epsilon _3+\phi_I\right) \sh\left(-\fra+\eps_3- \epsilon _4+\phi_I\right)
   \sh\left(-\fra+\eps_2- \epsilon _4+\phi_I\right) }{\sh\left(\fra+\epsilon _3-\phi_I\right) \sh\left(\fra+\epsilon _4-\phi_I\right) \sh\left(-\fra-\epsilon _1-2 \epsilon _3+\phi_I\right) \sh\left(-\fra-\epsilon _2-2 \epsilon
   _3+\phi_I\right)
  }\\
   &\times\frac{\sh\left(-\fra+\eps_1- \epsilon _4+\phi_I\right) \sh\left(-\fra+\eps_2- \epsilon _3+\phi
   _I\right) \sh\left(-\fra+\eps_1- \epsilon _3+\phi
   _I\right)}{ \sh\left(-\fra-\epsilon _1-2 \epsilon _4+\phi_I\right) \sh\left(-\fra-\epsilon
   _2-2 \epsilon _4+\phi_I\right) \sh\left(-\fra-\epsilon _1-\epsilon _3-\epsilon _4+\phi_I\right)
   \sh\left(-\fra-\epsilon _2-\epsilon _3-\epsilon _4+\phi_I\right)} .
\eea

Choosing the reference vector $\eta=\tilde{\eta}_{0}$, for level one, the poles picked up are
\bea
\phi_1=\fra+\eps_3,\quad \fra+\eps_4.
\eea

Let us next consider the next level. Assume we picked up the pole at $\phi_1=\fra+\eps_3$. For the next level, we can choose
\bea
\phi_2=\fra+\eps_3,\,\,\fra+\eps_4,\quad \phi_2=\phi_1-\eps_{1,2,3,4}.
\eea
The pole $\phi_2=\fra+\eps_3=\phi_1$ is canceled by the numerator of the D0-brane contribution. The poles $\phi_2=\phi_1-\eps_{1,2,3,4}=\fra,\fra+\eps_{3}-\eps_{1,2,4}$ are all canceled by the numerators of the framing node contribution. We thus have
\bea
(\phi_1,\phi_2)=(\fra+\eps_3,\fra+\eps_4).
\eea
Starting from $\phi_1=\fra+\eps_4$ also gives the same configuration.

For level three, the following poles are picked
\bea
\phi_3=\phi_1-\eps_{1,2,3,4},\quad \phi_2-\eps_{1,2,3,4}.
\eea
One will see that all of the poles except $\phi_3=\fra$ vanish. We thus have
\bea
(\phi_1,\phi_2,\phi_3)=(\fra+\eps_3,\fra+\eps_4,\fra).
\eea

For higher levels, similar to the previous examples, the partition function vanishes and thus we obtain
\bea
\mathcal{Z}^{\PT\tbar \JK}_{\{\lambda_{A}\}}[\mathfrak{q},\mu;q_{1,2,3,4}]=1+\fq\left(\mathcal{Z}_{(\eps_3)}+\mathcal{Z}_{(\eps_4)}\right)+\fq^{2}\mathcal{Z}_{(\eps_3,\eps_4)}+\fq^{3}\mathcal{Z}_{(\eps_3,\eps_4,0)},
\eea
where we denoted the JK-residue of the integrand without the Weyl group factor at $\fra+x,\fra+y,\fra+z$ as $\mathcal{Z}_{(x,y,z)}$.

Let us interpret the results. The poles $\fra+\eps_3,\fra+\eps_4$ correspond to the boxes of the PT3 counting for layer $1,2$, respectively. For layer one, since the origin is shifted to $\fra+2\eps_3$, if one performs the PT3 counting, the first box is $\fra+\eps_3$. For layer two, due to the shift $\fra+\eps_4+\eps_3$, the first box is $\fra+\eps_4$. Choosing the reference vector $\eta=\tilde{\eta}_0$ imposes the condition that the gravity is pointing the $(1,1,1,1)$ direction and thus boxes need to be supported from the positive direction. Since both boxes are supported, we can place both of them at level one.

Let us consider the next level. Assume we have a box at $\fra+\eps_3$. We trivially can place the boxes at $\fra+\eps_{4}$, because it does not break the melting rule. Following the PT3 counting rules, one would expect that we can place a box at $\fra$. However, this is not allowed, because it is not supported by any box from the positive direction of the 4-axis.

Similarly, assume we have a box at $\fra+\eps_4$. We cannot place a box at $\fra+\eps_4-\eps_3$ because it is not supported from the positive direction. Accordingly, we cannot place any boxes at the layer $l\geq 3$.

For level three, we can only additionally place a box at $\fra$, which does not break the melting rule. The allowed configuration is then $(\eps_3,\eps_4,0)$.

For higher levels, one would expect we can add a boxes at positions with negative coordinates of the 3-axis. However, such boxes are not supported from the positive direction of the 4-axis and thus they are not allowed. Therefore, we do not have any configurations for higher levels.

\paragraph{General case}
Given the above three examples, we can already deduce the PT4 box-counting rules for the situation when we have two surfaces with zero-dimensional intersection. The rules are as follows.

\begin{Rule}\label{rule:PT4twosurface-case2}
    When only two of the surfaces are nontrivial, and the intersection of them is zero-dimensional, the PT4 box counting rules and the PT4 partition function are given as follows.
\begin{itemize}
    \item We first start from the minimal solid partition in \eqref{eq:PT4_twosurface_2}. As mentioned, each layer is understood as a minimal plane partition with one-leg boundary conditions whose origin is shifted $\lambda_{12,l}$ in the positive direction of the 3-axis.
    \item Similar to usual PT3 counting, we then extend the one-leg boundary conditions to the negative direction and then remove the boxes at the positive quadrant. Since the origin is shifted, we obtain
    \bea
    \adjustbox{valign=c}{\includegraphics[width=12cm]{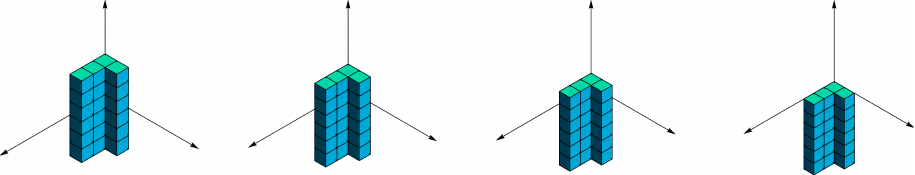}}\cdots
    \eea
    Such boxes give the hollow structure in which the boxes are placed.
    \item Boxes are stacked inside the hollow structure in a way such that the gravity is pointing $(1,1,1,1)$. Namely, boxes are supported from the positive directions.
    \item Since the surfaces $\lambda_{34}$ extends infinitely in the 4-direction, while the surfaces $\lambda_{12}$ terminate at $\ell(\lambda_{12})$ in the 4-direction, we can not place boxes at layers $l\geq \ell(\lambda_{12})+1$. This is because, due to the previous melting rule, the boxes also need to be supported from the positive direction of the 4-axis. So that we can place a box at the layer $\ell(\lambda_{12})+1$, we also need a box at layer $\ell(\lambda_{12})+2$, which inductively means we need infinite number of boxes extending in the 4-direction.

    We thus can only place boxes at the intersection of the two surfaces \eqref{eq:PT4-twosurface-zerodim-intersection} (the pink positions):
     \bea
    \adjustbox{valign=c}{\includegraphics[width=12cm]{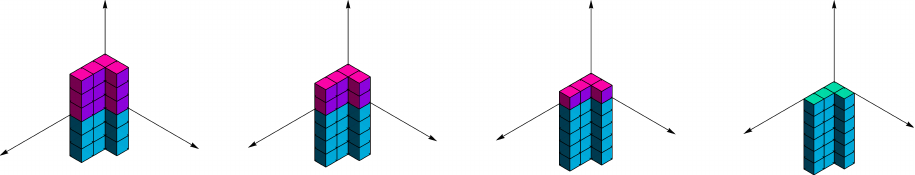}}\cdots
    \eea
    \item The maximal number of boxes we can add is $|\lambda_{12}|\times |\lambda_{34}|$ and thus the PT4 partition function terminates at level $|\lambda_{12}|\times |\lambda_{34}|$:
    \bea
    \mathcal{Z}^{\PT\tbar \JK}_{\{\lambda_{A}\}}[\mathfrak{q},\mu;q_{1,2,3,4}]=\sum_{k=0}^{|\lambda_{12}|\times |\lambda_{34}|}\fq^{k}\mathcal{Z}^{\PT\tbar \JK}_{\{\lambda_{A}\}}[k].
    \eea
\end{itemize}

\end{Rule}

\subsection{Three surfaces}\label{sec:PT4threesurface}
Fixing one of the surface to be spanning the 12-plane, we have 10 possible cases with three surfaces. They are classified in the following three situations.
\begin{itemize}
    \item \textbf{Case 1:} All of the intersections are D2-like and the surfaces can be embedded in a $\mathbb{C}^{3}$ subspace:
    \bea
        (\lambda_{12},\lambda_{23},\lambda_{31}),\quad (\lambda_{12},\lambda_{41},\lambda_{42}).
    \eea
    \item \textbf{Case 2:} All of the intersections are D2-like but the union of the surfaces span $\mathbb{C}^{4}$:
    \bea
        (\lambda_{12},\lambda_{23},\lambda_{42}),\quad (\lambda_{12},\lambda_{31},\lambda_{41}).
    \eea
    \item \textbf{Case 3:} One of the three intersections is a D0-like intersection. Namely, two out of the three surfaces are transverse with each other:
    \bea
    (\lambda_{12},\lambda_{23},\lambda_{41}),\quad (\lambda_{12},\lambda_{23},\lambda_{34}),\quad (\lambda_{12},\lambda_{31},\lambda_{34}),\\
    (\lambda_{12},\lambda_{31},\lambda_{24}),\quad (\lambda_{12},\lambda_{41},\lambda_{34}),\quad (\lambda_{12},\lambda_{24},\lambda_{34})
    \eea
\end{itemize}

\subsubsection{Case 1}
Using the symmetry we can focus on the three surfaces $(\lambda_{12},\lambda_{23},\lambda_{31})$, whose minimal solid partition is
\bea\label{eq:PT4_threesurface_1}
\adjustbox{valign=c}{\includegraphics[width=13cm]{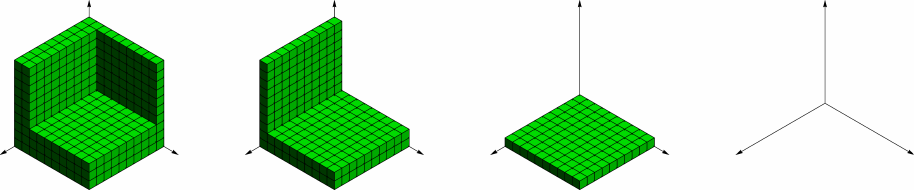}}
\eea
We denote the Young diagrams as
\bea
\lambda_{12}=(\lambda_{12,1},\ldots),\quad \lambda_{23}=(\lambda_{23,1},\lambda_{23,2},\ldots),\quad \lambda_{31}=(\lambda_{31,1},\lambda_{31,2},\ldots)
\eea
where $\lambda_{12,l},\lambda_{23,l},\lambda_{31,l}$ are the numbers of surfaces for each layer.

The framing node contribution is computed as
\bea
&\mathcal{Z}^{\D8\tbar\D4\tbar\D0}_{\{\lambda_{A}\}}(\fra,\frb,\phi_{I})=\mathcal{Z}^{\D8\tbar\D0}(\fra,\phi_I)\mathcal{Z}^{\D8'\tbar\D0}(\frb,\phi_I)\\
&\times\prod_{l=1}^{\infty}\left(\prod_{k=1}^{\lambda_{12,l}}\mathcal{Z}^{\overline{\D4}_{12}\tbar\D0}(\fra+(l-1)\eps_4+(k-1)\eps_3,\phi_I) \right. \prod_{i=1}^{\lambda_{23,l}}\mathcal{Z}^{\overline{\D4}_{23}\tbar\D0}(\fra+(l-1)\eps_4+(i-1)\eps_1+\lambda_{12,l}\eps_3,\phi_I)\\
&\times \left. \prod_{j=1}^{\lambda_{13,l}}\mathcal{Z}^{\overline{\D4}_{31}\tbar\D0}(\fra+(l-1)\eps_4+\lambda_{12,l}\eps_3+\lambda_{23,l}\eps_1+(j-1)\eps_2,\phi_I)    \right).
\eea
Similar to the case 1 of the two-surfaces case, using the infinite product expression of the D8-brane, we can rewrite it as
\bea
\mathcal{Z}^{\D8\tbar\D4\tbar\D0}_{\{\lambda_{A}\}}(\fra,\frb,\phi_{I})&=\mathcal{Z}^{\D8'\tbar\D0}(\frb,\phi_I)\mathcal{Z}^{\D8\tbar\D0}(\fra+N\eps_4,\phi_I)\\
&\times \prod_{l=1}^{N}\mathcal{Z}^{\D6_{\bar{4}}\tbar\D0}(\fra+(l-1)\eps_4+\lambda_{23,l}\eps_1+\lambda_{12,l}\eps_3+\lambda_{31,l}\eps_2,\phi_I)\\
&=\frac{\sh(\phi_I-\frb)}{\sh(\phi_I-\fra-N\eps_4)}\prod_{l=1}^{N}\frac{\sh(\phi_I-\fra-l\eps_4-\lambda_{23,l}\,\eps_1-\lambda_{12,l}\,\eps_3-\lambda_{31,l}\eps_2)}{\sh(\phi_I-\fra-(l-1)\eps_4-\lambda_{23,l}\,\eps_1-\lambda_{12,l}\,\eps_3-\lambda_{31,l}\eps_2)}
\eea
where $N=\max(\ell(\lambda_{12}),\ell(\lambda_{23}),\ell(\lambda_{31}))$. Note that again all of the poles correspond to the addable boxes of the minimal solid partition. 

The framing node implies that we can understand this setup as a stack of D8-D8$'$ brane and $N$ D6$_{\bar{4}}$-branes with their Coulomb branch parameters tuned. Picking the reference vector as $\eta=\eta_0$ and performing the JK-residue gives the following DT4 partition function
\bea
\mathcal{Z}^{\DT\tbar\JK}_{\{\lambda_{A}\}}[\fq,\mu;q_{1,2,3,4}]&=\PE\left[\frac{[q_{14}][q_{24}][q_{34}]}{[q_{1}][q_{2}][q_{3}][q_{4}]}\frac{[(\mu q_{4}^{-N}) q_{4}^{N}]}{[\fq (\mu^{1/2} q_{4}^{-N/2}) q_{4}^{N/2}][\fq (\mu^{-1/2} q_{4}^{N/2}) q_{4}^{-N/2}]}\right]\\
&=\mathcal{Z}^{\D8}[\fq,\mu;q_{1,2,3,4}]
\eea
which is similar to section~\ref{sec:twosurface-case1}. 

Moreover, since choosing the reference vector to be $\eta=\tilde{\eta}_{0}$ picks no poles, the PT partition function is trivial:
\bea
\mathcal{Z}^{\PT\tbar\JK}_{\{\lambda_{A}\}}[\fq,\mu;q_{1,2,3,4}]=1.
\eea

\subsubsection{Case 2}
Let us then move on to the situation when the union of surfaces span $\mathbb{C}^{4}$. We focus on $(\lambda_{12},\lambda_{23},\lambda_{42})$ whose minimal solid partition looks like
\bea
\adjustbox{valign=c}{\includegraphics[width=13cm]{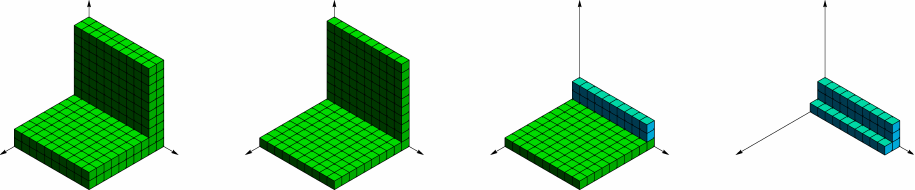}}\cdots 
\eea
To make the discussion concrete, let us focus on the example in the figure with
\bea
\lambda_{12}=(2,1,1),\quad \lambda_{23}=(2,1),\quad \lambda_{42}=(2,1,1).
\eea
The framing node contribution of this actually gives
\bea
\mathcal{Z}^{\D8\tbar\D4\tbar\D0}_{\{\lambda_{A}\}}(\fra,\frb,\phi_{I})&=\frac{\sh(\phi_I-\frb)\sh(\phi_I-\fra-\eps_{1^{2}3 4^{3}})\sh(\phi_I-\fra-\eps_{13^{3}4^{2}})\sh(\phi_I-\fra+2\eps_2+\eps_4)}{\sh(\phi_I-\fra-2\eps_1-2\eps_3)\sh(\phi_I-\fra-2\eps_1-3\eps_4)\sh(\phi_I-\fra-3\eps_3-2\eps_4)\sh(\phi_I-\fra-\eps_{134})}
\eea
where we used $\eps_{1^{a}2^{b}3^{c}4^{d}}=a\eps_1+b\eps_2+c\eps_3+d\eps_4$. Obviously all of the poles correspond to the position of the addable boxes and so after choosing the reference vector to be $\tilde{\eta}_{0}$, no poles are picked up and the PT4 partition function is trivial:
\bea
\mathcal{Z}^{\PT\tbar\JK}_{\{\lambda_{A}\}}[\fq,\mu;q_{1,2,3,4}]=1.
\eea

If we choose the reference vector to be $\eta=\eta_0$ instead, the DT4 partition actually becomes 
\bea
\mathcal{Z}^{\DT\tbar\JK}_{\{\lambda_{A}\}}[\fq,\mu;q_{1,2,3,4}]&=\mathcal{Z}^{\D8}[\fq,\mu;q_{1,2,3,4}].
\eea
One way to see this is to observe that we have the relation
\bea\label{eq:PT4-threesurface-case2-decomp}
\mathcal{Z}^{\D8\tbar\D4\tbar\D0}_{\{\lambda_{A}\}}(\fra,\frb,\phi_{I})&=\mathcal{Z}^{\D8\tbar\D0}(\fra+3\eps_4+2\eps_1,\phi_I)\mathcal{Z}^{\D8'\tbar\D0}(\frb,\phi_I)\\
&\times \mathcal{Z}^{\D6_{\bar{4}}\tbar\D0}(\fra+2\eps_1+2\eps_3,\phi_I)\mathcal{Z}^{\D6_{\bar{4}}\tbar\D0}(\fra+\eps_4+\eps_1+\eps_3,\phi_I)\mathcal{Z}^{\D6_{\bar{4}}\tbar\D0}(\fra+2\eps_4+\eps_1+\eps_3,\phi_I)\\
&\times \mathcal{Z}^{\D6_{\bar{1}}\tbar\D0}(\fra+2\eps_4+3\eps_3,\phi_I)\mathcal{Z}^{\D6_{\bar{1}}\tbar\D0}(\fra+3\eps_4+\eps_1+\eps_3,\phi_I).
\eea
Here we have a system with three D6$_{\bar{4}}$-branes, two D6$_{\bar{1}}$-branes and a pair of D8-D8$'$ brane with the distance $\frb-\fra-3\eps_4-2\eps_1$, which is a tetrahedron instanton system coupled with a pair of D8-branes. The total central charge is 
\bea
(\mu q_{4}^{-3}q_{1}^{-2})\times q_{4}^{3}\times q_{1}^{2}=\mu
\eea
and thus we obtain the magnificent four partition function.

Let us give a more intuitive way to see the decomposition into D6 and D8-branes. The D8-brane is understood as a positive quadrant of a four-dimensional region. In the $(1,3)$ decomposition, the D6-branes are understood as the positive quadrant of three-dimensional spaces in the four-dimensional space where we can stack boxes (D0-branes). In particular, the D6$_{\bar{4}}$-branes are understood as a three-dimensional cube spanning the 123-space:
\bea
\adjustbox{valign=c}{\includegraphics[width=3cm]{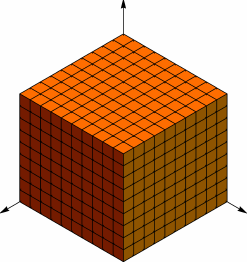}}
\eea
On the other hand, other D6$_{\bar{1,2,3}}$-branes are understood as two-dimensional surfaces extending semi-infinitely in the positive direction of the 4-axis:
\bea
\adjustbox{valign=c}{\includegraphics[width=12cm]{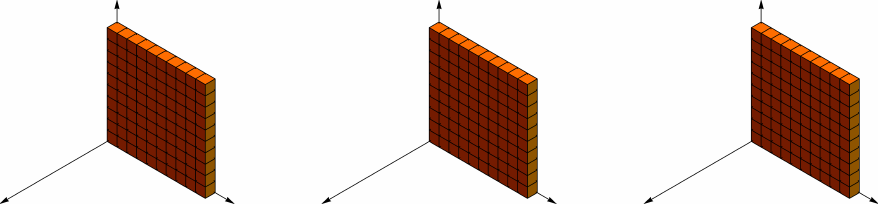}}\cdots
\eea
where this is an example of a D6$_{\bar{1}}$-brane. The smallest coordinate in the orange region represents the Coulomb branch parameter of the D6, D8-branes.

Given a minimal solid partition, the complement of it inside the positive quadrant of the four-dimensional region is the positions where we can add boxes. If such region can be decomposed into collection of D6 and D8-branes, the DT4 partition function is computable using the PE formula \eqref{eq:D8-PEformula}. The information we need to do so is the number of the D6-branes and the distance between the D8-D8$'$ branes.

For the example above, it is obvious that the complement for the first layer can be filled in by a
D6$_{\bar{4}}$-brane whose origin is $\fra+2\eps_1+2\eps_3$. For the second layer, we can use a D6$_{\bar{4}}$-brane with the origin $\fra+\eps_4+\eps_1+\eps_3$. For the higher layers, we need to be careful with the $\lambda_{24}$-surface. For the third layer, a candidate to fill the complement is to first place a D6$_{\bar{1}}$-brane at $\fra+2\eps_4+3\eps_3$ which fills the region $x_{1}=1, x_{2}\geq 1,x_{3}\geq 3, x_{4}\geq 3$.  The remaining regions in the third layer can then be filled by placing a D6$_{\bar{4}}$-brane at $\fra+2\eps_4+\eps_1+\eps_3$. For higher layers, note that the D6$_{\bar{1}}$-placed at the third layer also fills a part of the complement. The remaining regions can be filled by placing another D6$_{\bar{1}}$-brane at $\fra+2\eps_4+\eps_1+\eps_3$ that fills the region $x_{1}=2,x_{2}\geq 1, x_{3}\geq 2,x_{4}\geq 3$ and a D8-brane at $\fra+3\eps_4+2\eps_1$. In this way, we can observe the decomposition \eqref{eq:PT4-threesurface-case2-decomp}.

This phenomenon is actually general for generic $\lambda_{12},\lambda_{23},\lambda_{24}$ and the DT4 partition function is always the rank one magnificent four partition function.

The triviality of the PT4 partition function can be also observed in a more intuitive way. Using the symmetry, let us instead consider the surfaces $\lambda_{14},\lambda_{24},\lambda_{34}$:
\bea
\adjustbox{valign=c}{\includegraphics[width=13cm]{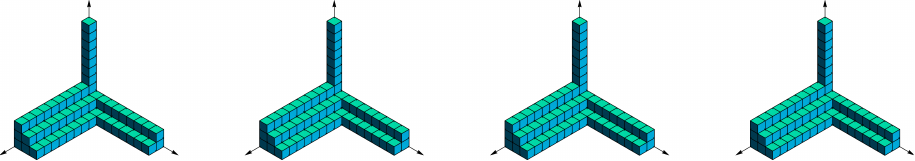}}\cdots
\eea
Each layer is understood as a minimal plane partition with $\lambda_{14},\lambda_{24},\lambda_{34}$ as boundary conditions and they are extending semi-infinitely in the positive direction of the 4-axis. We then use the standard prescription to obtain the PT3 counting for each layer:
\bea
\adjustbox{valign=c}{\includegraphics[width=13cm]{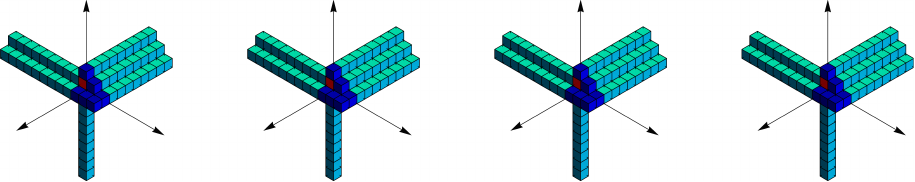}}\cdots
\eea
However, since now the boxes need to be supported also from the positive direction of the 4-axis, we cannot place any boxes in the hollow structure above. Thus, the PT4 partition function must be trivial.

Combining with the previous situation (case 1), we have the following proposition.
\begin{proposition}
    When only three of the surface boundary conditions are nontrivial and all of the intersections are D2-like, the PT4 partition function is trivial:
    \bea
\mathcal{Z}^{\PT\tbar\JK}_{\{\lambda_{A}\}}[\fq,\mu;q_{1,2,3,4}]=1.
\eea

\end{proposition}

\subsubsection{Case 3}
Let us then move on to the case when one of the three intersections is a D0-like intersection. Using the symmetry, we choose the three surfaces $\lambda_{12},\lambda_{13},\lambda_{34}$. The intersection of the 12 and 34 surfaces are the D0-like. For this case, there are situations when the PT4 partition function is nontrivial.

To make the discussion concrete, let us consider the following examples.

\paragraph{Example 1: $\lambda_{12}=1+q_{3}+q_{4},\,\lambda_{13}=1+q_{4},\,\lambda_{34}=1$}The minimal solid partition is
\bea
\adjustbox{valign=c}{\includegraphics[width=13cm]{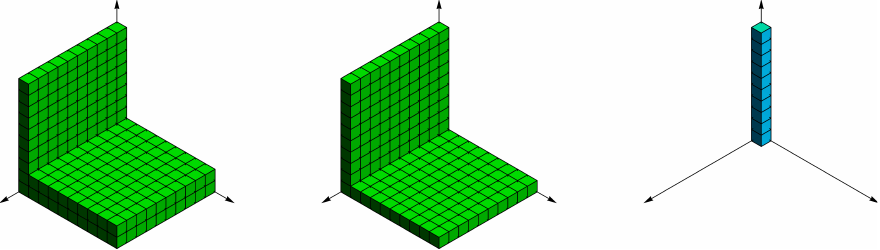}}\cdots
\eea
The framing node contribution is
\bea
\mathcal{Z}^{\D8\tbar\D4\tbar\D0}_{\{\lambda_{A}\}}(\fra,\mathfrak{b},\phi_{I})&=\frac{\sh\left(\phi _I-\frb\right) \sh\left(-\fra-\epsilon _1-\epsilon _2-2 \epsilon _4+\phi _I\right)}{\sh\left(-\fra-\epsilon _2-2 \epsilon
   _3+\phi _I\right) \sh\left(-\fra-\epsilon _1-2 \epsilon _4+\phi _I\right)}\\
   &\times \frac{ \sh\left(-\fra-\epsilon _2-\epsilon
   _3-2 \epsilon _4+\phi _I\right) \sh\left(-\fra-\epsilon _2-2 \epsilon _3-\epsilon _4+\phi _I\right)}{ \sh\left(-\fra-\epsilon _2-2 \epsilon _4+\phi _I\right)
   \sh\left(-\fra-\epsilon _2-\epsilon _3-\epsilon _4+\phi _I\right)}.
\eea
Choosing the reference vector $\eta=\tilde{\eta}_0$, obviously no poles are picked up and so the partition function is trivial
\bea
\mathcal{Z}^{\PT\tbar \JK}_{\{\lambda_{A}\}}[\mathfrak{q},\mu;q_{1,2,3,4}]=1.
\eea

\paragraph{Example 2: $\lambda_{12}=1+q_{4},\, \lambda_{13}=\lambda_{34}=1$} The minimal solid partition is
\bea
\adjustbox{valign=c}{\includegraphics[width=10cm]{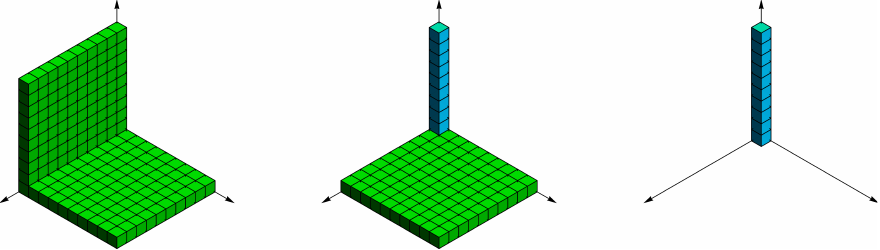}}\cdots
\eea
The framing node contribution is computed as
\bea
\mathcal{Z}^{\D8\tbar\D4\tbar\D0}_{\{\lambda_{A}\}}(\fra,\mathfrak{b},\phi_{I})&=\frac{\sh(\mathfrak{b}-\phi_{I})}{\sh(\fra+\epsilon_{4}-\phi_{I})}\frac{\sh(\phi_{I}-\fra)\sh(\phi_{I}-\fra-\epsilon_{4}+\epsilon_{1,2,3})}{\sh(\phi_{I}-\fra-\epsilon_{134})\sh(\phi_{I}-\fra-\epsilon_{23})\sh(\phi_{I}-\fra-\epsilon_{14^{2}})\sh(\phi_{I}-\fra-\epsilon_{24^{2}})}.
\eea
Choosing the reference vector as $\eta=\tilde{\eta}_0$, the poles picked up are
\bea
\phi_{I}=\fra+\epsilon_{4},\quad \phi_{I}-\phi_{J}=-\epsilon_{1,2,3,4}.
\eea
For the first level, we can take
\bea
\phi_{1}=\fra+\epsilon_{4}.
\eea
For next level, we have
\bea
\phi_{2}=\phi_{1}-\epsilon_{1,2,3,4}=\fra,\fra+\epsilon_{4}-\epsilon_{1,2,3}
\eea
but all of them cancel with the numerator
\bea
\prod_{I=1}^{k}\sh(\phi_{I}-\fra)\sh(\phi_{I}-\fra-\epsilon_{4}+\epsilon_{1,2,3}).
\eea
Therefore, the only nontrivial JK-residue is
\bea
\mathcal{Z}^{\PT\tbar\JK}_{\{\lambda_{A}\}}[1]&=-\frac{\sh(-\epsilon_{14,24,34})}{\sh(-\epsilon_{1,2,3,4})}\underset{\phi_{1}=\fra+\epsilon_{4}}{\Res}\mathcal{Z}^{\D8\tbar\D4\tbar\D0}_{\{\lambda_{A}\}}(\fra,\mathfrak{b},\phi_{1})\\
&=\frac{\sh(\epsilon_{12})\sh(\mathfrak{b}-\fra+\epsilon_{123})}{\sh(\epsilon_{13})\sh(\epsilon_{1}+2\epsilon_{2}+2\epsilon_{3})}=\frac{[q_{123}\mu][q_{12}]}{[q_{13}][q_{1}q_{2}^{2}q_{3}^{2}]},
\eea
which gives
\bea
\mathcal{Z}^{\PT\tbar \JK}_{\{\lambda_{A}\}}[\mathfrak{q},\mu;q_{1,2,3,4}]=1+\mathfrak{q}\frac{[q_{123}\mu][q_{12}]}{[q_{13}][q_{1}q_{2}^{2}q_{3}^{2}]}.
\eea
This indeed matches with \cite[Example 6.18.(2)]{Bae:2024bpx}.



\paragraph{Example 3: $\lambda_{12}=1+q_{3}+q_{4},\,\lambda_{13}=1+q_{4},\,\lambda_{34}=1+q_{2}$}
The minimal solid partition is
\bea
\adjustbox{valign=c}{\includegraphics[width=13cm]{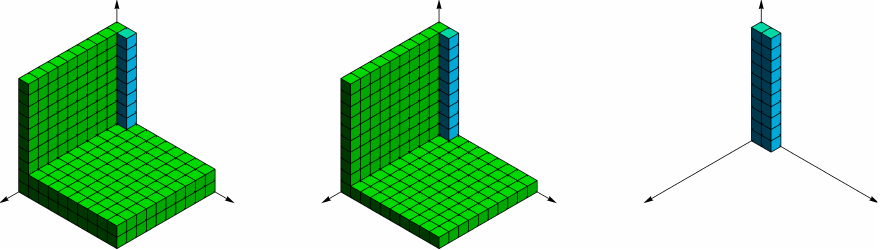}}\cdots
\eea
and the corresponding framing node contribution is
\bea
\mathcal{Z}^{\D8\tbar\D4\tbar\D0}_{\{\lambda_{A}\}}(\fra,\mathfrak{\frb},\phi_{I})&=\frac{\sh\left(\phi_I-\frb\right) \sh\left(-\fra-\epsilon _2+\phi_I\right) \sh\left(-\fra-\epsilon _1-2 \epsilon _2-2 \epsilon _3+\phi
   _I\right)  }{ \sh\left(\fra+\epsilon _2+\epsilon _3-\phi_I\right) \sh\left(\fra+\epsilon
   _2+\epsilon _4-\phi_I\right) \sh\left(-\fra-\eps_{1}-\eps_{2}-\eps_{3}-\eps_{4}+\phi_I\right) }\\
   &\times \frac{\sh\left(-\fra-\epsilon _3+\phi_I\right) \sh\left(-\fra-\epsilon _1-2 \epsilon _2-2 \epsilon _4+\phi_I\right)\sh\left(-\fra-2
   \epsilon _2-\epsilon _3-2 \epsilon _4+\phi_I\right)}{\sh\left(-\fra-2 \epsilon _2-2 \epsilon _3+\phi_I\right) \sh\left(-\fra-\epsilon _1-\epsilon _2-2 \epsilon
   _3+\phi_I\right)\sh\left(-\fra-\epsilon _1-2 \epsilon _4+\phi_I\right)}\\
   &\times \frac{ \sh\left(-\fra-\epsilon _4+\phi_I\right) \sh\left(-\fra-2 \epsilon _2-2 \epsilon
   _3-\epsilon _4+\phi_I\right)}{ \sh\left(-\fra-2 \epsilon _2-2 \epsilon _4+\phi_I\right)
   \sh\left(-\fra-2 \epsilon _2-\epsilon _3-\epsilon _4+\phi_I\right)}.
\eea
Note that the pole at $\phi_I=\fra=\fra+\eps_{1234}$ is understood as the addable pole of the minimal solid partition.

We can perform the JK-residue formalism explicitly as the previous example. Let us only list down the poles giving non-zero JK-residues. At level one, the poles picked up are
\bea
\phi_1=\fra+\eps_{23},\quad \fra+\eps_{24
}.
\eea
For level two, the poles picked up are
\bea
(\phi_1,\phi_2)=(\fra+\eps_{23},\fra+\eps_{24}).
\eea
For level three, we have
\bea
(\phi_1,\phi_2,\phi_3)&=(\fra+\eps_{23},\fra+\eps_{24},\fra+\eps_2).
\eea
For higher levels, the residue vanishes and thus the PT4 partition function terminates at finite terms:
\bea
\mathcal{Z}^{\PT\tbar \JK}_{\{\lambda_{A}\}}[\mathfrak{q},\mu;q_{1,2,3,4}]=1+\fq\left(\mathcal{Z}_{(\eps_{23})}+\mathcal{Z}_{(\eps_{24})}\right)+\fq^{2}\mathcal{Z}_{(\eps_{23},\eps_{24})}+\fq^{3}\mathcal{Z}_{(\eps_{23},\eps_{24},\eps_2)}.
\eea

\paragraph{General case}
The above examples have an intuitive combinatorial structure which is just a generalization of Rule~\ref{rule:PT4twosurface-case2}. The difference is the existence of the 13-surface and that the D0-like intersections coming from the 12 and 34-surfaces may be included in it. For Example 1, the intersections of the 12, 34 surfaces are $\{\fra,\fra+\eps_3,\fra+\eps_4\}$ and all of them also belong to the 13-surface. For Example 2, the intersections of the 12, 34 surfaces are $\{\fra,\fra+\eps_4\}$ and $\fra$ is included in the 13-surface. For Example 3, the D0-like intersections are $\{\fra,\fra+\eps_{3},\fra+\eps_{2},\fra+\eps_{23},\fra+\eps_{4},\fra+\eps_{34},\fra+\eps_{24}\}$ and only $\{\fra+\eps_{23},\fra+\eps_{24},\fra+\eps_{2}\}$ are not included in the 13-surface. Looking at the results of the non-zero JK-residues, one can observe that the D0-like intersections that do not belong to the 13-surface only appears. 

This phenomenon can be understood easily by decomposing the minimal solid partition to an infinite sequence of minimal plane partitions with boundary conditions. We denote the 12,13-surfaces as $\lambda_{12}=(\lambda_{12,l}),\,\,\lambda_{13}=(\lambda_{13,l})$, where again each component corresponds to the number of surfaces at layer $l$. The Young diagram $\lambda_{34}$ is a constant Young diagram for each layer. Due to the existence of the 13-surface, we can construct a Young diagram $\tilde{\lambda}^{(l)}_{34}$ for each layer that do not have any intersections with the 13-surface:
\bea
\tilde{\lambda}^{(l)}_{34}=(\tilde{\lambda}^{(l)}_{34,1},\tilde{\lambda}^{(l)}_{34,2},\ldots,\tilde{\lambda}^{(l)}_{34,i},\ldots),\quad \tilde{\lambda}^{(l)}_{34,i}=\max(\lambda_{34,1}^{(l)}-\lambda_{13,l},0).
\eea
In other words, we are cutting of from the original Young diagram $\lambda_{34}$ the part including the 13-surface. Each layer then can be understood as a minimal plane partition with one-leg boundary condition $\tilde{\lambda}_{34}^{(l)}$ whose origin is $\fra+(l-1)\eps_4+\lambda_{13,l}\eps_2+\lambda_{12,l}\eps_3$. Indeed the framing node contribution obeys
\bea
\mathcal{Z}^{\D8\tbar\D4\tbar\D0}_{\{\lambda_{A}\}}(\fra,\mathfrak{\frb},\phi_{I})=\mathcal{Z}^{\D8'\tbar\D0}(\frb,\phi_I)\prod_{l=1}^{\infty}\mathcal{Z}^{\D6_{\bar{4}}\tbar\D2\tbar\D0}_{\varnothing\varnothing\tilde{\lambda}_{34}^{(l)}}(\fra+(l-1)\eps_4+\lambda_{13,l}\eps_2+\lambda_{12,l}\eps_3,\phi_I).
\eea
We then can use the standard prescription to obtain the PT3 configurations for each layers, but additionally we have the gravity pointing the $(1,1,1,1)$-direction (see Cond.~\ref{cond:PT4}). We extend the minimal plane partition to the negative direction and then remove the positive quadrant boxes and obtain a hollow structure extending in the negative direction. Boxes are stacked in it by obeying the melting rule. 

Similar to Rule~\ref{rule:PT4twosurface-case2}, the PT configurations terminate at some level because for suitable large $l$, we cannot place any boxes at the positions $\fra+(l-1)\eps_{4}-\eps_{3}k$ for $k>0$. Thus, the PT4 partition function terminates.

\begin{Rule}\label{rule:PT4threesurface-case3}
    When only three of the surfaces are nontrivial and one of the three intersections is D0-like, the PT4 box counting rules and the PT4 partition function are given as follows. Without loosing generality, we consider the case when $\lambda_{12,34,13}$ are nontrivial.
\begin{itemize}
    \item We first start from the minimal solid partition. We can define a sequence of Young diagrams $\tilde{\lambda}_{34}^{(l)}$:
    \bea
\tilde{\lambda}^{(l)}_{34}=(\tilde{\lambda}^{(l)}_{34,1},\tilde{\lambda}^{(l)}_{34,2},\ldots,\tilde{\lambda}^{(l)}_{34,i},\ldots),\quad \tilde{\lambda}^{(l)}_{34,i}=\max(\lambda_{34,1}^{(l)}-\lambda_{13,l},0)
    \eea
     In the $(1,3)$-description, each layer is understood as a minimal plane partition with one-leg $\tilde{\lambda}^{(l)}_{34}$ whose origin is shifted by the 12, 13-surfaces to $\fra+(l-1)\eps_4+\lambda_{13,l}\eps_2+\lambda_{12,l}\eps_3$.

    \item Similar to usual PT3 counting, we then extend the one-leg boundary conditions to the negative direction and then remove the boxes at the positive quadrant. Such boxes give the hollow structure in which the boxes are placed.
    \item Boxes are stacked inside the hollow structure in a way such that the gravity is pointing $(1,1,1,1)$. Namely, boxes are supported from the positive directions.
    \item Boxes can be stacked only at the positions
    \bea
    \fra+\eps_{1}(i-1)+\eps_{2}(j-1)+\eps_{3}(k-1)+\eps_{4}(l-1),\quad (i,j)\in\tilde{\lambda}^{(l)}_{34},\,\,1\leq k \leq \lambda_{12,l}.
    \eea

    \item The maximal number of boxes we can add is $\sum_{l=1}^{\infty}|\lambda_{12,l}|\times |\tilde{\lambda}^{(l)}_{34}|$, which is finite and thus the PT4 partition function terminates at this level
    \bea
    \mathcal{Z}^{\PT\tbar \JK}_{\{\lambda_{A}\}}[\mathfrak{q},\mu;q_{1,2,3,4}]=\sum_{k=0}^{N}\fq^{k}\mathcal{Z}^{\PT\tbar \JK}_{\{\lambda_{A}\}}[k],\quad N=\sum_{l=1}^{\infty}|\lambda_{12,l}|\times |\tilde{\lambda}^{(l)}_{34}|.
    \eea
\end{itemize}

\end{Rule}

\section{DT/PT correspondence}\label{sec:DTPTcorrespondence}
The equivariant DT4 vertex and PT4 vertex are related by an overall factor coming from the magnificent four partition function. Such phenomenon is called the DT/PT correspondence \cite{Pandharipande:2007sq,Pandharipande:2007kc,Toda:2008ASPM,Toda:2010JAMS,Bridgeland:2011JAMS,Stoppa:2011BSMF,Toda2016HallAI,Kononov:2019fni,Jenne2020TheCP,Jenne:2021irh,Kuhn:2023koa,Cao:2014bca,Cao:2014bca,Cao:2019tnw,Cao:2019tvv,liu20234foldpandharipandethomasvertex,Bae:2024bpx}. 

In the JK-residue formalism, the computations using either $\eta=\eta_0$ or $\eta=\tilde{\eta}_0$ are related with each other. We have confirmed them for various examples up to three-instantons. 
\subsection{Rank one DT/PT correspondence}
The DT/PT correspondence was already observed when the PT partition functions are trivial in the one, two, three surfaces cases (see for example \eqref{eq:twosurface-DTPT}). Even when the partition function is nontrivial, for both the leg and surface boundary conditions, we have the following DT/PT correspondence in our notations.

\begin{conjecture}
Let $\pi_{1,2,3,4}$ be plane partitions and then we have
    \bea
    \mathcal{Z}^{\DT\tbar\JK}_{\pi_1\pi_2\pi_3\pi_4}[\fq,\mu;q_{1,2,3,4}]=\text{MF}[\mu]\,
    \mathcal{Z}^{\PT\tbar\JK}_{\pi_1\pi_2\pi_3\pi_4}[\fq,\mu;q_{1,2,3,4}].
    \eea
\end{conjecture}

\begin{conjecture}
    Let $\{\lambda_{A}\}_{A\in\six}$ be Young diagrams and then we have
    \bea
\mathcal{Z}^{\DT\tbar\JK}_{\{\lambda_{A}\}}[\fq,\mu;q_{1,2,3,4}]=\text{MF}[\mu]\,
    \mathcal{Z}^{\PT\tbar\JK}_{\{\lambda_{A}\}}[\fq,\mu;q_{1,2,3,4}]
    \eea
\end{conjecture}

We note that although we do not discuss in this paper, one can take specific limits of $\mu$ to decouple the parameter $\mu$ and we still have the DT/PT correspondence. In this paper, we always keep the parameter $\mu$.

\paragraph{Wall-crossing}
From the JK-residue viewpoint, the fact that the result depends on the reference vector is a consequence of the wall-crossing phenomenon \cite{Hori:2014tda}. In the contour integral formalism, residues at the asymptotics are the origin of it. We leave a detailed discussion for future work but let us briefly discuss it for the one-leg one box case at one instanton level. See \cite{Kimura-Noshita-PT3} for a similar discussion for the PT3 case.

Consider the integral
\bea
&\left(\frac{\sh(-\epsilon_{14,24,34})}{\sh(-\epsilon_{1,2,3,4})}\right)\oint_{\mathcal{C}} \frac{d\phi}{2\pi  i}\mathcal{Z}_{\varnothing\varnothing\varnothing\,\scube}^{\D8\tbar\D2\tbar\D0}(\fra,\mathfrak{b},\phi)\\
=&\frac{(1-q_{12})(1-q_{13})(1-q_{23})}{(1-q_{1})(1-q_{2})(1-q_{3})(1-q_{123})}\oint_{\mathcal{C}}\frac{dz}{2\pi i z}\mu^{1/2}\frac{(z-\mu^{-1})(z-q_{12,13,23})}{(z-q_{4}^{-1})(z-q_{1,2,3})}\\
\eqqcolon &\oint_{\mathcal{C}} \frac{dz}{2\pi i z} g(z)
\eea
where in the second line we used the multiplicative notation and $e^{\phi-\fra}=z,e^{\frb-\fra}=\mu$ and $\mathcal{C}$ is some contour traversed counter-clockwise. Since we have the condition $q_{1}q_{2}q_{3}q_{4}=1$, we need to be careful with the analytic region and at least one of the parameters need to obey $|q_{a}|>1$. Let us assume that the parameters obey $|q_{1,2,3}|<|q_{4}^{-1}|<1$ and that the contour\footnote{One may also take the contour as $|z|=1$, and the discussion is the same. The difference is that some of the poles for the DT4 (PT4) side might not be included in the region $|z|<1$ ($|z|>1$) respectively, and thus the right hand side of equations \eqref{eq:wallcrossing-DT4side}, \eqref{eq:wallcrossing-PT4side} cannot be written as $\oint_{\eta_{0}}\frac{dz}{2\pi i z}g(z)$ $\oint_{\tilde{\eta}_{0}}\frac{dz}{2\pi i z}g(z)$ anymore. Even so, \eqref{eq:wallcrossing-DTPT} still holds and we can obtain the DT/PT correspondence for level one.
} $\mathcal{C}$ contains $q_{1,2,3}$ but not $q_{4}^{-1}$. 

Taking the poles inside the region surrounded by $\mathcal{C}$ gives
\bea\label{eq:wallcrossing-DT4side}
\oint_{\mathcal{C}} \frac{dz}{2\pi i z} g(z)&=\underset{z=0}{\Res}\left[g(z)\frac{dz}{z}\right]+\underset{z=q_{1}}{\Res}\left[g(z)\frac{dz}{z}\right]+\underset{z=q_{2}}{\Res}\left[g(z)\frac{dz}{z}\right]+\underset{z=q_{3}}{\Res}\left[g(z)\frac{dz}{z}\right]\\
&=\underset{z=0}{\Res}\left[g(z)\frac{dz}{z}\right]+\oint_{\eta_0}\frac{dz}{2\pi i z}g(z).
\eea
Picking the poles outside the region instead gives
\bea\label{eq:wallcrossing-PT4side}
\oint_{\mathcal{C}} \frac{dz}{2\pi i z} g(z)&=-\underset{z=\infty}{\Res}\left[g(z)\frac{dz}{z}\right]-\underset{z=q_{4}^{-1}}{\Res}\left[g(z)\frac{dz}{z}\right]\\
&=-\underset{z=\infty}{\Res}\left[g(z)\frac{dz}{z}\right]+\oint_{\tilde{\eta}_0}\frac{dz}{2\pi i z}g(z).
\eea
Since the residue is the same, we have 
\bea\label{eq:wallcrossing-DTPT}
\oint_{{\eta}_0}\frac{dz}{2\pi i z}g(z)-\oint_{\tilde{\eta}_0}\frac{dz}{2\pi i z}g(z)&=-\underset{z=\infty}{\Res}\left[g(z)\frac{dz}{z}\right]-\underset{z=0}{\Res}\left[g(z)\frac{dz}{z}\right]\\
&=\lim_{z\rightarrow \infty}g(z)-\lim_{z\rightarrow 0}g(z)\\
&=\frac{[q_{14}][q_{24}][q_{34}][\mu]}{[q_{1}][q_{2}][q_{3}][q_{4}]}
\eea
which indeed matches with the one-instanton contribution of the magnificent four.

One can perform this computation and study the asymptotics of the framing node contributions for the leg and surface boundary conditions generally and can confirm that at one-instanton level we always have the DT/PT correspondence.

\subsection{Rank \texorpdfstring{$n$}{n} DT/PT correspondence}
Let us move on to higher rank generalizations, where we have multiple pairs of D8-D8$'$ branes in the setup. We assume that we have $n$ pairs of D8-D8$'$ branes. 

Let us start from the leg boundary conditions first. For each pair of D8-D8$'$ branes, we can associate a set of plane partitions to each leg, which we denote as $\vec{\pi}_{a}=(\pi_{a}^{(\alpha)})\,(a\in\four)$. The framing node contribution is
\bea
\prod_{\alpha=1}^{n}\mathcal{Z}^{\D8\tbar\D2\tbar\D0}_{\pi_{1}^{(\alpha)}\pi_{2}^{(\alpha)}\pi_{3}^{(\alpha)}\pi_{4}^{(\alpha)}}(\fra_{\alpha},\frb_{\alpha},\phi_{I}).
\eea
The DT partition function is then defined as
\bea
\mathcal{Z}^{\DT\tbar\JK}_{\vec{\pi}_1\vec{\pi}_2\vec{\pi}_3\vec{\pi}_4}[\fq,q_{1,2,3,4}]=\sum_{k=0}^{\infty}\fq^{k}\mathcal{Z}^{\DT\tbar\JK}_{\vec{\pi}_1\vec{\pi}_2\vec{\pi}_3\vec{\pi}_4}[k]
\eea
where
\bea
\mathcal{Z}^{\DT\tbar\JK}_{\vec{\pi}_1\vec{\pi}_2\vec{\pi}_3\vec{\pi}_4}[k]&=\frac{1}{k!}\left(\frac{\sh(-\epsilon_{14,24,34})}{\sh(-\epsilon_{1,2,3,4})}\right)^{k}\oint_{\eta_{0}} \prod_{I=1}^{k}\frac{d\phi_{I}}{2\pi i}\prod_{I=1}^{k}\prod_{\alpha=1}^{n}\mathcal{Z}^{\D8\tbar\D2\tbar\D0}_{\pi_{1}^{(\alpha)}\pi_{2}^{(\alpha)}\pi_{3}^{(\alpha)}\pi_{4}^{(\alpha)}}(\fra_{\alpha},\frb_{\alpha},\phi_{I})\prod_{I<J}^{k}\mathcal{Z}^{\D0\tbar\D0}(\phi_{I},\phi_{J}).
\eea
Choosing the reference vector $\eta=\eta_{0}$, the poles picked up are classified by $n$-tuple of solid partitions with the leg boundary conditions. Generally, it depends on the parameters $\vec{v}=(v_{\alpha})_{\alpha=1}^{n}=(e^{\fra_{\alpha}})_{\alpha=1}^{n}$ and $\vec{w}=(w_{\alpha})_{\alpha=1}^{n}=(e^{\frb_{\alpha}})_{\alpha=1}^{n}$, but we omit the dependence of it.

On the other hand, the PT partition function is defined as
\bea
\mathcal{Z}^{\PT\tbar\JK}_{\vec{\pi}_1\vec{\pi}_2\vec{\pi}_3\vec{\pi}_4}[\fq,q_{1,2,3,4}]=\sum_{k=0}^{\infty}\fq^{k}\mathcal{Z}^{\PT\tbar\JK}_{\vec{\pi}_1\vec{\pi}_2\vec{\pi}_3\vec{\pi}_4}[k]
\eea
where
\bea
\mathcal{Z}^{\PT\tbar\JK}_{\vec{\pi}_1\vec{\pi}_2\vec{\pi}_3\vec{\pi}_4}[k]&=\frac{1}{k!}\left(\frac{\sh(-\epsilon_{14,24,34})}{\sh(-\epsilon_{1,2,3,4})}\right)^{k}\oint_{\tilde{\eta}_{0}} \prod_{I=1}^{k}\frac{d\phi_{I}}{2\pi i}\prod_{I=1}^{k}\prod_{\alpha=1}^{n}\mathcal{Z}^{\D8\tbar\D2\tbar\D0}_{\pi_{1}^{(\alpha)}\pi_{2}^{(\alpha)}\pi_{3}^{(\alpha)}\pi_{4}^{(\alpha)}}(\fra_{\alpha},\frb_{\alpha},\phi_{I})\prod_{I<J}^{k}\mathcal{Z}^{\D0\tbar\D0}(\phi_{I},\phi_{J}).
\eea
Choosing the reference vector $\eta=\tilde{\eta}_0$, the poles picked up are understood as multiple PT4 configurations. When three and four legs are nontrivial, the integrand contains second or third order poles and thus evaluating the residue includes higher derivatives acting on the integrand. Such derivatives also act on the contribution coming from the other D8-branes and so the evaluation will be complicated.

Even for higher rank DT and PT countings, we have the DT/PT correspondence.

\begin{conjecture}
    Let $\vec{\pi}_{a}$ be $n$-tuples of plane partitions for the four legs and $\vec{v},\vec{w}$ be the flavor fugacities of the $n$ D8 and D8$'$-branes. We then have
    \bea    
    \mathcal{Z}^{\DT\tbar\JK}_{\vec{\pi}_1\vec{\pi}_2\vec{\pi}_3\vec{\pi}_4}[\fq,q_{1,2,3,4}]=\text{MF}\left[\prod_{\alpha=1}^{n}\frac{w_{\alpha}}{v_{\alpha}}\right]\mathcal{Z}^{\PT\tbar\JK}_{\vec{\pi}_1\vec{\pi}_2\vec{\pi}_3\vec{\pi}_4}[\fq,q_{1,2,3,4}].
    \eea    
\end{conjecture}

For the surface boundary conditions, we denote the surfaces as $\vec{\lambda}_{A}=(\lambda_A^{(\alpha)})_{\alpha=1,\ldots,n}$. The framing node contribution is
\bea
\prod_{\alpha=1}^{n}\mathcal{Z}^{\D8\tbar\D4\tbar\D0}_{\{\lambda^{(\alpha)}_{A}\}}(\fra_{\alpha},\frb_{\alpha},\phi_{I}).
\eea
The DT partition function is
\bea
\mathcal{Z}^{\DT\tbar\JK}_{\{\vec{\lambda}_A\}}[\fq,q_{1,2,3,4}]=\sum_{k=0}^{\infty}\fq^{k}\mathcal{Z}^{\DT\tbar\JK}_{\{\vec{\lambda}_A\}}[k]
\eea
where
\bea
\mathcal{Z}^{\DT\tbar\JK}_{\{\vec{\lambda}_A\}}[k]&=\frac{1}{k!}\left(\frac{\sh(-\epsilon_{14,24,34})}{\sh(-\epsilon_{1,2,3,4})}\right)^{k}\oint_{\eta_{0}} \prod_{I=1}^{k}\frac{d\phi_{I}}{2\pi i}\prod_{I=1}^{k}\prod_{\alpha=1}^{n}\mathcal{Z}^{\D8\tbar\D4\tbar\D0}_{\{\lambda^{(\alpha)}_{A}\}}(\fra_{\alpha},\frb_{\alpha},\phi_{I})\prod_{I<J}^{k}\mathcal{Z}^{\D0\tbar\D0}(\phi_{I},\phi_{J}).
\eea
The poles are classified by $n$-tuple of solid partitions with surface boundary conditions.

The PT partition function is defined similarly as
\bea
\mathcal{Z}^{\PT\tbar\JK}_{\{\vec{\lambda}_A\}}[\fq,q_{1,2,3,4}]=\sum_{k=0}^{\infty}\fq^{k}\mathcal{Z}^{\PT\tbar\JK}_{\{\vec{\lambda}_A\}}[k]
\eea
where
\bea
\mathcal{Z}^{\PT\tbar\JK}_{\{\vec{\lambda}_A\}}[k]&=\frac{1}{k!}\left(\frac{\sh(-\epsilon_{14,24,34})}{\sh(-\epsilon_{1,2,3,4})}\right)^{k}\oint_{\tilde{\eta}_{0}} \prod_{I=1}^{k}\frac{d\phi_{I}}{2\pi i}\prod_{I=1}^{k}\prod_{\alpha=1}^{n}\mathcal{Z}^{\D8\tbar\D4\tbar\D0}_{\{\lambda^{(\alpha)}_{A}\}}(\fra_{\alpha},\frb_{\alpha},\phi_{I})\prod_{I<J}^{k}\mathcal{Z}^{\D0\tbar\D0}(\phi_{I},\phi_{J}).
\eea
We will obtain $n$-tuples of PT4 counting with surface boundary conditions.

The DT/PT correspondence also holds.
\begin{conjecture}
        Let $\vec{\lambda}_{A}$ be $n$-tuples of plane partitions for the six surfaces and $\vec{v},\vec{w}$ be the flavor fugacities of the $n$ D8 and D8$'$-branes. We then have
    \bea    
    \mathcal{Z}^{\DT\tbar\JK}_{\{\vec{\lambda}_A\}}[\fq,q_{1,2,3,4}]=\text{MF}\left[\prod_{\alpha=1}^{n}\frac{w_{\alpha}}{v_{\alpha}}\right]\mathcal{Z}^{\PT\tbar\JK}_{\{\vec{\lambda}_A\}}[\fq,q_{1,2,3,4}].
    \eea  
\end{conjecture}

\subsection{Introducing anti-fundamental multiplets}
Following the discussion in \cite[section~3.8]{Kimura-Noshita-PT3}, we can also introduce anti-fundamental chiral and Fermi multiplets to the supersymmetric quantum mechanics and compute the Witten index:
\bea
\mathcal{Z}^{\overline{\D8}\tbar\D0}(\fra,\phi_I)=\frac{1}{\sh(\fra-\phi_I)},\quad \mathcal{Z}^{\overline{\D8}'\tbar\D0}(\fra,\phi_I)=\sh(\fra-\phi_I)
\eea
where we denoted the corresponding D8-branes using the notation $\overline{\D8}$, $\overline{\D8}'$. The physical implication of this setup is yet to be studied and we will not discuss it in this paper. The quiver diagram for the 1d $\mathcal{N}=2$ SQM is written as
\bea\label{eq:2SUSYquiver-magnificent-supergroup}
\begin{tikzpicture}[decoration={markings,mark=at position 0.7 with {\arrow{latex}}}]
 \tikzset{
        box/.style={draw, minimum width=0.7cm, minimum height=0.7cm, text centered,thick},
        ->-/.style={decoration={
        markings,mark=at position #1 with {\arrow[scale=1.5]{>}}},postaction={decorate},line width=0.5mm},
        -<-/.style={decoration={
        markings,
        mark=at position #1 with {\arrow[scale=1.5]{<}}},postaction={decorate},line width=0.5mm}    
    }
\begin{scope}[xshift=4cm]
    \draw[fill=black!10!white,thick](0,0) circle(0.4cm);
    \node at (0,0) {$k$};
    \node[box,fill=black!10!white] at (+0.8,1.6) {$n|n$};
    \draw[postaction={decorate},thick] (0.7,1.25)--(0.1,0.4);
    \draw[postaction={decorate},thick,red] (0.9,1.25)--(0.25,0.35);

    \node[box,fill=black!10!white] at (-0.9,1.6) {$m|m$};
     \draw[postaction={decorate},thick] (-0.1,0.4)--(-0.7,1.25);
    \draw[postaction={decorate},thick,red] (-0.25,0.35)--(-0.9,1.25);

    \foreach \ang in {90,145,215,270} {
    \begin{scope}[rotate=\ang]
        \chiralarc[postaction={decorate},thick](0,0.5)(-45:225:0.22:0.65)
    \end{scope}
    }
    \foreach \ang in {90,145,270} {
    \begin{scope}[rotate=\ang]
    \fermiarc[postaction={decorate},thick](0,0.5)(-45:225:0.1:0.5)
    \end{scope}
    \node[left] at (0.3,0.8) {$\mathsf{I}$};
    
    \node[right] at (0.6,0.8) {\textcolor{red}{$\Lambda_{\mathsf{I}}$}};

    \node[right] at (-0.4,0.9) {$\mathsf{J}$};
    
    \node[left] at (-0.6,0.8) {\textcolor{red}{$\Lambda_{\mathsf{J}}$}};

    \node[left] at (-1.5,0) {$\mathsf{B}_{2},\textcolor{red}{\Lambda_{2}}$};
    \node[right] at (1.6,0) {$\mathsf{B}_{1},\textcolor{red}{\Lambda_{1}}$};
    \node[below left] at (-0.9,-1){$\mathsf{B}_{3},\textcolor{red}{\Lambda_{3}}$};
    \node[below right] at (0.9,-1){$\mathsf{B}_{4}$};
    \draw[fill=black!10!white,thick](0,0) circle(0.4cm);
    \node at (0,0) {$k$};
    
    }
\end{scope}
\end{tikzpicture}
\eea

Let us first start from a setup when we do not have the D8-D8$'$ branes, which gives the following framing node contribution
\bea
\mathcal{Z}^{\overline{\D8}\tbar\D0}(\fra,\phi_I)
\mathcal{Z}^{\overline{\D8}'\tbar\D0}(\frb,\phi_I).
\eea
Choosing the reference vector to be $\eta=\eta_{0}$, no poles are picked up and the partition function is trivial. Instead, choosing $\eta=\tilde{\eta}_0$ picks the poles from the hyperplanes
\bea
\fra-\phi_I=0,\quad \phi_I-\phi_J=\eps_{1,2,3,4}.
\eea
Recursively, one can show that the poles are classified by solid partitions, but this time, the coordinates are
\bea
\fra-(i-1)\eps_1-(j-1)\eps_2-(k-1)\eps_3-(l-1)\eps_4.
\eea
Namely, the solid partition extends in the negative direction.

Higher rank generalizations are obtained by including $n$ pairs of D8-D8$'$ branes and $m$ pairs of $\overline{\D8}\tbar\overline{\D8}'$ branes.\footnote{One can also consider different number of D8 ($\overline{\D8}$) and D8$'$ ($\overline{\D8}'$) branes but in this paper we only focus when the numbers of them are the same.} The framing node contribution is 
\bea
\prod_{\alpha=1}^{n}\mathcal{Z}^{\D8\tbar\D0}(\fra_{\alpha},\phi_I)\mathcal{Z}^{\D8'\tbar\D0}(\frb_{\alpha},\phi_I)\prod_{\beta=1}^{m}\mathcal{Z}^{\overline{\D8}\tbar\D0}(\frc_{\beta},\phi_I)\mathcal{Z}^{\overline{\D8}'\tbar\D0}(\frd_{\beta},\phi_I).
\eea
and the partition function is defined as
\bea
\mathcal{Z}^{\D8,+}_{n|m}[\fq,q_{1,2,3,4}]&=\sum_{k=0}^{\infty}\fq^{k}
\mathcal{Z}^{\D8,+}_{n|m}[k]
\eea
where the $k$-instanton sector is
\bea
\mathcal{Z}^{\D8,+}_{n|m}[k]&=\frac{1}{k!}\left(\frac{\sh(-\epsilon_{14,24,34})}{\sh(-\epsilon_{1,2,3,4})}\right)^{k}\oint_{\eta_{0}} \prod_{I=1}^{k}\frac{d\phi_{I}}{2\pi i}\prod_{I=1}^{k}\prod_{\alpha=1}^{n}\mathcal{Z}^{\D8\tbar\D0}(\fra_{\alpha},\phi_I)\mathcal{Z}^{\D8'\tbar\D0}(\frb_{\alpha},\phi_I)\\
&\times\prod_{I=1}^{k}\prod_{\beta=1}^{m}\mathcal{Z}^{\overline{\D8}\tbar\D0}(\frc_{\beta},\phi_I)\mathcal{Z}^{\overline{\D8}'\tbar\D0}(\frd_{\beta},\phi_I)\prod_{I<J}^{k}\mathcal{Z}^{\D0\tbar\D0}(\phi_{I},\phi_{J}).
\eea
The positive sign $+$ denotes the fact that the reference vector is chosen to be $\eta=\eta_0$. For this case, the JK-residue formalism picks the poles coming from the D8-D8$'$-branes and they are classified by multiple solid partitions extending in the positive direction. The $\overline{\D8}\tbar\overline{\D8}'$-branes simply play the roles of extra matter contributions.

We can also define a different partition function
\bea
\mathcal{Z}^{\D8,-}_{n|m}[\fq,q_{1,2,3,4}]&=\sum_{k=0}^{\infty}\fq^{k}
\mathcal{Z}^{\D8,-}_{n|m}[k]
\eea
where the $k$-instanton sector is 
\bea
\mathcal{Z}^{\D8,-}_{n|m}[k]&=\frac{1}{k!}\left(\frac{\sh(-\epsilon_{14,24,34})}{\sh(-\epsilon_{1,2,3,4})}\right)^{k}\oint_{\tilde{\eta}_{0}} \prod_{I=1}^{k}\frac{d\phi_{I}}{2\pi i}\prod_{I=1}^{k}\prod_{\alpha=1}^{n}\mathcal{Z}^{\D8\tbar\D0}(\fra_{\alpha},\phi_I)\mathcal{Z}^{\D8'\tbar\D0}(\frb_{\alpha},\phi_I)\\
&\times\prod_{I=1}^{k}\prod_{\beta=1}^{m}\mathcal{Z}^{\overline{\D8}\tbar\D0}(\frc_{\beta},\phi_I)\mathcal{Z}^{\overline{\D8}'\tbar\D0}(\frd_{\beta},\phi_I)\prod_{I<J}^{k}\mathcal{Z}^{\D0\tbar\D0}(\phi_{I},\phi_{J}).
\eea
The negative sign $-$ denotes the fact that the reference vector is $\eta=\tilde{\eta}_0$. This time, the poles picked up come from the $\overline{\D8}\tbar\overline{\D8}'$-branes and they are classified by multiple solid partitions extending in the negative direction. Note again that for this case, the $\D8\tbar\D8'$-branes play the roles of extra matter contributions.

An interesting property is that we also have a ``DT/PT correspondence" between these two partition functions.

\begin{conjecture}\label{thm:MF-supergroup-relation}
    We have the following identity
    \bea
\mathcal{Z}^{\D8,+}_{n|m}[\fq,q_{1,2,3,4}]
=\text{MF}\left[\prod_{\alpha=1}^{n}\frac{w_{\alpha}}{v_{\alpha}}\prod_{\beta=1}^{m}\frac{y_{\beta}}{x_{\beta}}\right]
\mathcal{Z}^{\D8,-}_{n|m}[\fq,q_{1,2,3,4}]
    \eea
where
\bea
v_{\alpha}=e^{\fra_{\alpha}},\quad w_{\alpha}=e^{\frb_{\alpha}},\quad x_{\beta}=e^{\frc_{\beta}},\quad y_{\beta}=e^{\frd_{\beta}}.
\eea
\end{conjecture}

\begin{corollary}
    When $m=0$, the partition function $\mathcal{Z}^{\D8,-}_{n|m}[\fq,q_{1,2,3,4}]$ is trivial and we have
    \bea
\mathcal{Z}^{\D8,+}_{n|0}[\fq,q_{1,2,3,4}]=\text{MF}\left[\prod_{\alpha=1}^{n}\frac{w_{\alpha}}{v_{\alpha}}\right]
    \eea
    which is the PE-formula of the rank $n$ magnificent four partition function discussed in \cite{Nekrasov:2018xsb}.
\end{corollary}
This comes from the fact that when $m=0$, we do not have any anti-fundamental multiplets and thus choosing the reference vector $\eta=\tilde{\eta}_0$ picks no poles and the partition function is trivial. 
\begin{corollary}
    When $n=0$, the partition function $\mathcal{Z}^{\D8,+}_{n|m}[\fq,q_{1,2,3,4}]$ is trivial and we have
    \bea
\mathcal{Z}^{\D8,-}_{0|m}[\fq,q_{1,2,3,4}]=\text{MF}\left[\prod_{\alpha=1}^{m}\frac{x_{\beta}}{y_{\beta}}\right],
    \eea
    which is also the rank $m$ magnificent four partition function. Note that the ratio of the flavor fugacities of the $\overline{\D8}$-branes are opposite from that of the previous corollary.
\end{corollary}
Similar to the previous case, when $n=0$, we do not have any fundamental multiplets and choosing the reference vector $\eta=\eta_0$ picks no poles and the partition function of the positive part is trivial. From Conj.~\ref{thm:MF-supergroup-relation}, we have
\bea
1=\text{MF}\left[\prod_{\alpha=1}^{m}\frac{y_{\beta}}{x_{\beta}}\right]\mathcal{Z}^{\D8,-}_{0|m}[\fq,q_{1,2,3,4}].
\eea
Using the reflection property \eqref{eq:MF-reflection}, we obtain the claim.

If we tune the parameters as
\bea
\frac{w_{\alpha}}{v_{\alpha}}=q_{4},\quad \frac{y_{\beta}}{x_{\beta}}=q_{4}^{-1}
\eea
we obtain the pure $\U(n|m)$ gauge theory obtained from $n$ D6$_{\bar{4}}$-branes and $m$ $\overline{\D6}_{\bar{4}}$-branes discussed in \cite[section~3.8.3]{Kimura-Noshita-PT3}. Note that for the $\overline{\D8}$-$\overline{\D8}'$ branes, the tuning parameters are the inverse. Instead, tuning the parameters as
\bea
\left\{\left.\frac{w_{\alpha}}{v_{\alpha}}\right| \alpha=1,\ldots,n\right\}\rightarrow \{q_{1},q_{2},q_{3},q_{4}\},\quad \left\{\left.\frac{y_{\beta}}{x_{\beta}}\right| \beta=1,\ldots,m\right\} \rightarrow \{q_{1}^{-1},q_{2}^{-1},q_{3}^{-1},q_{4}^{-1}\},
\eea
we obtain the tetrahedron version of the supergroup theory.

\paragraph{Leg and surface boundary conditions}
We can also perform the same generalization for the leg and surface boundary conditions. We start from the leg boundary conditions. We introduce
\bea
\mathcal{Z}^{\D8\tbar\D2\tbar\D0,+}_{\pi_1\pi_2\pi_3\pi_4}(\fra,\frb,\phi_I)&=
\mathcal{Z}^{\D8\tbar\D2\tbar\D0}_{\pi_1\pi_2\pi_3\pi_4}(\fra,\frb,\phi_I)\\
\mathcal{Z}^{\D8\tbar\D2\tbar\D0,-}_{\pi_1\pi_2\pi_3\pi_4}(\fra,\frb,\phi_I)&=\left.
\mathcal{Z}^{\D8\tbar\D2\tbar\D0}_{\pi_1\pi_2\pi_3\pi_4}(\fra,\frb,\phi_I)\right|_{\substack{\eps_a\rightarrow -\eps_{a}\\ \sh(x)\rightarrow -\sh(-x)}}
\eea
where the negative part is obtained by reversing the sign of the $\eps$-parameters and changing the fundamental/antifundamental multiplets to antifundamental/fundamental multiplets.

We introduce $n$ pairs of D8-D8$'$ branes and $m$ pairs of $\overline{\D8}$-$\overline{\D8}'$ branes and denote the boundary conditions as $(\vec{\pi}_{1,\pm},\vec{\pi}_{2,\pm},\vec{\pi}_{3,\pm},\vec{\pi}_{4,\pm})$, respectively. Let us consider the framing node contribution 
\bea
\prod_{\alpha=1}^{n}\mathcal{Z}^{\D8\tbar\D2\tbar\D0,+}_{\pi_{1,+}^{(\alpha)}\pi_{2,+}^{(\alpha)}\pi_{3,+}^{(\alpha)}\pi_{4,+}^{(\alpha)}}(\fra_{\alpha},\frb_{\alpha},\phi_{I})\prod_{\beta=1}^{m}\mathcal{Z}^{\D8\tbar\D2\tbar\D0,-}_{\pi_{1,-}^{(\beta)}\pi_{2,-}^{(\beta)}\pi_{3,-}^{(\beta)}\pi_{4,-}^{(\beta)}}(\frc_{\beta},\frd_{\beta},\phi_{I}).
\eea

We define the partition function as
\bea
\mathcal{Z}^{+}_{\vec{\pi}_{1,\pm}\vec{\pi}_{2,\pm}\vec{\pi}_{3,\pm}\vec{\pi}_{4,\pm}}[\fq,q_{1,2,3,4}]=\sum_{k=0}^{\infty}\fq^{k}\mathcal{Z}^{+}_{\vec{\pi}_{1,\pm}\vec{\pi}_{2,\pm}\vec{\pi}_{3,\pm}\vec{\pi}_{4,\pm}}[k]
\eea
where
\bea
\mathcal{Z}^{+}_{\vec{\pi}_{1,\pm}\vec{\pi}_{2,\pm}\vec{\pi}_{3,\pm}\vec{\pi}_{4,\pm}}[k]&=\frac{1}{k!}\left(\frac{\sh(-\epsilon_{14,24,34})}{\sh(-\epsilon_{1,2,3,4})}\right)^{k}\oint_{\eta_{0}} \prod_{I=1}^{k}\frac{d\phi_{I}}{2\pi i}\prod_{I=1}^{k}\prod_{\alpha=1}^{n}\mathcal{Z}^{\D8\tbar\D2\tbar\D0,+}_{\pi_{1,+}^{(\alpha)}\pi_{2,+}^{(\alpha)}\pi_{3,+}^{(\alpha)}\pi_{4,+}^{(\alpha)}}(\fra_{\alpha},\frb_{\alpha},\phi_{I})\\
&\times\prod_{I=1}^{k}\prod_{\beta=1}^{m}\mathcal{Z}^{\D8\tbar\D2\tbar\D0,-}_{\pi_{1,-}^{(\beta)}\pi_{2,-}^{(\beta)}\pi_{3,-}^{(\beta)}\pi_{4,-}^{(\beta)}}(\frc_{\beta},\frd_{\beta},\phi_{I})\prod_{I<J}^{k}\mathcal{Z}^{\D0\tbar\D0}(\phi_{I},\phi_{J}).
\eea
Choosing the reference vector $\eta=\eta_0$, the poles picked up from the D8-D8$'$ part is the rank $n$ DT4 counting with $(\vec{\pi}_{1,+},\vec{\pi}_{2,+},\vec{\pi}_{3,+},\vec{\pi}_{4,+})$, while the poles picked up from the $\overline{\D8}\tbar\overline{\D8}'$ part is the PT4 counting with $(\vec{\pi}_{1,-},\vec{\pi}_{2,-},\vec{\pi}_{3,-},\vec{\pi}_{4,-})$. Strictly speaking, the PT4 counting appearing here is the PT4 counting with the coordinates $\eps_a\rightarrow -\eps_a$ compared to the discussion in the previous sections.\footnote{In \cite{Kimura-Noshita-PT3}, this was called the conjugate PT counting.} The arising partition function is a mixed of DT4 and PT4 countings and so we call this the (DT|PT) 4-vertex or the (DT|PT) partition function following \cite{Kimura-Noshita-PT3}.

We can define a different partition function with the reference vector $\tilde{\eta}_0$ as
\bea
\mathcal{Z}^{-}_{\vec{\pi}_{1,\pm}\vec{\pi}_{2,\pm}\vec{\pi}_{3,\pm}\vec{\pi}_{4,\pm}}[\fq,q_{1,2,3,4}]=\sum_{k=0}^{\infty}\fq^{k}\mathcal{Z}^{-}_{\vec{\pi}_{1,\pm}\vec{\pi}_{2,\pm}\vec{\pi}_{3,\pm}\vec{\pi}_{4,\pm}}[k]
\eea
where
\bea
\mathcal{Z}^{-}_{\vec{\pi}_{1,\pm}\vec{\pi}_{2,\pm}\vec{\pi}_{3,\pm}\vec{\pi}_{4,\pm}}[k]&=\frac{1}{k!}\left(\frac{\sh(-\epsilon_{14,24,34})}{\sh(-\epsilon_{1,2,3,4})}\right)^{k}\oint_{\tilde{\eta}_{0}} \prod_{I=1}^{k}\frac{d\phi_{I}}{2\pi i}\prod_{I=1}^{k}\prod_{\alpha=1}^{n}\mathcal{Z}^{\D8\tbar\D2\tbar\D0,+}_{\pi_{1,+}^{(\alpha)}\pi_{2,+}^{(\alpha)}\pi_{3,+}^{(\alpha)}\pi_{4,+}^{(\alpha)}}(\fra_{\alpha},\frb_{\alpha},\phi_{I})\\
&\times\prod_{I=1}^{k}\prod_{\beta=1}^{m}\mathcal{Z}^{\D8\tbar\D2\tbar\D0,-}_{\pi_{1,-}^{(\beta)}\pi_{2,-}^{(\beta)}\pi_{3,-}^{(\beta)}\pi_{4,-}^{(\beta)}}(\frc_{\beta},\frd_{\beta},\phi_{I})\prod_{I<J}^{k}\mathcal{Z}^{\D0\tbar\D0}(\phi_{I},\phi_{J}).
\eea
This time, the poles picked up from the D8-D8$'$ part is the rank $n$ PT4 counting with $(\vec{\pi}_{1,+},\vec{\pi}_{2,+},\vec{\pi}_{3,+},\vec{\pi}_{4,+})$, while the poles picked up from the $\overline{\D8}\tbar\overline{\D8}'$ part is the DT4 counting with $(\vec{\pi}_{1,-},\vec{\pi}_{2,-},\vec{\pi}_{3,-},\vec{\pi}_{4,-})$. Strictly speaking the DT4 counting here is the one whose coordinates of boxes are reversed by $\eps_a\rightarrow -\eps_a$. We call this the (PT|DT) 4-vertex or the (PT|DT) partition function.

We also have the DT/PT correspondence for these vertices.
\begin{conjecture}
The DT/PT correspondence of the (DT|PT) and (PT|DT) vertices is
\bea
\mathcal{Z}^{+}_{\vec{\pi}_{1,\pm}\vec{\pi}_{2,\pm}\vec{\pi}_{3,\pm}\vec{\pi}_{4,\pm}}[\fq,q_{1,2,3,4}]=\text{MF}\left[\prod_{\alpha=1}^{n}\frac{w_{\alpha}}{v_{\alpha}}\prod_{\beta=1}^{m}\frac{y_{\beta}}{x_{\beta}}\right]\mathcal{Z}^{-}_{\vec{\pi}_{1,\pm}\vec{\pi}_{2,\pm}\vec{\pi}_{3,\pm}\vec{\pi}_{4,\pm}}[\fq,q_{1,2,3,4}]
\eea
where
\bea
v_{\alpha}=e^{\fra_{\alpha}},\quad w_{\alpha}=e^{\frb_{\alpha}},\quad x_{\beta}=e^{\frc_{\beta}},\quad y_{\beta}=e^{\frd_{\beta}}.
\eea
\end{conjecture}

For the surface boundary conditions, we similarly introduce
\bea
\mathcal{Z}^{\D8\tbar\D4\tbar\D0,+}_{\{\lambda_A\}}(\fra,\frb,\phi_{I})&=
\mathcal{Z}^{\D8\tbar\D4\tbar\D0}_{\{\lambda_A\}}(\fra,\frb,\phi_{I}),\\
\mathcal{Z}^{\D8\tbar\D4\tbar\D0,-}_{\{\lambda_A\}}(\fra,\frb,\phi_{I})&=\left.
\mathcal{Z}^{\D8\tbar\D4\tbar\D0}_{\{\lambda_A\}}(\fra,\frb,\phi_{I})\right|_{\substack{\eps_a\rightarrow -\eps_a\\\sh(x)\rightarrow -\sh(-x)} }.
\eea
Denoting the surface boundary conditions for the $n$ pairs of the D8-D8$'$ branes and $m$ pairs of the $\overline{\D8}\tbar\overline{\D8}'$ branes as $\vec{\lambda}_{A,\pm}$, the framing node contribution is
\bea
\prod_{\alpha=1}^{n}\mathcal{Z}^{\D8\tbar\D4\tbar\D0,+}_{\{\lambda^{(\alpha)}_{A,+}\}}(\fra_{\alpha},\frb_{\alpha},\phi_{I})\prod_{\beta=1}^{m}\mathcal{Z}^{\D8\tbar\D4\tbar\D0,-}_{\{\lambda^{(\beta)}_{A,-}\}}(\frc_{\alpha},\frd_{\alpha},\phi_{I}).
\eea
We can define the (DT|PT) vertex
\bea
\mathcal{Z}^{+}_{\{\vec{\lambda}_{A,\pm}\}}[\fq,q_{1,2,3,4}]=\sum_{k=0}^{\infty}\fq^{k}\mathcal{Z}^{+}_{\{\vec{\lambda}_{A,\pm}\}}[k]
\eea
where
\bea
\mathcal{Z}^{+}_{\{\vec{\lambda}_{A,\pm}\}}[k]&=\frac{1}{k!}\left(\frac{\sh(-\epsilon_{14,24,34})}{\sh(-\epsilon_{1,2,3,4})}\right)^{k}\oint_{\eta_{0}} \prod_{I=1}^{k}\frac{d\phi_{I}}{2\pi i}\prod_{I=1}^{k}\prod_{\alpha=1}^{n}\mathcal{Z}^{\D8\tbar\D4\tbar\D0,+}_{\{\lambda^{(\alpha)}_{A,+}\}}(\fra_{\alpha},\frb_{\alpha},\phi_{I})\\
&\times\prod_{\beta=1}^{m}\mathcal{Z}^{\D8\tbar\D4\tbar\D0,-}_{\{\lambda^{(\beta)}_{A,-}\}}(\frc_{\alpha},\frd_{\alpha},\phi_{I})\prod_{I<J}^{k}\mathcal{Z}^{\D0\tbar\D0}(\phi_{I},\phi_{J}).
\eea
The poles picked up from the D8-D8$'$ part is the DT4 configurations with surface boundary conditions $\{\vec{\lambda}_{A,+}\}$, while the poles picked up from the $\overline{\D8}\tbar\overline{\D8}'$ part is the PT4 configurations with $\{\vec{\lambda}_{A,-}\}$.

Similarly, the (PT|DT) vertex is defined as
\bea
\mathcal{Z}^{-}_{\{\vec{\lambda}_{A,\pm}\}}[\fq,q_{1,2,3,4}]=\sum_{k=0}^{\infty}\fq^{k}\mathcal{Z}^{-}_{\{\vec{\lambda}_{A,\pm}\}}[k]
\eea
where
\bea
\mathcal{Z}^{-}_{\{\vec{\lambda}_{A,\pm}\}}[k]&=\frac{1}{k!}\left(\frac{\sh(-\epsilon_{14,24,34})}{\sh(-\epsilon_{1,2,3,4})}\right)^{k}\oint_{\tilde{\eta}_{0}} \prod_{I=1}^{k}\frac{d\phi_{I}}{2\pi i}\prod_{I=1}^{k}\prod_{\alpha=1}^{n}\mathcal{Z}^{\D8\tbar\D4\tbar\D0,+}_{\{\lambda^{(\alpha)}_{A,+}\}}(\fra_{\alpha},\frb_{\alpha},\phi_{I})\\
&\times\prod_{\beta=1}^{m}\mathcal{Z}^{\D8\tbar\D4\tbar\D0,-}_{\{\lambda^{(\beta)}_{A,-}\}}(\frc_{\alpha},\frd_{\alpha},\phi_{I})\prod_{I<J}^{k}\mathcal{Z}^{\D0\tbar\D0}(\phi_{I},\phi_{J}).
\eea
The poles picked up from the D8-D8$'$ part gives the PT4 counting, while the $\overline{\D8}\tbar\overline{\D8}'$ part gives the DT4 counting.

\begin{conjecture}
The DT/PT correspondence of the (DT|PT) and (PT|DT) vertices is
\bea
\mathcal{Z}^{+}_{\{\vec{\lambda}_{A,\pm}\}}[\fq,q_{1,2,3,4}]=\text{MF}\left[\prod_{\alpha=1}^{n}\frac{w_{\alpha}}{v_{\alpha}}\prod_{\beta=1}^{m}\frac{y_{\beta}}{x_{\beta}}\right]\mathcal{Z}^{-}_{\{\vec{\lambda}_{A,\pm}\}}[\fq,q_{1,2,3,4}]
\eea
where
\bea
v_{\alpha}=e^{\fra_{\alpha}},\quad w_{\alpha}=e^{\frb_{\alpha}},\quad x_{\beta}=e^{\frc_{\beta}},\quad y_{\beta}=e^{\frd_{\beta}}.
\eea
\end{conjecture}

\section{Conclusion and discussion}\label{sec:conclusion}
We gave a way to determine the framing node contribution and choose the contour to evaluate the partition function. We proposed that choosing the reference vector $\eta=\eta_0$ gives the DT-side while choosing $\eta=\tilde{\eta}_0$ gives the PT-side and have shown them explicitly for various examples for both the leg and surface boundary conditions. Although the framing node contribution we used was enough to perform practical computations, finding a combinatorial formula as the one given in~\cite{Kimura-Noshita-PT3} will be helpful for future use. How to determine the $J,E$-terms physically is also left for future work.

When we have only one-leg or two-legs, the poles appearing from the JK-residue formalism are always first order poles. For the three-legs case, second order poles appear similar to the PT3 case. For the four-legs case, third order poles appear and the computation is much more complicated. In this paper, we did not give a combinatorial way to understand and perform PT4 counting. Such kind of aspects need to be understood and is left for future work. The JK-residue formalism we showed might help the understanding. Moreover, we hope to produce computations for more examples elsewhere.

For the surface boundary conditions, new phenomena appeared. Depending on how the surfaces are placed, the PT4 vertex sometimes become trivial and sometimes become non-trivial but with only finite terms. The computations indeed reproduced results given in \cite{Bae:2022pif,Bae:2024bpx}. Although in this paper, we only focused on cases when we have one, two, and three-surfaces only, generalizations to four, five, and six surfaces are straightforward. Giving a complete classification and formulas for such cases should not be tedious compared to the leg boundary conditions.

Furthermore, one can combine the leg and surface boundary conditions generally using our formalism, though we omitted it to make the paper concise. For the following example
\bea
\adjustbox{valign=c}{\includegraphics[width=13cm]{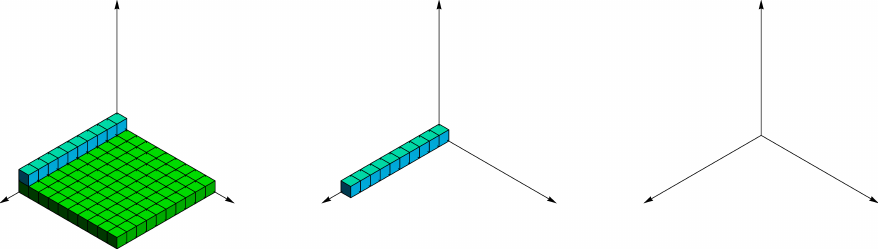}}
\eea
the character (see the notation in section~\ref{sec:DT4vertexnotation}) is 
\bea
\widetilde{\pi}_{1}=\frac{1}{1-q_{2}}+q_{3}+q_{4},\quad \widetilde{\pi}_{2}=\frac{1}{1-q_{1}},\quad \widetilde{\pi}_{3}=\widetilde{\pi}_{4}=\emptyset
\eea
and the framing node contribution is
\bea
\mathcal{Z}^{\D8\tbar\D4\tbar\D2\tbar\D0}_{\{\pi_{a},\lambda_{A}\}_{a\in\four}^{A\in\six}}(\fra,\frb,\phi_I)&= \mathcal{Z}^{\D8\tbar\D0}(\fra,\phi_I)\mathcal{Z}^{\D8'\tbar\D0}(\frb,\phi_I)\mathcal{Z}^{\overline{\D4}_{12}\tbar\D0}(\fra,\phi_I)\mathcal{Z}^{\overline{\D2}_{1}\tbar\D0}(\fra+\eps_3,\phi_I)\mathcal{Z}^{\overline{\D2}_{1}\tbar\D0}(\fra+\eps_4,\phi_I)\\
&=\frac{\sh(\phi_I-\frb)\sh(\phi_I-\fra-\eps_{2}-2\eps_{3})\sh(\phi_I-\fra-\eps_2-2\eps_4)}{\sh(\phi_I-\fra-\eps_{34})\sh(\phi_I-\fra-\eps_{24})\sh(\phi_I-\fra-2\eps_4)\sh(\phi_I-\fra-\eps_{23})}\\
&\times \frac{\sh(\phi_I-\fra-\eps_3-2\eps_4)\sh(\phi_I-\fra-2\eps_3-\eps_4)\sh(\phi_I-\fra+\eps_1)^2}{\sh(\phi_I-\fra-2\eps_{3})\sh(-\phi_I+\fra-\eps_{1}+\eps_{4})\sh(-\phi_I+\fra-\eps_1+\eps_3)}
\eea
Choosing the reference vector $\eta=\eta_{0}$ picks the poles
\bea
\phi_1=\fra+\eps_{34},\,\,\fra+\eps_{24},\,\,\fra+2\eps_{4},\,\,\fra+\eps_{23},\,\,\fra+2\eps_3
\eea
while choosing the reference vector $\eta=\tilde{\eta}_0$ picks the poles
\bea
\phi_1=\fra+\eps_{3}-\eps_{1},\quad \fra+\eps_4-\eps_1
\eea
at one-instanton level. Studying higher levels, boxes will be added on the minimal solid partition for the DT4 side. For the PT4 side, boxes will be placed at the positions $\{\fra+\eps_{3}-i\eps_{1}\}_{i=1}^{\infty}$ and $\{\fra+\eps_{4}-i\eps_{1}\}_{i=1}^{\infty}$ obeying the Cond.~\ref{cond:PT4}. Namely, for each 123-layer, we have a PT3 configuration shifted by the 12-surface. After evaluating the residues, one can also easily confirm that the DT/PT correspondence is satisfied. A complete classification of the PT configurations for general cases when we have both the leg and surface boundary conditions is interesting. The PT vertex for surface boundary conditions coming from the JK-residue formalism here is the PT$_{0}$ vertex in \cite{Bae:2022pif,Bae:2024bpx}. Understanding the derivation of PT$_{1}$ vertex discussed there is also interesting. See also a recent paper \cite{Diaconescu:2025jct}, which fixes all signs for the case of a single surface.

We then discussed the DT/PT correspondence of the PT4 vertices. For low levels, they have been explicitly confirmed. Understanding the wall-crossing phenomenon from the supersymmetric quantum mechanics viewpoint is interesting. For the one-instanton level, it was obvious that the wall-crossing phenomenon occurred due to nontrivial residues at the asymptotics. Studying it for higher levels is also interesting. It was also shown recently in \cite{Cao:2023lon,Piazzalunga:2023qik,Kimura:2024xpr} that the one-leg PT4 vertex appears as vortex partition functions for 3d $\mathcal{N}=2$ gauge theories appearing from D2-branes parallel to a D8-brane. The full PT4 vertex should appear from intersecting D2-branes inside a D8-brane. Studying the vortex partition functions and the wall-crossing of them might be also interesting. The PT4 vertex for surface boundary conditions are expected to appear from a setup with D4-branes inside a D8-brane. Understanding the corresponding partition function and the relation with the PT4 vertex is also interesting.

We also proposed generalizations of the DT4 and PT4 counting by additionally introducing anti-fundamental multiplets. Understanding the physical origin of these anti-fundamental multiplets is also interesting. 

Although, we did not discuss the BPS/CFT correspondence of the PT4 partition functions, one can follow the strategy in \cite{Kimura:2023bxy,Kimura:2024xpr,Kimura:2024osv,Kimura-Noshita-PT3} and study the free field realizations and the associated $qq$-characters. For the three-legs case, the second order pole similar gives derivatives of the operators as \cite{Kimura-Noshita-PT3}. For the four-legs case, the situation is much more complicated and the third order pole introduce higher derivatives. This will be one of the new part for the 4-fold case. Another new part for the 4-fold case is the $qq$-character of the PT4 vertex with surface boundary conditions. Since the partition function terminates in finite terms, the same applies to the $qq$-characters. Understanding the algebraic interpretation of them might be interesting.

\acknowledgments
The authors thank Jiaqun Jiang, Satoshi Nawata, Jiahao Zheng for discussions on the JK-residue formalism.
The work of TK was supported by CNRS through MITI interdisciplinary programs, EIPHI Graduate School (No.~ANR-17-EURE-0002) and Bourgogne-Franche-Comté region. GN is supported by JSPS KAKENHI Grant-in-Aid for JSPS fellows Grant No.~JP25KJ0124.

\section*{Data availability}
The manuscript has no associated data.

\section*{Statements and Declarations}
On behalf of all authors, the corresponding author states that there is no
conflict of interest and data sharing is not applicable to this article as no datasets were generated or analyzed during the current study.

\appendix
\section{JK-residue for degenerate poles}\label{app:JK-residue-degenerate}
In this section, we briefly review the JK-residue formalism when we have degenerate poles \cite{Jeffrey1993LocalizationFN,Szenes2003ToricRA,Brion1999ArrangementOH} (see also \cite{Benini:2013xpa,Benini:2013nda,Hori:2014tda,Hwang:2014uwa}).

 For a degenerate pole, we can associate a set of charge vectors as $Q_{\ast}=\{Q_{1},\ldots,Q_{n}\}$ with $n>k$. We then choose a sequence of $k$ linearly independent charge vectors $Q_{j_{1}},\ldots Q_{j_{k}}$ from $Q_{\ast}$ and construct a flag $F$:
\bea
\{0\}\subset F_{1}\subset \cdots \subset F_{k}=\mathbb{R}^{k},\quad F_{i}=\text{span}\{Q_{j_{1}},\ldots,Q_{j_{k}}\}
\eea
The sequence $Q_{j_{1}},\ldots,Q_{j_{k}}$ is called the basis $\mathcal{B}(F,Q_{\ast})$ of the flag and it is not unique generally, but for a given flag we choose one of them. From the flag $F$ and its basis $\mathcal{B}(F,Q_{\ast})$, a sequence of vectors is defined as
\bea
\kappa(F,Q_{\ast})=(\kappa_1,\ldots,\kappa_{k}),\quad \kappa_{i}=\sum_{\substack{Q\in Q_{\ast}\\ Q\in F_{a}}} Q.
\eea
The JK residue is then given as
\bea\label{eq:JKresidue-iterative}
\underset{\phi=\phi_{\ast}}{\text{JK-}\text{Res}}(Q(\phi_{\ast}),\eta)\mathcal{Z}(\phi)=\sum_{F}\delta(F,\eta)\frac{\text{sign}\det(\kappa(F,Q_{\ast}))}{\det \mathcal{B}(F,Q_{\ast})}\underset{\delta_{k}=0}{\Res}\cdots\underset{\delta_{1}=0}{\Res}\left.\mathcal{Z}(\phi)\right|_{Q_{i}(\phi)+f_{i}(\fra,\eps_a)=\delta_{a}}
\eea
where $\delta(F,\eta)$ is $1$ when the reference vector $\eta$ is included in the closed cone spanned by $\kappa(F,Q_{\ast})$ and $0$ otherwise and the residue is performed in the order $\delta_{1},\ldots,\delta_{k}$. Note also that the sum is taken over all possible flags. One can also confirm that the iterative residue given here reduces back to the definition for the non-degenerate poles discussed above.

\bibliographystyle{utphys}
\bibliography{Worigami}

@article{Diaconescu:2025jct,
    author = "Diaconescu, Duiliu-Emanuel and Piazzalunga, Nicolo",
    title = "{Coulomb branch localization, quasimaps, and surface counting in Calabi--Yau fourfolds}",
    eprint = "2508.11512",
    archivePrefix = "arXiv",
    primaryClass = "math.AG",
    month = "8",
    year = "2025"
}

@article{Noshita:2025bzg,
    author = "Noshita, Go",
    title = "{Gauge Origami and BPS/CFT correspondence}",
    eprint = "2502.07573",
    archivePrefix = "arXiv",
    primaryClass = "hep-th",
    type = "{Doctoral Dissertation}",
school="{University of Tokyo}",
year = "2025"
}

@article{Bao:2024ygr,
    author = "Bao, Jiakang and Seong, Rak-Kyeong and Yamazaki, Masahito",
    title = "{The origin of Calabi-Yau crystals in BPS states counting}",
    eprint = "2401.02792",
    archivePrefix = "arXiv",
    primaryClass = "hep-th",
    doi = "10.1007/JHEP03(2024)140",
    journal = "JHEP",
    volume = "03",
    pages = "140",
    year = "2024"
}

@article{DelZotto:2021gzy,
    author = "Del Zotto, Michele and Nekrasov, Nikita and Piazzalunga, Nicolo' and Zabzine, Maxim",
    title = "{Playing With the Index of M-Theory}",
    eprint = "2103.10271",
    archivePrefix = "arXiv",
    primaryClass = "hep-th",
    doi = "10.1007/s00220-022-04479-7",
    journal = "Commun. Math. Phys.",
    volume = "396",
    number = "2",
    pages = "817--865",
    year = "2022"
}

@article{Kimura-Noshita-PT3,
    author = "Kimura, Taro and Noshita, Go",
    title = "{Gauge origami and quiver W-algebras IV: Pandharipande--Thomas $qq$-characters}",
    eprint = "2508.12125",
    archivePrefix = "arXiv",
    primaryClass = "hep-th",
    month = "8",
    year = "2025"
}

@article{Kimura:2024osv,
    author = "Kimura, Taro and Noshita, Go",
    title = "{Gauge origami and quiver W-algebras III: Donaldson--Thomas $qq$-characters}",
    eprint = "2411.01987",
    archivePrefix = "arXiv",
    primaryClass = "hep-th",
    doi = "10.1007/JHEP03(2025)050",
    journal = "JHEP",
    volume = "03",
    pages = "050",
    year = "2025"
}

@article{Hori:2014tda,
    author = "Hori, Kentaro and Kim, Heeyeon and Yi, Piljin",
    title = "{Witten Index and Wall Crossing}",
    eprint = "1407.2567",
    archivePrefix = "arXiv",
    primaryClass = "hep-th",
    reportNumber = "KIAS-P14039",
    doi = "10.1007/JHEP01(2015)124",
    journal = "JHEP",
    volume = "01",
    pages = "124",
    year = "2015"
}

@article{Cordova:2014oxa,
    author = "Cordova, Clay and Shao, Shu-Heng",
    title = "{An Index Formula for Supersymmetric Quantum Mechanics}",
    eprint = "1406.7853",
    archivePrefix = "arXiv",
    primaryClass = "hep-th",
    doi = "10.5427/jsing.2016.15b",
    journal = "J. Singul.",
    volume = "15",
    pages = "14--35",
    year = "2016"
}

@article{Szenes2003ToricRA,
  title={{Toric reduction and a conjecture of Batyrev and Materov}},
  author={Andr{\'a}s Szenes and Mich{\`e}le Vergne},
  eprint = {math/0306311},
  primaryclass = {math.AT},
  archiveprefix = {arXiv},
  journal={Invent. Math.},
  year={2003},
  volume={158},
  pages={453-495},
  doi = {10.1007/s00222-004-0375-2}
}

@article{Witten:2000mf,
    author = "Witten, Edward",
    title = "{BPS Bound states of D0-D6 and D0-D8 systems in a B field}",
    eprint = "hep-th/0012054",
    archivePrefix = "arXiv",
    doi = "10.1088/1126-6708/2002/04/012",
    journal = "JHEP",
    volume = "04",
    pages = "012",
    year = "2002"
}

@article{Nekrasov:2015wsu,
    author = "Nekrasov, Nikita",
    title = "{BPS/CFT correspondence: non-perturbative Dyson-Schwinger equations and qq-characters}",
    eprint = "1512.05388",
    archivePrefix = "arXiv",
    primaryClass = "hep-th",
    doi = "10.1007/JHEP03(2016)181",
    journal = "JHEP",
    volume = "03",
    pages = "181",
    year = "2016"
}

@article{Nekrasov:2016qym,
    author = "Nekrasov, Nikita",
    title = "{BPS/CFT correspondence II: Instantons at crossroads, moduli and compactness theorem}",
    eprint = "1608.07272",
    archivePrefix = "arXiv",
    primaryClass = "hep-th",
    doi = "10.4310/ATMP.2017.v21.n2.a4",
    journal = "Adv. Theor. Math. Phys.",
    volume = "21",
    pages = "503--583",
    year = "2017"
}

@article{Nekrasov:2016ydq,
    author = "Nekrasov, Nikita",
    title = "{BPS/CFT Correspondence III: Gauge Origami partition function and qq-characters}",
    eprint = "1701.00189",
    archivePrefix = "arXiv",
    primaryClass = "hep-th",
    doi = "10.1007/s00220-017-3057-9",
    journal = "Commun. Math. Phys.",
    volume = "358",
    number = "3",
    pages = "863--894",
    year = "2018"
}

@article{Nekrasov:2016gud,
    author = "Nekrasov, Nikita and Prabhakar, Naveen S.",
    title = "{Spiked Instantons from Intersecting D-branes}",
    eprint = "1611.03478",
    archivePrefix = "arXiv",
    primaryClass = "hep-th",
    doi = "10.1016/j.nuclphysb.2016.11.014",
    journal = "Nucl. Phys. B",
    volume = "914",
    pages = "257--300",
    year = "2017"
}

@article{Nekrasov:2017rqy,
    author = "Nekrasov, Nikita",
    title = "{BPS/CFT correspondence IV: sigma models and defects in gauge theory}",
    eprint = "1711.11011",
    archivePrefix = "arXiv",
    primaryClass = "hep-th",
    doi = "10.1007/s11005-018-1115-7",
    journal = "Lett. Math. Phys.",
    volume = "109",
    number = "3",
    pages = "579--622",
    year = "2019"
}

@article{Nekrasov:2017gzb,
    author = "Nekrasov, Nikita",
    title = "{BPS/CFT correspondence V: BPZ and KZ equations from qq-characters}",
    eprint = "1711.11582",
    archivePrefix = "arXiv",
    primaryClass = "hep-th",
    month = "11",
    year = "2017"
}

@article{Nekrasov:2017cih,
    author = "Nekrasov, Nikita",
    title = "{Magnificent four}",
    eprint = "1712.08128",
    archivePrefix = "arXiv",
    primaryClass = "hep-th",
    doi = "10.4310/ATMP.2020.v24.n5.a4",
    journal = "Adv. Theor. Math. Phys.",
    volume = "24",
    number = "5",
    pages = "1171--1202",
    year = "2020"
}

@article{Nekrasov:2018xsb,
    author = "Nekrasov, Nikita and Piazzalunga, Nicol\`o",
    title = "{Magnificent Four with Colors}",
    eprint = "1808.05206",
    archivePrefix = "arXiv",
    primaryClass = "hep-th",
    doi = "10.1007/s00220-019-03426-3",
    journal = "Commun. Math. Phys.",
    volume = "372",
    number = "2",
    pages = "573--597",
    year = "2019"
}

@article{Pomoni:2021hkn,
    author = "Pomoni, Elli and Yan, Wenbin and Zhang, Xinyu",
    title = "{Tetrahedron Instantons}",
    eprint = "2106.11611",
    archivePrefix = "arXiv",
    primaryClass = "hep-th",
    reportNumber = "DESY 21-087, DESY-21-087",
    doi = "10.1007/s00220-022-04376-z",
    journal = "Commun. Math. Phys.",
    volume = "393",
    number = "2",
    pages = "781--838",
    year = "2022"
}

@article{Cao:2023lon,
    author = "Cao, Yalong and Zhao, Gufang",
    title = "{Quasimaps to quivers with potentials}",
    eprint = "2306.01302",
    archivePrefix = "arXiv",
    primaryClass = "math.AG",
    reportNumber = "RIKEN-iTHEMS-Report-23",
    month = "6",
    year = "2023"
}

@article{Pomoni:2023nlf,
    author = "Pomoni, Elli and Yan, Wenbin and Zhang, Xinyu",
    title = "{Probing M-theory with tetrahedron instantons}",
    eprint = "2306.06005",
    archivePrefix = "arXiv",
    primaryClass = "hep-th",
    reportNumber = "DESY-23-073",
    doi = "10.1007/JHEP11(2023)177",
    journal = "JHEP",
    volume = "11",
    pages = "177",
    year = "2023"
}

@article{Fasola:2023ypx,
    author = "Fasola, Nadir and Monavari, Sergej",
    title = "{Tetrahedron instantons in Donaldson-Thomas theory}",
    journal = {Adv. Math.},
    eprint = "2306.07145",
    archivePrefix = "arXiv",
    primaryClass = "math.AG",
    volume = {462},
    ISSN = {0001-8708},
    DOI = {10.1016/j.aim.2024.110099},
    publisher = {Elsevier BV},
    year = {2025},
    month = feb,
    pages = {110099}
}

@article{Nekrasov:2023nai,
    author = "Nekrasov, Nikita and Piazzalunga, Nicol{\`o}",
    title = "{Global Magni4icence, or: 4G Networks}",
    eprint = "2306.12995",
    archivePrefix = "arXiv",
    primaryClass = "hep-th",
    doi = "10.3842/SIGMA.2024.106",
    journal = "SIGMA",
    volume = "20",
    pages = "106",
    year = "2024"
}

@article{Cao:2017swr,
    author = "Cao, Yalong and Kool, Martijn",
    title = "{Zero-dimensional Donaldson\textendash{}Thomas invariants of Calabi\textendash{}Yau 4-folds}",
    eprint = "1712.07347",
    archivePrefix = "arXiv",
    primaryClass = "math.AG",
    doi = "10.1016/j.aim.2018.09.011",
    journal = "Adv. Math.",
    volume = "338",
    pages = "601--648",
    year = "2018"
}

@article{Cao:2019tvv,
    author = "Cao, Yalong and Kool, Martijn and Monavari, Sergej",
    title = "{K-Theoretic DT/PT Correspondence for Toric Calabi\textendash{}Yau 4-Folds}",
    eprint = "1906.07856",
    archivePrefix = "arXiv",
    primaryClass = "math.AG",
    doi = "10.1007/s00220-022-04472-0",
    journal = "Commun. Math. Phys.",
    volume = "396",
    number = "1",
    pages = "225--264",
    year = "2022"
}

@article{Rapcak:2018nsl,
    author = "Rap\v{c}\'ak, Miroslav and Soibelman, Yan and Yang, Yaping and Zhao, Gufang",
    title = "{Cohomological Hall algebras, vertex algebras and instantons}",
    eprint = "1810.10402",
    archivePrefix = "arXiv",
    primaryClass = "math.QA",
    doi = "10.1007/s00220-019-03575-5",
    journal = "Commun. Math. Phys.",
    volume = "376",
    number = "3",
    pages = "1803--1873",
    year = "2019"
}

@article{Rapcak:2020ueh,
    author = "Rapcak, Miroslav and Soibelman, Yan and Yang, Yaping and Zhao, Gufang",
    title = "{Cohomological Hall algebras and perverse coherent sheaves on toric Calabi{\textendash}Yau $3$-folds}",
    eprint = "2007.13365",
    archivePrefix = "arXiv",
    primaryClass = "math.QA",
    doi = "10.4310/CNTP.2023.v17.n4.a2",
    journal = "Commun. Num. Theor. Phys.",
    volume = "17",
    number = "4",
    pages = "847--939",
    year = "2023"
}

@article{Galakhov:2021xum,
    author = "Galakhov, Dmitry and Li, Wei and Yamazaki, Masahito",
    title = "{Shifted quiver Yangians and representations from BPS crystals}",
    eprint = "2106.01230",
    archivePrefix = "arXiv",
    primaryClass = "hep-th",
    doi = "10.1007/JHEP08(2021)146",
    journal = "JHEP",
    volume = "08",
    pages = "146",
    year = "2021"
}

@article{Iqbal:2007ii,
      author         = "Iqbal, Amer and Kozcaz, Can and Vafa, Cumrun",
      title          = "{The Refined topological vertex}",
      journal        = "JHEP",
      volume         = "10",
      year           = "2009",
      pages          = "069",
      doi            = "10.1088/1126-6708/2009/10/069",
      eprint         = "hep-th/0701156",
      archivePrefix  = "arXiv",
      primaryClass   = "hep-th",
      SLACcitation   = "%%CITATION = HEP-TH/0701156;%%"
}

@article{Awata:2008ed,
      author         = "Awata, Hidetoshi and Kanno, Hiroaki",
      title          = "{Refined BPS state counting from Nekrasov's formula and
                        Macdonald functions}",
      journal        = "Int. J. Mod. Phys.",
      volume         = "A24",
      year           = "2009",
      pages          = "2253-2306",
      doi            = "10.1142/S0217751X09043006",
      eprint         = "0805.0191",
      archivePrefix  = "arXiv",
      primaryClass   = "hep-th",
      SLACcitation   = "%%CITATION = ARXIV:0805.0191;%%"
}

@article{Aganagic:2003db,
      author         = "Aganagic, Mina and Klemm, Albrecht and Marino, Marcos and
                        Vafa, Cumrun",
      title          = "{The Topological vertex}",
      journal        = "Commun. Math. Phys.",
      volume         = "254",
      year           = "2005",
      pages          = "425-478",
      doi            = "10.1007/s00220-004-1162-z",
      eprint         = "hep-th/0305132",
      archivePrefix  = "arXiv",
      primaryClass   = "hep-th",
      reportNumber   = "CALT-68-2439, HUTP-03-A032, HU-EP-03-24,
                        CERN-TH-2003-111",
      SLACcitation   = "%%CITATION = HEP-TH/0305132;%%"
}

@article{Okounkov:2003sp,
      author         = "Okounkov, Andrei and Reshetikhin, Nikolai and Vafa,
                        Cumrun",
      title          = "{Quantum Calabi-Yau and classical crystals}",
      journal        = "Prog. Math.",
      volume         = "244",
      year           = "2006",
      pages          = "597",
      doi            = "10.1007/0-8176-4467-9_16",
      eprint         = "hep-th/0309208",
      archivePrefix  = "arXiv",
      primaryClass   = "hep-th",
      reportNumber   = "HUTP-03-A061",
      SLACcitation   = "%%CITATION = HEP-TH/0309208;%%"
}

@article{Kimura:2023bxy,
    author = "Kimura, Taro and Noshita, Go",
    title = "{Gauge origami and quiver W-algebras}",
    eprint = "2310.08545",
    archivePrefix = "arXiv",
    primaryClass = "hep-th",
    doi = "10.1007/JHEP05(2024)208",
    journal = "JHEP",
    volume = "05",
    pages = "208",
    year = "2024"
}

@article{Kimura:2024xpr,
    author = "Kimura, Taro and Noshita, Go",
    title = "{Gauge origami and quiver W-algebras II: Vertex function and beyond quantum $q$-Langlands correspondence}",
    eprint = "2404.17061",
    archivePrefix = "arXiv",
    primaryClass = "hep-th",
    month = "4",
    year = "2024"
}

@article{Billo:2021xzh,
    author = "Bill\`o, M. and Frau, M. and Fucito, F. and Gallot, L. and Lerda, A. and Morales, J. F.",
    title = "{On the D(\textendash{}1)/D7-brane systems}",
    eprint = "2101.01732",
    archivePrefix = "arXiv",
    primaryClass = "hep-th",
    doi = "10.1007/JHEP04(2021)096",
    journal = "JHEP",
    volume = "04",
    pages = "096",
    year = "2021"
}

@article{Borisov:2017GT,
	doi = {10.2140/gt.2017.21.3231},
	year = 2017,  
	publisher = {Mathematical Sciences Publishers},
	volume = {21},
	number = {6},
	pages = {3231--3311},
	author = {Dennis Borisov and Dominic Joyce},
	title = {Virtual fundamental classes for moduli spaces of sheaves on Calabi--Yau four-folds},
    eprint = {1504.00690},
    archiveprefix = {arXiv},
    primaryclass = {math.AG},
	journal = {Geom. Topol.}
}

@article{Oh:2020rnj,
    author = "Oh, Jeongseok and Thomas, Richard P.",
    title = "{Counting sheaves on Calabi\textendash{}Yau 4-folds, I}",
    eprint = "2009.05542",
    archivePrefix = "arXiv",
    primaryClass = "math.AG",
    doi = "10.1215/00127094-2022-0059",
    journal = "Duke Math. J.",
    volume = "172",
    number = "7",
    pages = "1333--1409",
    year = "2023"
}

@article{Piazzalunga:2023qik,
    author = "Piazzalunga, Nicolo",
    title = "{The one-legged K-theoretic vertex of fourfolds from 3d gauge theory}",
    eprint = "2306.12405",
    archivePrefix = "arXiv",
    primaryClass = "hep-th",
    month = "6",
    year = "2023"
}

@article{Szabo:2023ixw,
    author = "Szabo, Richard J. and Tirelli, Michelangelo",
    title = "{Instanton counting and Donaldson{\textendash}Thomas theory on toric Calabi{\textendash}Yau four-orbifolds}",
    eprint = "2301.13069",
    archivePrefix = "arXiv",
    primaryClass = "hep-th",
    reportNumber = "EMPG-23-01",
    doi = "10.4310/ATMP.2023.v27.n6.a2",
    journal = "Adv. Theor. Math. Phys.",
    volume = "27",
    number = "6",
    pages = "1665--1757",
    year = "2023"
}

@article{Benini:2013xpa,
    author = "Benini, Francesco and Eager, Richard and Hori, Kentaro and Tachikawa, Yuji",
    title = "{Elliptic Genera of 2d ${\mathcal{N}}$ = 2 Gauge Theories}",
    eprint = "1308.4896",
    archivePrefix = "arXiv",
    primaryClass = "hep-th",
    reportNumber = "IPMU-13-0146, UT-13-29",
    doi = "10.1007/s00220-014-2210-y",
    journal = "Commun. Math. Phys.",
    volume = "333",
    number = "3",
    pages = "1241--1286",
    year = "2015"
}

@article{Benini:2013nda,
    author = "Benini, Francesco and Eager, Richard and Hori, Kentaro and Tachikawa, Yuji",
    title = "{Elliptic genera of two-dimensional N=2 gauge theories with rank-one gauge groups}",
    eprint = "1305.0533",
    archivePrefix = "arXiv",
    primaryClass = "hep-th",
    reportNumber = "IPMU-13-0082, UT-13-17",
    doi = "10.1007/s11005-013-0673-y",
    journal = "Lett. Math. Phys.",
    volume = "104",
    pages = "465--493",
    year = "2014"
}

@article{Jeffrey1993LocalizationFN,
  title={Localization for nonabelian group actions},
  author={Lisa C. Jeffrey and Frances Kirwan},
  journal={Topology},
  year={1993},
  volume={34},
  pages={291-327},
  doi = {10.1016/0040-9383(94)00028-J}
}

@article{Monavari:2022rtf,
    author = "Monavari, Sergej",
    title = "{Canonical vertex formalism in DT theory of toric Calabi-Yau 4-folds}",
    eprint = "2203.11381",
    archivePrefix = "arXiv",
    primaryClass = "math.AG",
    doi = "10.1016/j.geomphys.2022.104466",
    journal = "J. Geom. Phys.",
    volume = "174",
    pages = "104466",
    year = "2022"
}

@article{Nekrasov:2014nea,
    author = "Nekrasov, Nikita and Okounkov, Andrei",
    title = "{Membranes and sheaves.}",
    eprint = "1404.2323",
    archivePrefix = "arXiv",
    primaryClass = "math.AG",
    doi = "10.14231/AG-2016-015",
    journal = "Algebr.  Geom.",
    volume = "3",
    number = "3",
    pages = "320--369",
    year = "2016"
}

@article{Maulik:2003rzb,
    author = "Maulik, D. and Nekrasov, N. and Okounkov, A. and Pandharipande, R.",
    title = "{Gromov\textendash{}Witten theory and Donaldson\textendash{}Thomas theory, I}",
    eprint = "math/0312059",
    archivePrefix = "arXiv",
    reportNumber = "ITEP-TH-61-03, IHES-M-03-67",
    doi = "10.1112/S0010437X06002302",
    journal = "Compos. Math.",
    volume = "142",
    number = "05",
    pages = "1263--1285",
    year = "2006"
}

@article{Maulik:2004txy,
    author = "Maulik, D. and Nekrasov, N. and Okounkov, A. and Pandharipande, R.",
    title = "{Gromov\textendash{}Witten theory and Donaldson\textendash{}Thomas theory, II}",
    eprint = "math/0406092",
    archivePrefix = "arXiv",
    doi = "10.1112/S0010437X06002314",
    journal = "Compos. Math.",
    volume = "142",
    number = "05",
    pages = "1286--1304",
    year = "2006"
}

@article{Hwang:2014uwa,
    author = "Hwang, Chiung and Kim, Joonho and Kim, Seok and Park, Jaemo",
    title = "{General instanton counting and 5d SCFT}",
    eprint = "1406.6793",
    archivePrefix = "arXiv",
    primaryClass = "hep-th",
    reportNumber = "SNUTP14-006",
    doi = "10.1007/JHEP07(2015)063",
    journal = "JHEP",
    volume = "07",
    pages = "063",
    year = "2015",
    note = "[Addendum: JHEP 04, 094 (2016)]"
}

@article{Pandharipande:2007kc,
    author = "Pandharipande, R. and Thomas, R. P.",
    title = "{Curve counting via stable pairs in the derived category}",
    eprint = "0707.2348",
    archivePrefix = "arXiv",
    primaryClass = "math.AG",
    doi = "10.1007/s00222-009-0203-9",
    journal = "Invent. Math.",
    volume = "178",
    pages = "407--447",
    year = "2009"
}

@article{Pandharipande:2007sq,
    author = "Pandharipande, R. and Thomas, Richard P.",
    title = "{The 3\textendash{}fold vertex via stable pairs}",
    eprint = "0709.3823",
    archivePrefix = "arXiv",
    primaryClass = "math.AG",
    doi = "10.2140/gt.2009.13.1835",
    journal = "Geom. Topol.",
    volume = "13",
    number = "4",
    pages = "1835--1876",
    year = "2009"
}

@article{Jenne2020TheCP,
    author = "Jenne, Helen and Webb, Gautam",
    title = "{The Combinatorial PT-DT Correspondence}",
    journal = {Sémin. Lothar. Comb.},
    volume = "85B",
    pages = "89",
    eprint = "2012.08484",
    archivePrefix = "arXiv",
    primaryClass = "math.CO",
    url = "https://hdl.handle.net/1794/26871",
    year = "2021"
}

@article{Kim:2024vci,
    author = "Kim, Sung-Soo and Li, Xiaobin and Nawata, Satoshi and Yagi, Futoshi",
    title = "{Freezing and BPS jumping}",
    eprint = "2403.12525",
    archivePrefix = "arXiv",
    primaryClass = "hep-th",
    doi = "10.1007/JHEP05(2024)340",
    journal = "JHEP",
    volume = "05",
    pages = "340",
    year = "2024"
}

@article{Nawata:2023aoq,
    author = "Nawata, Satoshi and Pan, Yiwen and Zheng, Jiahao",
    title = "{Class $\mathcal{S}$ theories on $S^2$}",
    eprint = "2310.07965",
    archivePrefix = "arXiv",
    primaryClass = "hep-th",
    doi = "10.1103/PhysRevD.109.105015",
    journal = "Phys. Rev. D",
    volume = "109",
    number = "10",
    pages = "105015",
    year = "2024"
}

@article{Bae:2022pif,
    author = "Bae, Younghan and Kool, Martijn and Park, Hyeonjun",
    title = "{Counting surfaces on Calabi-Yau 4-folds I: Foundations}",
    eprint = "2208.09474",
    archivePrefix = "arXiv",
    primaryClass = "math.AG",
    month = "8",
    year = "2022"
}

@article{Bae:2024bpx,
    author = "Bae, Younghan and Kool, Martijn and Park, Hyeonjun",
    title = "{Counting surfaces on Calabi-Yau 4-folds II: $\mathrm{DT}$-$\mathrm{PT}_0$ correspondence}",
    eprint = "2402.06526",
    archivePrefix = "arXiv",
    primaryClass = "math.AG",
    month = "2",
    year = "2024"
}

@article{Toda2016HallAI,
  title = {{Hall algebras in the derived category and higher-rank DT invariants}},
  ISSN = {2214-2584},
  DOI = {10.14231/ag-2020-008},
  journal = {Algebraic Geometry},
  publisher = {Foundation Compositio Mathematica},
  author = {Toda,  Yukinobu},
  year = {2020},
  pages = {240–262},
  eprint = {1601.07519},
  archiveprefix = {arXiv},
  primaryclass = {math.AG}
}

@article{Brion1999ArrangementOH,
  title={{Arrangement of hyperplanes. I. Rational functions and Jeffrey-Kirwan residue}},
  author={Michel Brion and Mich{\`e}le Vergne},
  journal={Ann. Sci. Éc. Norm. Supér.},
  year={1999},
  volume={32},
  pages={715-741},
  doi={10.1016/S0012-9593(01)80005-7}
}

@article{Cao:2014bca,
    author = "Cao, Yalong and Leung, Naichung Conan",
    title = "{Donaldson-Thomas theory for Calabi-Yau 4-folds}",
    eprint = "1407.7659",
    archivePrefix = "arXiv",
    primaryClass = "math.AG",
    month = "7",
    year = "2014"
}

@article{Cao:2019tnw,
    author = "Cao, Yalong and Kool, Martijn",
    title = "{Curve counting and DT/PT correspondence for Calabi-Yau 4-folds}",
    eprint = "1903.12171",
    archivePrefix = "arXiv",
    primaryClass = "math.AG",
    doi = "10.1016/j.aim.2020.107371",
    journal = "Adv. Math.",
    volume = "375",
    pages = "107371",
    year = "2020"
}

@article{Jenne:2021irh,
    author = "Jenne, Helen and Webb, Gautam and Young, Benjamin",
    title = "{Double-dimer condensation and the PT-DT correspondence}",
    eprint = "2109.11773",
    archivePrefix = "arXiv",
    primaryClass = "math.CO",
    month = "9",
    year = "2021"
}

@article{Toda:2010JAMS,
  title = {{Curve counting theories via stable objects I. DT/PT correspondence}},
  volume = {23},
  ISSN = {1088-6834},
  DOI = {10.1090/s0894-0347-10-00670-3},
  number = {4},
  journal = {J. Amer. Math. Soc.},
  publisher = {American Mathematical Society (AMS)},
  author = {Toda,  Yukinobu},
  year = {2010},
  pages = {1119–1157},
  eprint = {0902.4371},
  primaryclass = {math.AG},
  archiveprefix = {arXiv}
}

@article{Toda:2008ASPM,
  title = {Generating functions of stable pair invariants via wall-crossings in derived categories},
  ISSN = {0920-1971},
  DOI = {10.2969/aspm/05910389},
  booktitle = {New Developments in Algebraic Geometry,  Integrable Systems and Mirror Symmetry (RIMS,  Kyoto,  2008)},
  pages = {389--434},
  publisher = {Mathematical Society of Japan},
  author = {Toda, Yukinobu},
  pages = {389–434},
  journal = {Adv. Stud. Pure Math.},
  eprint = {0806.0062},
  primaryclass = {math.AG},
  archiveprefix = {arXiv}
}

@article{Bridgeland:2011JAMS,
  title = {Hall algebras and curve-counting invariants},
  volume = {24},
  ISSN = {1088-6834},
  DOI = {10.1090/s0894-0347-2011-00701-7},
  number = {4},
  journal = {J. Amer. Math. Soc.},
  publisher = {American Mathematical Society (AMS)},
  author = {Bridgeland,  Tom},
  year = {2011},
  pages = {969–998},
  eprint = {1002.4374},
  primaryclass = {math.AG},
  archiveprefix = {arXiv}
}

@article{Stoppa:2011BSMF,
  title = {{Hilbert schemes and stable pairs: GIT and derived category wall crossings}},
  volume = {139},
  ISSN = {2102-622X},
  DOI = {10.24033/bsmf.2610},
  number = {3},
  journal = {Bull. Soc. Math. Fr.},
  publisher = {Societe Mathematique de France},
  author = {Stoppa,  Jacopo and Thomas,  Richard P.},
  year = {2011},
  pages = {297–339},
  eprint = {0903.1444},
  primaryclass = {math.AG},
  archiveprefix = {arXiv}
}

@article{Kuhn:2023koa,
    author = "Kuhn, Nikolas and Liu, Henry and Thimm, Felix",
    title = "{The 3-fold K-theoretic DT/PT vertex correspondence holds}",
    eprint = "2311.15697",
    archivePrefix = "arXiv",
    primaryClass = "math.AG",
    month = "11",
    year = "2023"
}

@article{Kononov:2019fni,
    author = "Kononov, Ya. and Okounkov, A. and Osinenko, A.",
    title = "{The 2-Leg Vertex in K-theoretic DT Theory}",
    eprint = "1905.01523",
    archivePrefix = "arXiv",
    primaryClass = "math-ph",
    doi = "10.1007/s00220-021-03936-z",
    journal = "Commun. Math. Phys.",
    volume = "382",
    number = "3",
    pages = "1579--1599",
    year = "2021"
}

@article{liu20234foldpandharipandethomasvertex,
      title={{The 4-fold Pandharipande--Thomas vertex}}, 
      author={Henry Liu},
      year={2023},
      eprint={2306.12923},
      archivePrefix={arXiv},
      primaryClass={math.AG}
}

@article{Cao:2019fqq,
    author = "Cao, Yalong and Maulik, Davesh and Toda, Yukinobu",
    title = "{Stable pairs and Gopakumar{\textendash}Vafa type invariants for Calabi{\textendash}Yau 4-folds}",
    eprint = "1902.00003",
    archivePrefix = "arXiv",
    primaryClass = "math.AG",
    doi = "10.4171/jems/1110",
    journal = "J. Eur. Math. Soc.",
    volume = "24",
    number = "2",
    pages = "527--581",
    year = "2021"
}

@phdthesis{Monavari:2022umi,
    author = "Monavari, Sergej",
    title = "{Equivariant Enumerative Geometry and Donaldson-Thomas Theory}",
    doi = "10.33540/1119",
    school = "Utrecht U.",
    year = "2022"
}

@article{Kool:2025qou,
    author = "Kool, M. and Rennemo, J. V.",
    title = "{Proof of a magnificent conjecture}",
    eprint = "2507.02852",
    archivePrefix = "arXiv",
    primaryClass = "math.AG",
    month = "7",
    year = "2025"
}

@article{Thomas:1998uj,
    author = "Thomas, R. P.",
    title = "{A Holomorphic Casson invariant for Calabi-Yau three folds, and bundles on K3 fibrations}",
    eprint = "math/9806111",
    archivePrefix = "arXiv",
    journal = "J. Diff. Geom.",
    volume = "54",
    number = "2",
    pages = "367--438",
    year = "2000"
}
\end{document}